\documentclass[12pt]{article}
\input{amssym.def}
\input{amssym}
\usepackage[dvips]{color}
\usepackage{epsfig}
\usepackage{hyperref}

\normalbaselines
\makeatletter
\@addtoreset{equation}{section}
\makeatother

\def\nn{\nonumber}
\def\be{\begin{equation}}
\def\ee{\end{equation}}
\def\ba{\begin{eqnarray}}
\def\ea{\end{eqnarray}}
\def\lb{\label}
\def\ni{\noindent}

\def\bra#1{\langle #1|}
\def\ket#1{|#1\rangle}

\def\blue#1{\color{blue}{#1}}
\def\red#1{\color{red}{#1}}
\def\green#1{\color{green}{#1}}

\begin{document}

\title{Density profiles in the raise and peel model with and without a wall.
Physics and combinatorics.
}
\author{Francisco C. Alcaraz$^1$, Pavel Pyatov$^2$, and Vladimir Rittenberg$^3$
\\[5mm] {\small\it
$^1$Instituto de F\'{\i}sica de S\~{a}o Carlos, Universidade de S\~{a}o Paulo, Caixa Postal 369, }\\
{\small\it 13560-590, S\~{a}o Carlos, SP, Brazil}\\
{\small\it $^2$Bogoliubov Laboratory of Theoretical Physics, JINR
141980 Dubna, }\\
{\small\it Moscow Region, Russia}\\
{\small\it$^{3}$Physikalisches Institut, Universit\"at Bonn,
  Nussallee 12, 53115 Bonn, Germany}
}
\date{\today}
\maketitle
\footnotetext[1]{\tt alcaraz@if.sc.usp.br}
\footnotetext[2]{\tt pyatov@theor.jinr.ru}
\footnotetext[3]{\tt vladimir@th.physik.uni-bonn.de}

\begin{abstract}
We consider the raise and peel model of a one-dimensional fluctuating
interface in the presence of an attractive wall. The model can also
describe a pair annihilation process in a disordered unquenched media with
a source at one end of the system.
For the stationary states, several density profiles are studied using Monte Carlo
simulations. We point out a deep connection between some profiles seen in the presence
of the wall and in its absence. Our results are discussed  in the
context of conformal invariance ($c = 0$ theory). We discover some unexpected values for
the critical  exponents, which are obtained using combinatorial
methods.
 We have
solved known (Pascal's hexagon) and new (split-hexagon) bilinear recurrence relations.
The solutions of these equations are interesting on their own since they give
information on certain classes of alternating sign matrices.
\end{abstract}

\newpage


\bigskip

\section{Introduction}

The raise and peel model (RPM)  is discussed in several papers \cite{GNPR,ALR}
(for a review see \cite{AR2}). This
is a one-parameter dependent stochastic model of a one-dimensional fluctuating
interface. The RPM can be seen as a model for
wetting in which the adsorption is local (the interface raises) and the
desorption is nonlocal (the interface is peeled). The parameter is given
by the ratio of the adsorption and desorption rates. When the two rates are
equal (the Razumov-Stroganov point) the Hamiltonian which describes the time
evolution of the system is conformal invariant and the stationary states have
remarkable combinatorial properties. The origin of these properties can be
traced back to the definition of the model in terms of generators of  the
Temperley-Lieb (TL) algebra at the semigroup point. This algebra has two
representations which are relevant for us: one in terms of link patterns (which makes the connection
with the interface model) and another one in terms of an integrable XXZ
quantum chain with $L$ sites (which allows one to obtain the spectrum of the
Hamiltonian and its conformal properties).

The fluctuating interface is described by RSOS (Dyck) paths which can be
interpreted as clusters of tiles deposited on a substrate. The heights
at both ends of the system vanish. The clusters touch each other at
contact points. In \cite{G}, a conjecture for the average number
of clusters at the Razumov-Stroganov point in the stationary states was
given. According to this conjecture, the average size of a cluster increases
like $L^{1/3}$ ($L$ is the size of the system, $L$ even). In the present paper we give
combinatorial arguments in favour of this conjecture. We also show that the average size of the cluster
at one of the ends of the system also increases like $L^{1/3}$. This is in
contrast to a model in which the RSOS paths are taken with equal probabilities
when one obtains $L^{1/2}$.

An interesting result obtained in \cite{ALR} using Monte Carlo simulations is
the behavior of the density of contact points in the finite-size scaling
limit. It has the functional form expected from conformal invariance for a
one-point function with an unexpected exponent. Further details on this
observation will be given in the next sections.

In the present paper we extend the RPM at the Razumov-Stroganov point in
two ways. Firstly we consider the case when the size of the system $L$
is an odd number. This implies a change of the configuration space of the
model, one has ballot paths with fixed ends: one at height $1$ the other one
at height zero. We call them one-step Dyck paths. The model is still
based on the TL algebra at the
semigroup point.  We also consider the RPM in the presence of a wall (RPMW). The
model is based on the one-boundary TL algebra \cite{MS,NRG} at the semigroup point.
The configuration space is now given by ballot paths \cite{Sh} with one end
fixed at zero and the other one free, corresponding to the wall. In the bulk
the rules for adsorption and desorption as in the original RPM but at
the wall one has a new adsorption process with an arbitrary fixed rate $a$
(for convenience the bulk rates are taken equal to one).

For both extensions conformal invariance is obeyed and the stationary
states have new combinatorial properties.

There is an alternative interpretation \cite{AR1} of  the RPMW.
One can see the stochastic process as a pair annihilation processes in an
unquenched disordered media (see Appendix A for detailed explanations). In this
description, one has clusters of tiles which either touch
each other or are separated by impurities (defects). There are no
empty sites. The defects can either hop over a cluster peeling its surface
 or they can annihilate in pairs if
the defects are on neighboring sites. The clusters change in shape and number
like in the usual RPM. The rates of all these processes are fixed at the value one.
At one end of the system one has a source of defects which acts with a rate
$a$. The source acts as follows: if a cluster touches the end of the
system, a layer of the cluster is desorbed and a defect is added at
each of the two ends of the peeled cluster. In the stationary states, the
average number of defects corresponds to the average height of the ballot
paths at the wall.

In the "defects" interpretation of the model, the RPM for $L$ odd can be
seen as a system in which one impurity (defect) hops at large distances in
an unquenched disordered media. During the hop, the media is changed (the
cluster over which the defects hops, is peeled).
\medskip

This paper has two distinctive parts: a physical one and a mathematical
one. Physics is in the first part of the paper (Sections 2-4), mathematics
is concentrated in the second part. The two parts can be read independently.

In Section 2 we present the stochastic models describing the RPM models
with  or without a wall. This presentation is suitable for
Monte Carlo simulations. We also define some relevant observables. The
alternative description of the model in terms of defects is given in
Appendix A.

In Section 3 we present without proof some exact results for the
stationary states obtained in the last part of the paper. We give the
expressions for the average number of clusters for any size of the system. In
particular we give the large $L$ behaviour. These expressions are
relevant because they are the integrated quantities of the density of contact
points, the latter being obtained from Monte Carlo simulations. In particular we derive
the values of the critical exponents.

We show a surprising identity between the
probability density function (PDF) to have the first cluster at distance $x$
from the wall  for the RPMW
and the PDF to have defect
at distance $x$ from the boundary for the RPM for an odd number of sites.
As shown in Section 4 this PDF has remarkable properties in the finite-size
 scaling limit.

We also give some conjectures for the  average values of the number of sites
where adsorption can take place. From their values one can derive the
average number of tiles desorbed (desorption takes place through
avalanches \cite{ALR}).

In Section 4 we present the results obtained from Monte Carlo
simulations. Firstly we remind the reader about the observation \cite{ALR} that
in the stationary state of the RPM ($L$ even) the density of contact points,
in the finite-size scaling limit is given by the expression expected from
conformal invariance with an exponent $1/3$ while we should expect an
exponent $2/3$ (the one-point function of an operator with conformal
spin vanishes).

We next consider the presence of the wall in the model with a boundary
rate $a$ and study the density of contact points profile in the
finite-size scaling limit. It is independent of the boundary rate $a$ and has
a functional form suggested by conformal invariance with the same exponent
$1/3$ as  in the absence of the wall.

We also study the density of defects  profile and find that in the finite-size scaling
limit it has the functional dependence expected from conformal invariance with (again!)
an
unexpected exponent.

In the finite-size scaling limit, the probability distribution function to have a cluster at
distance $x/L$ from the wall stays, surprisingly, a probability density function
(its integrated value is equal to one) and is independent of $a$ {\it i.e.} is
universal. The functional dependence of this  PDF
is similar to what is expected from conformal invariance for a one-point function.
No explanation for this observation was found.

In Section 5 we present the derivation of the RPM and RPMW in continuum time
using the Temperley-Lieb algebra
\cite{TL} and its one-boundary extension \cite{MS}
at the semigroup points and make the connection with the spin one-half
XXZ quantum chain. Based on this connection and the integrability of the
chain, one can show that our systems are conformal invariant. This section is a review
of known results.

In Section 6 we present firstly  the bilinear recurrence relations called the Pascal's
hexagon. The relations are not new but the solutions, specified by the boundary
conditions are. We next propose some new bilinear relations that we called the
split-hexagon relations and give their solutions for certain boundary conditions.
This section is pure mathematics.

In Sections 7 and 8 we make the contact with physics and derive the results enumerated
in Section 3. Firstly in Section 7, for the RPM  we derive the probabilities to have
$k$ clusters for a system of size $L$ ($L$ even and odd).

In Section 8 we consider the RPMW. For $L$ odd we show a remarkable connection between
the RPMW  and the RPM for the same value of $L$.
In particular one can show that independently of a value of the boundary rate $a$
all properties of
clusters, except the first, are identical.
 The case of the RPMW with $L$ even and arbitrary values of $a$ is more subtle. Nevertheless
for the case $a =1$ we were able
to derive the probability of having $k$ clusters in a system of size $L$.

Our conclusions are presented in Section 9.

\section{Raise and peel models with and without a wall}

We consider a one-dimensional lattice  with $(L + 1)$ sites. An interface
is formed by attaching at each site non-negative integer heights $h_i$ which
obey the restricted solid-on-solid (RSOS) rules:
\be
\lb{2.1}
h_{i+1} - h_i = \pm 1,\quad h_i\geq 0, \quad  i = 0,1,\dots,L.
\ee
We will consider three kind of interfaces, each with different
configuration spaces depending on the conditions at the boundaries (the
values of $h_0$ and $h_L$):
\begin{itemize}
\item[a)]
Dyck path configurations.\\
One takes $L$ even and $h_0 = h _L = 0$.
There are
\be
\lb{2.2}
C_L = \frac{L!}{(L/2)!(L/2+1)!}
\ee
configurations of this kind. An example of such a configuration for
$L =16$ is shown in Figure\,1.
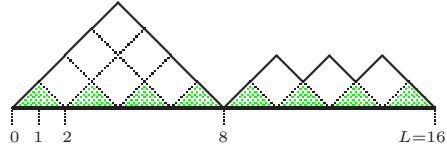
\begin{figure}[t]
\begin{center}
\begin{picture}(200,45)(0,-10)
{\green{
\qbezier[8](3,11)(7,15)(11,19)
\qbezier[7](5,11)(8.5,14.5)(12,18)
\qbezier[6](7,11)(10,14)(13,17)
\qbezier[5](9,11)(11.5,13.5)(14,16)
\qbezier[4](11,11)(13,13)(15,15)
\qbezier[3](13,11)(14.5,12.5)(16,14)
\qbezier[2](15,11)(16,12)(17,13)
\qbezier[1](17,11)(17.5,11.5)(18,12)
\qbezier[8](23,11)(27,15)(31,19)
\qbezier[7](25,11)(28.5,14.5)(32,18)
\qbezier[6](27,11)(30,14)(33,17)
\qbezier[5](29,11)(31.5,13.5)(34,16)
\qbezier[4](31,11)(33,13)(35,15)
\qbezier[3](33,11)(34.5,12.5)(36,14)
\qbezier[2](35,11)(36,12)(37,13)
\qbezier[1](37,11)(37.5,11.5)(38,12)
\qbezier[8](43,11)(47,15)(51,19)
\qbezier[7](45,11)(48.5,14.5)(52,18)
\qbezier[6](47,11)(50,14)(53,17)
\qbezier[5](49,11)(51.5,13.5)(54,16)
\qbezier[4](51,11)(53,13)(55,15)
\qbezier[3](53,11)(54.5,12.5)(56,14)
\qbezier[2](55,11)(56,12)(57,13)
\qbezier[1](57,11)(57.5,11.5)(58,12)
\qbezier[8](63,11)(67,15)(71,19)
\qbezier[7](65,11)(68.5,14.5)(72,18)
\qbezier[6](67,11)(70,14)(73,17)
\qbezier[5](69,11)(71.5,13.5)(74,16)
\qbezier[4](71,11)(73,13)(75,15)
\qbezier[3](73,11)(74.5,12.5)(76,14)
\qbezier[2](75,11)(76,12)(77,13)
\qbezier[1](77,11)(77.5,11.5)(78,12)
\qbezier[8](83,11)(87,15)(91,19)
\qbezier[7](85,11)(88.5,14.5)(92,18)
\qbezier[6](87,11)(90,14)(93,17)
\qbezier[5](89,11)(91.5,13.5)(94,16)
\qbezier[4](91,11)(93,13)(95,15)
\qbezier[3](93,11)(94.5,12.5)(96,14)
\qbezier[2](95,11)(96,12)(97,13)
\qbezier[1](97,11)(97.5,11.5)(98,12)
\qbezier[8](103,11)(107,15)(111,19)
\qbezier[7](105,11)(108.5,14.5)(112,18)
\qbezier[6](107,11)(110,14)(113,17)
\qbezier[5](109,11)(111.5,13.5)(114,16)
\qbezier[4](111,11)(113,13)(115,15)
\qbezier[3](113,11)(114.5,12.5)(116,14)
\qbezier[2](115,11)(116,12)(117,13)
\qbezier[1](117,11)(117.5,11.5)(118,12)
\qbezier[8](123,11)(127,15)(131,19)
\qbezier[7](125,11)(128.5,14.5)(132,18)
\qbezier[6](127,11)(130,14)(133,17)
\qbezier[5](129,11)(131.5,13.5)(134,16)
\qbezier[4](131,11)(133,13)(135,15)
\qbezier[3](133,11)(134.5,12.5)(136,14)
\qbezier[2](135,11)(136,12)(137,13)
\qbezier[1](137,11)(137.5,11.5)(138,12)
\qbezier[8](143,11)(147,15)(151,19)
\qbezier[7](145,11)(148.5,14.5)(152,18)
\qbezier[6](147,11)(150,14)(153,17)
\qbezier[5](149,11)(151.5,13.5)(154,16)
\qbezier[4](151,11)(153,13)(155,15)
\qbezier[3](153,11)(154.5,12.5)(156,14)
\qbezier[2](155,11)(156,12)(157,13)
\qbezier[1](157,11)(157.5,11.5)(158,12)
}}
\qbezier[30](20,10)(35,25)(50,40)
\qbezier[20](40,10)(50,20)(60,30)
\qbezier[10](60,10)(65,15)(70,20)
\qbezier[10](100,10)(105,15)(110,20)
\qbezier[10](120,10)(125,15)(130,20)
\qbezier[10](140,10)(145,15)(150,20)
\qbezier[10](10,20)(15,15)(20,10)
\qbezier[10](90,20)(95,15)(100,10)
\qbezier[10](110,20)(115,15)(120,10)
\qbezier[10](130,20)(135,15)(140,10)
\qbezier[20](20,30)(30,20)(40,10)
\qbezier[30](30,40)(45,25)(60,10)
{\thicklines
\put(0,10){\line(1,0){160}}
\put(0,10){\line(1,1){40}}
\put(40,50){\line(1,-1){40}}
\put(80,10){\line(1,1){20}}
\put(100,30){\line(1,-1){10}}
\put(110,20){\line(1,1){10}}
\put(120,30){\line(1,-1){10}}
\put(130,20){\line(1,1){10}}
\put(140,30){\line(1,-1){20}}
}
\qbezier[5](0,4)(0,7)(0,10)
\qbezier[5](10,4)(10,7)(10,10)
\qbezier[5](20,4)(20,7)(20,10)
\qbezier[5](80,4)(80,7)(80,10)
\qbezier[5](160,4)(160,7)(160,10)
\put(-1,-3){$\scriptscriptstyle 0$}
\put(8,-3){$\scriptscriptstyle 1$}
\put(19,-3){$\scriptscriptstyle 2$}
\put(78,-3){$\scriptscriptstyle 8$}
\put(146,-3){$\scriptscriptstyle L=16$}
\end{picture}
\parbox{15cm}{
\caption{\small (Color online) A configuration of the interface given by a Dyck path with three
contact points and two clusters for a lattice size $L = 16$. The
size of the leftmost cluster is equal to 8. The substrate is also shown
(the dashed region).}
}
\end{center}
\label{fig1}
\end{figure}

To characterize the interface, it is useful to define several quantities.
A {\em contact point}
\label{contactpoint}
is a site $j$ where $h_j = 0$. A {\em cluster} is the domain
between two consecutive contact points. The {\em size} of the leftmost cluster is $j$, if
$j\neq 0$ is the smallest number for which $h_j = 0$.

It is useful to visualize the interface as a film of tiles (tilted
squares) deposited on a {\em substrate} defined by the Dyck path
$h_i =
\frac{1 - (-1)^i}{2}$ ($i=0,1,\ldots,L$).
In this picture the clusters can be seen as
droplets of a fluid deposited on the substrate.

\item[b)] One-step Dyck paths configurations.\\
One takes $L$ odd,  $h_0 = 1$ and $h_L = 0$.
There are
$C_{L+1}$
configurations of this kind. An example of such a configuration for $L = 15$
is shown in Figure\,2.
\begin{figure}[t]
\begin{center}
\begin{picture}(230,80)(0,-10)
{\green{
\qbezier[1](10,18)(10.5,18.5)(11,19)
\qbezier[2](10,16)(11,17)(12,18)
\qbezier[3](10,14)(11.5,15.5)(13,17)
\qbezier[4](10,12)(12,14)(14,16)
\qbezier[4](11,11)(13,13)(15,15)
\qbezier[3](13,11)(14.5,12.5)(16,14)
\qbezier[2](15,11)(16,12)(17,13)
\qbezier[1](17,11)(17.5,11.5)(18,12)
\qbezier[8](23,11)(27,15)(31,19)
\qbezier[7](25,11)(28.5,14.5)(32,18)
\qbezier[6](27,11)(30,14)(33,17)
\qbezier[5](29,11)(31.5,13.5)(34,16)
\qbezier[4](31,11)(33,13)(35,15)
\qbezier[3](33,11)(34.5,12.5)(36,14)
\qbezier[2](35,11)(36,12)(37,13)
\qbezier[1](37,11)(37.5,11.5)(38,12)
\qbezier[8](43,11)(47,15)(51,19)
\qbezier[7](45,11)(48.5,14.5)(52,18)
\qbezier[6](47,11)(50,14)(53,17)
\qbezier[5](49,11)(51.5,13.5)(54,16)
\qbezier[4](51,11)(53,13)(55,15)
\qbezier[3](53,11)(54.5,12.5)(56,14)
\qbezier[2](55,11)(56,12)(57,13)
\qbezier[1](57,11)(57.5,11.5)(58,12)
\qbezier[8](63,11)(67,15)(71,19)
\qbezier[7](65,11)(68.5,14.5)(72,18)
\qbezier[6](67,11)(70,14)(73,17)
\qbezier[5](69,11)(71.5,13.5)(74,16)
\qbezier[4](71,11)(73,13)(75,15)
\qbezier[3](73,11)(74.5,12.5)(76,14)
\qbezier[2](75,11)(76,12)(77,13)
\qbezier[1](77,11)(77.5,11.5)(78,12)
\qbezier[8](83,11)(87,15)(91,19)
\qbezier[7](85,11)(88.5,14.5)(92,18)
\qbezier[6](87,11)(90,14)(93,17)
\qbezier[5](89,11)(91.5,13.5)(94,16)
\qbezier[4](91,11)(93,13)(95,15)
\qbezier[3](93,11)(94.5,12.5)(96,14)
\qbezier[2](95,11)(96,12)(97,13)
\qbezier[1](97,11)(97.5,11.5)(98,12)
\qbezier[8](103,11)(107,15)(111,19)
\qbezier[7](105,11)(108.5,14.5)(112,18)
\qbezier[6](107,11)(110,14)(113,17)
\qbezier[5](109,11)(111.5,13.5)(114,16)
\qbezier[4](111,11)(113,13)(115,15)
\qbezier[3](113,11)(114.5,12.5)(116,14)
\qbezier[2](115,11)(116,12)(117,13)
\qbezier[1](117,11)(117.5,11.5)(118,12)
\qbezier[8](123,11)(127,15)(131,19)
\qbezier[7](125,11)(128.5,14.5)(132,18)
\qbezier[6](127,11)(130,14)(133,17)
\qbezier[5](129,11)(131.5,13.5)(134,16)
\qbezier[4](131,11)(133,13)(135,15)
\qbezier[3](133,11)(134.5,12.5)(136,14)
\qbezier[2](135,11)(136,12)(137,13)
\qbezier[1](137,11)(137.5,11.5)(138,12)
\qbezier[8](143,11)(147,15)(151,19)
\qbezier[7](145,11)(148.5,14.5)(152,18)
\qbezier[6](147,11)(150,14)(153,17)
\qbezier[5](149,11)(151.5,13.5)(154,16)
\qbezier[4](151,11)(153,13)(155,15)
\qbezier[3](153,11)(154.5,12.5)(156,14)
\qbezier[2](155,11)(156,12)(157,13)
\qbezier[1](157,11)(157.5,11.5)(158,12)
}}
\qbezier[30](20,10)(35,25)(50,40)
\qbezier[20](40,10)(50,20)(60,30)
\qbezier[10](60,10)(65,15)(70,20)
\qbezier[10](100,10)(105,15)(110,20)
\qbezier[10](10,20)(15,15)(20,10)
\qbezier[10](90,20)(95,15)(100,10)
\qbezier[20](20,30)(30,20)(40,10)
\qbezier[30](30,40)(45,25)(60,10)
{\thicklines
\put(10,10){\line(1,0){150}}
\put(10,10){\line(0,1){10}}
\put(10,20){\line(1,1){30}}
\put(40,50){\line(1,-1){40}}
\put(80,10){\line(1,1){20}}
\put(100,30){\line(1,-1){20}}
\put(120,10){\line(1,1){10}}
\put(130,20){\line(1,-1){10}}
\put(140,10){\line(1,1){10}}
\put(150,20){\line(1,-1){10}}
}
\qbezier[5](10,4)(10,7)(10,10)
\qbezier[5](20,4)(20,7)(20,10)
\qbezier[5](80,4)(80,7)(80,10)
\qbezier[6](160,4)(160,7)(160,10)
\put(8,-3){$\scriptscriptstyle 0$}
\put(19,-3){$\scriptscriptstyle 1$}
\put(78,-3){$\scriptscriptstyle 7$}
\put(146,-3){$\scriptscriptstyle L=15$}
\end{picture}
\parbox{14.4cm}{
\caption{\small (Color online) A one-step Dyck path configuration with four contact points
and four clusters for $L = 15$. The leftmost cluster has size 7.}
}
\end{center}
\label{fig2}
\end{figure}
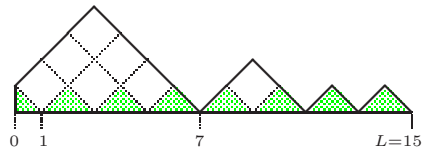

The one-step Dyck paths can be mapped onto configurations with clusters
and one defect (impurity). In order to obtain these configurations one
draws a horizontal line at height one through the first (leftmost)
cluster starting at $h_0 = 1$. If the first cluster has size  $x$ ($h_x = 0$),
the horizontal line intersects the cluster at $(x - 1)$. One puts a
defect at the point $(x-1/2)$ and one lowers the cluster by  one unit:
$h_j\rightarrow  (h_j - 1)$ $\forall\,j = 0,1, \dots ,x-1$.
This procedure is illustrated in Figure~3 for the
configuration shown in Figure\,2.
\begin{figure}[t]
\begin{center}
\begin{picture}(185,60)(0,-10)
\qbezier[10](70,20)(75,15)(80,10)
\qbezier[8](10,10)(10,15)(10,20)
\qbezier[60](10,20)(40,20)(70,20)
\put(75,15){\red{\circle*{5}}}
\put(40,20){{\vector(0,-1){7}}}
\put(20,20){{\vector(0,-1){7}}}
\put(60,20){{\vector(0,-1){7}}}
{\thicklines
\put(10,10){\line(1,0){150}}
\put(10,20){\line(1,1){30}}
\put(40,50){\line(1,-1){30}}
\put(80,10){\line(1,1){20}}
\put(100,30){\line(1,-1){20}}
\put(120,10){\line(1,1){10}}
\put(130,20){\line(1,-1){10}}
\put(140,10){\line(1,1){10}}
\put(150,20){\line(1,-1){10}}
}
\qbezier[5](10,4)(10,7)(10,10)
\qbezier[5](20,4)(20,7)(20,10)
\qbezier[5](80,4)(80,7)(80,10)
\qbezier[6](160,4)(160,7)(160,10)
\put(8,-3){$\scriptscriptstyle 0$}
\put(19,-3){$\scriptscriptstyle 1$}
\put(78,-3){$\scriptscriptstyle 7$}
\put(146,-3){$\scriptscriptstyle L=15$}
\end{picture}
\raisebox{3mm}{$\Rrightarrow$}\hspace{8mm}
\begin{picture}(200,60)(0,-10)
\put(75,10){\red{\circle*{5}}}
{\thicklines
\put(10,10){\line(1,0){60}}
\put(80,10){\line(1,0){80}}
\put(10,10){\line(1,1){30}}
\put(40,40){\line(1,-1){30}}
\put(80,10){\line(1,1){20}}
\put(100,30){\line(1,-1){20}}
\put(120,10){\line(1,1){10}}
\put(130,20){\line(1,-1){10}}
\put(140,10){\line(1,1){10}}
\put(150,20){\line(1,-1){10}}
}
\qbezier[5](10,4)(10,7)(10,10)
\qbezier[5](20,4)(20,7)(20,10)
\qbezier[5](70,4)(70,7)(70,10)
\qbezier[5](80,4)(80,7)(80,10)
\qbezier[6](160,4)(160,7)(160,10)
\put(8,-3){$\scriptscriptstyle 0$}
\put(19,-3){$\scriptscriptstyle 1$}
\put(68,-3){$\scriptscriptstyle 6$}
\put(78,-3){$\scriptscriptstyle 7$}
\put(146,-3){$\scriptscriptstyle L=15$}
\end{picture}
\parbox{14.4cm}{
\caption{\small (Color online) Mapping the one-step Dyck path shown in Figure\ref{fig2} onto a
configuration with one defect and clusters.}}

\end{center}
\label{fig3}
\end{figure}
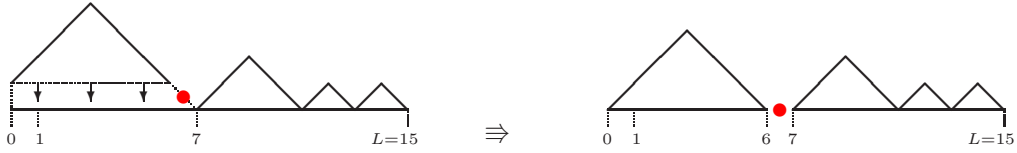

\item[c)] Ballot paths.\\
One takes (for both $L$ even and odd) $h_L = 0$ and $h_0$ free ($h_0=L,L-2,\dots, 0\mbox{~or~}1$).\footnote{
In the usual definition \cite{Sh}, a ballot path is defined by taking $h_0$ fixed.
}
There are
\be
\lb{2.3}
{L\choose \lfloor L/2\rfloor}
\ee
configurations of this kind (here $\lfloor n\rfloor$ is the integer part of $n$).
An example of such a configuration for
$L =14$ is shown in Figure\,4.
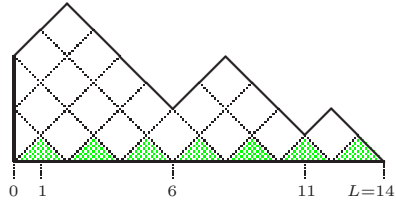
\begin{figure}[t]
\begin{center}
\begin{picture}(180,70)(0,-10)
{\green{
\qbezier[8](13,11)(17,15)(21,19)
\qbezier[7](15,11)(18.5,14.5)(22,18)
\qbezier[6](17,11)(20,14)(23,17)
\qbezier[5](19,11)(21.5,13.5)(24,16)
\qbezier[4](21,11)(23,13)(25,15)
\qbezier[3](23,11)(24.5,12.5)(26,14)
\qbezier[2](25,11)(26,12)(27,13)
\qbezier[1](27,11)(27.5,11.5)(28,12)
\qbezier[8](33,11)(37,15)(41,19)
\qbezier[7](35,11)(38.5,14.5)(42,18)
\qbezier[6](37,11)(40,14)(43,17)
\qbezier[5](39,11)(41.5,13.5)(44,16)
\qbezier[4](41,11)(43,13)(45,15)
\qbezier[3](43,11)(44.5,12.5)(46,14)
\qbezier[2](45,11)(46,12)(47,13)
\qbezier[1](47,11)(47.5,11.5)(48,12)
\qbezier[8](53,11)(57,15)(61,19)
\qbezier[7](55,11)(58.5,14.5)(62,18)
\qbezier[6](57,11)(60,14)(63,17)
\qbezier[5](59,11)(61.5,13.5)(64,16)
\qbezier[4](61,11)(63,13)(65,15)
\qbezier[3](63,11)(64.5,12.5)(66,14)
\qbezier[2](65,11)(66,12)(67,13)
\qbezier[1](67,11)(67.5,11.5)(68,12)
\qbezier[8](73,11)(77,15)(81,19)
\qbezier[7](75,11)(78.5,14.5)(82,18)
\qbezier[6](77,11)(80,14)(83,17)
\qbezier[5](79,11)(81.5,13.5)(84,16)
\qbezier[4](81,11)(83,13)(85,15)
\qbezier[3](83,11)(84.5,12.5)(86,14)
\qbezier[2](85,11)(86,12)(87,13)
\qbezier[1](87,11)(87.5,11.5)(88,12)
\qbezier[8](93,11)(97,15)(101,19)
\qbezier[7](95,11)(98.5,14.5)(102,18)
\qbezier[6](97,11)(100,14)(103,17)
\qbezier[5](99,11)(101.5,13.5)(104,16)
\qbezier[4](101,11)(103,13)(105,15)
\qbezier[3](103,11)(104.5,12.5)(106,14)
\qbezier[2](105,11)(106,12)(107,13)
\qbezier[1](107,11)(107.5,11.5)(108,12)
\qbezier[8](113,11)(117,15)(121,19)
\qbezier[7](115,11)(118.5,14.5)(122,18)
\qbezier[6](117,11)(120,14)(123,17)
\qbezier[5](119,11)(121.5,13.5)(124,16)
\qbezier[4](121,11)(123,13)(125,15)
\qbezier[3](123,11)(124.5,12.5)(126,14)
\qbezier[2](125,11)(126,12)(127,13)
\qbezier[1](127,11)(127.5,11.5)(128,12)
\qbezier[8](133,11)(137,15)(141,19)
\qbezier[7](135,11)(138.5,14.5)(142,18)
\qbezier[6](137,11)(140,14)(143,17)
\qbezier[5](139,11)(141.5,13.5)(144,16)
\qbezier[4](141,11)(143,13)(145,15)
\qbezier[3](143,11)(144.5,12.5)(146,14)
\qbezier[2](145,11)(146,12)(147,13)
\qbezier[1](147,11)(147.5,11.5)(148,12)
}}
\qbezier[30](10,30)(25,45)(40,60)
\qbezier[40](10,10)(30,30)(50,50)
\qbezier[30](30,10)(45,25)(60,40)
\qbezier[20](50,10)(60,20)(70,30)
\qbezier[30](70,10)(85,25)(100,40)
\qbezier[20](90,10)(100,20)(110,30)
\qbezier[10](110,10)(115,15)(120,20)
\qbezier[10](130,10)(135,15)(140,20)
\qbezier[20](10,30)(20,20)(30,10)
\qbezier[40](10,50)(30,30)(50,10)
\qbezier[50](20,60)(45,35)(70,10)
\qbezier[20](70,30)(80,20)(90,10)
\qbezier[30](80,40)(95,25)(110,10)
\qbezier[10](120,20)(125,15)(130,10)
{\thicklines
\put(10,10){\line(1,0){140}}
\put(10,10){\line(0,1){40}}
\put(10,50){\line(1,1){20}}
\put(30,70){\line(1,-1){40}}
\put(70,30){\line(1,1){20}}
\put(90,50){\line(1,-1){30}}
\put(120,20){\line(1,1){10}}
\put(130,30){\line(1,-1){20}}
}
\qbezier[5](10,4)(10,7)(10,10)
\qbezier[5](20,4)(20,7)(20,10)
\qbezier[5](70,4)(70,7)(70,10)
\qbezier[5](120,4)(120,7)(120,10)
\qbezier[6](150,4)(150,7)(150,10)
\put(8,-3){$\scriptscriptstyle 0$}
\put(19,-3){$\scriptscriptstyle 1$}
\put(68,-3){$\scriptscriptstyle 6$}
\put(117,-3){$\scriptscriptstyle 11$}
\put(136,-3){$\scriptscriptstyle L=14$}
\end{picture}
\parbox{15cm}{
\caption{\small (Color online) A ballot path for $L = 14$. One has one contact point and one
cluster of size 14.}
}
\end{center}
\label{fig4}
\end{figure}

The ballot paths configurations can be seen as droplets of a fluid in the
presence of a wet wall.

Similar to the case of one-step Dyck paths, one can map the ballot paths
onto configurations with clusters and defects (impurities). We will call
them configurations with defects. The mapping follows  the same
procedure as the one used for one-step Dyck paths and is illustrated in
Figure\,5. Notice that the height at the origin $h_0$ is equal to the number of
defects.
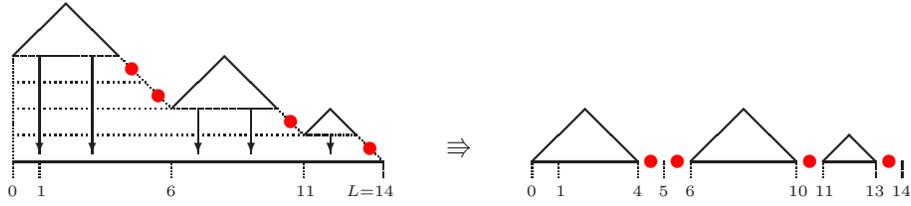
\begin{figure}[t]
\begin{center}
\begin{picture}(170,85)(0,-10)
\qbezier[10](140,20)(145,15)(150,10)
\qbezier[10](110,30)(115,25)(120,20)
\qbezier[20](50,50)(60,40)(70,30)
\qbezier[32](10,10)(10,30)(10,50)
\qbezier[55](10,20)(65,20)(120,20)
\qbezier[20](120,20)(130,20)(140,20)
\qbezier[40](70,30)(90,30)(110,30)
\qbezier[30](10,30)(40,30)(70,30)
\qbezier[25](10,40)(35,40)(60,40)
\qbezier[40](10,50)(30,50)(50,50)
\put(145,15){\red{\circle*{5}}}
\put(115,25){\red{\circle*{5}}}
\put(65,35){\red{\circle*{5}}}
\put(55,45){\red{\circle*{5}}}
\put(20,50){{\vector(0,-1){37}}}
\put(40,50){{\vector(0,-1){37}}}
\put(80,30){{\vector(0,-1){17}}}
\put(100,30){{\vector(0,-1){17}}}
\put(130,20){{\vector(0,-1){7}}}
{\thicklines
\put(10,10){\line(1,0){140}}
\put(10,50){\line(1,1){20}}
\put(30,70){\line(1,-1){20}}
\put(70,30){\line(1,1){20}}
\put(90,50){\line(1,-1){20}}
\put(120,20){\line(1,1){10}}
\put(130,30){\line(1,-1){10}}
}
\qbezier[5](10,4)(10,7)(10,10)
\qbezier[5](20,4)(20,7)(20,10)
\qbezier[5](70,4)(70,7)(70,10)
\qbezier[5](120,4)(120,7)(120,10)
\qbezier[6](150,4)(150,7)(150,10)
\put(8,-3){$\scriptscriptstyle 0$}
\put(19,-3){$\scriptscriptstyle 1$}
\put(68,-3){$\scriptscriptstyle 6$}
\put(117,-3){$\scriptscriptstyle 11$}
\put(136,-3){$\scriptscriptstyle L=14$}
\end{picture}
\raisebox{8mm}{$\Rrightarrow$}\hspace{3mm}
\begin{picture}(160,80)(0,-10)
\put(145,10){\red{\circle*{5}}}
\put(115,10){\red{\circle*{5}}}
\put(65,10){\red{\circle*{5}}}
\put(55,10){\red{\circle*{5}}}
{\thicklines
\put(10,10){\line(1,0){40}}
\put(70,10){\line(1,0){40}}
\put(120,10){\line(1,0){20}}
\put(10,10){\line(1,1){20}}
\put(30,30){\line(1,-1){20}}
\put(70,10){\line(1,1){20}}
\put(90,30){\line(1,-1){20}}
\put(120,10){\line(1,1){10}}
\put(130,20){\line(1,-1){10}}
}
\qbezier[5](10,4)(10,7)(10,10)
\qbezier[5](20,4)(20,7)(20,10)
\qbezier[5](50,4)(50,7)(50,10)
\qbezier[5](60,4)(60,7)(60,10)
\qbezier[5](70,4)(70,7)(70,10)
\qbezier[5](120,4)(120,7)(120,10)
\qbezier[5](110,4)(110,7)(110,10)
\qbezier[6](150,4)(150,7)(150,10)
\qbezier[6](140,4)(140,7)(140,10)
\put(8,-3){$\scriptscriptstyle 0$}
\put(19,-3){$\scriptscriptstyle 1$}
\put(48,-3){$\scriptscriptstyle 4$}
\put(58,-3){$\scriptscriptstyle 5$}
\put(68,-3){$\scriptscriptstyle 6$}
\put(107,-3){$\scriptscriptstyle 10$}
\put(117,-3){$\scriptscriptstyle 11$}
\put(136,-3){$\scriptscriptstyle 13$}
\put(146,-3){$\scriptscriptstyle 14$}
\end{picture}
\parbox{15cm}{
\caption{\small (Color online) The mapping of the ballot path shown in Figure\,4 onto a
configuration with four defects.}
}
\end{center}
\label{fig5}
\end{figure}

The configurations with defects can also be seen as an interface in which
the droplets have empty spaces between them.
\end{itemize}

In the raise and peel model (RPM) the dynamics of the interface is
described in a transparent way in the language of tiles (tilted squares)
which cover the area between the interface and the substrate. We consider
the interface separating a film of tiles deposited on the substrate from a
rarefied gas of tiles. The interface can be Dyck path, one-step Dyck path or
ballot path configurations (the dynamics for defect configurations is
discussed in Appendix A).

The evolution of the system in discrete time (Monte Carlo steps) is
given by the following rules. With a probability $P_i= 1/(L-1)$ a tile from
the gas hits the site $i,\;\; i=1,\dots ,L-1$. Depending on the value of the slope
$s_i=(h_{i+1}-h_{i-1})/2$ at the site $i$, the following processes can occur:
\begin{itemize}
\item[i)] $s_i=0$ and $h_i > h_{i-1}$. 

The tile hits a local peak and is reflected.
\item[ii)] $s_i=0$ and $h_i < h_{i-1}$. 

The tile hits a local minimum. With a probability 1 the tile
is adsorbed ($h_i \mapsto h_i+2$).
\item[iii)] $s_i=1$.

With probability 1 the tile is reflected after triggering
the desorption of a layer of tiles from the segment
($h_j>h_i=h_{i+b},\; j=i+1,\ldots , i+b-1$),
{\it i.e.} $h_j \mapsto h_j-2$ for $j=i+1,\dots ,i+b-1$. This layer contains
$b-1$ tiles (this is always an odd number). For an example see Figure~6.
\item[iv)] $s_i=-1$.

With probability 1 the tile is reflected after triggering
the desorption of a layer of tiles belonging to the segment
($h_j>h_i=h_{i-b},\; j=i-b+1,\ldots, i-1$),
{\it i.e.}  $h_j \mapsto h_j-2$ for $j=i-b+1,...,i-1$.
\end{itemize}
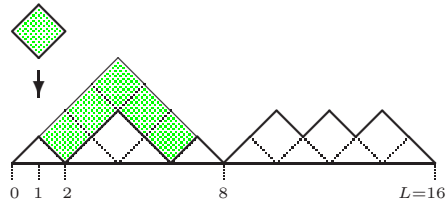
\begin{figure}[t]
\begin{center}
\begin{picture}(200,75)(0,-10)
{\green{
\qbezier[8](2,60)(6,64)(10,68)
\qbezier[8](3,59)(7,63)(11,67)
\qbezier[8](4,58)(8,62)(12,66)
\qbezier[8](5,57)(9,61)(13,65)
\qbezier[8](6,56)(10,60)(14,64)
\qbezier[8](7,55)(11,59)(15,63)
\qbezier[8](8,54)(12,58)(16,62)
\qbezier[8](9,53)(13,57)(17,61)
\qbezier[8](10,52)(14,56)(18,60)
\qbezier[8](12,20)(16,24)(20,28)
\qbezier[8](13,19)(17,23)(21,27)
\qbezier[8](14,18)(18,22)(22,26)
\qbezier[8](15,17)(19,21)(23,25)
\qbezier[8](16,16)(20,20)(24,24)
\qbezier[8](17,15)(21,19)(25,23)
\qbezier[8](18,14)(22,18)(26,22)
\qbezier[8](19,13)(23,17)(27,21)
\qbezier[8](20,12)(24,16)(28,20)
\qbezier[8](22,30)(26,34)(30,38)
\qbezier[8](23,29)(27,33)(31,37)
\qbezier[8](24,28)(28,32)(32,36)
\qbezier[8](25,27)(29,31)(33,35)
\qbezier[8](26,26)(30,30)(34,34)
\qbezier[8](27,25)(31,29)(35,33)
\qbezier[8](28,24)(32,28)(36,32)
\qbezier[8](29,23)(33,27)(37,31)
\qbezier[8](30,22)(34,26)(38,30)
\qbezier[8](52,20)(56,24)(60,28)
\qbezier[8](53,19)(57,23)(61,27)
\qbezier[8](54,18)(58,22)(62,26)
\qbezier[8](55,17)(59,21)(63,25)
\qbezier[8](56,16)(60,20)(64,24)
\qbezier[8](57,15)(61,19)(65,23)
\qbezier[8](58,14)(62,18)(66,22)
\qbezier[8](59,13)(63,17)(67,21)
\qbezier[8](60,12)(64,16)(68,20)
\qbezier[8](42,30)(46,34)(50,38)
\qbezier[8](43,29)(47,33)(51,37)
\qbezier[8](44,28)(48,32)(52,36)
\qbezier[8](45,27)(49,31)(53,35)
\qbezier[8](46,26)(50,30)(54,34)
\qbezier[8](47,25)(51,29)(55,33)
\qbezier[8](48,24)(52,28)(56,32)
\qbezier[8](49,23)(53,27)(57,31)
\qbezier[8](50,22)(54,26)(58,30)
\qbezier[8](32,40)(36,44)(40,48)
\qbezier[8](33,39)(37,43)(41,47)
\qbezier[8](34,38)(38,42)(42,46)
\qbezier[8](35,37)(39,41)(43,45)
\qbezier[8](36,36)(40,40)(44,44)
\qbezier[8](37,35)(41,39)(45,43)
\qbezier[8](38,34)(42,38)(46,42)
\qbezier[8](39,33)(43,37)(47,41)
\qbezier[8](40,32)(44,36)(48,40)}}
\qbezier[30](20,10)(35,25)(50,40)
\qbezier[20](40,10)(50,20)(60,30)
\qbezier[10](60,10)(65,15)(70,20)
\qbezier[10](100,10)(105,15)(110,20)
\qbezier[10](120,10)(125,15)(130,20)
\qbezier[10](140,10)(145,15)(150,20)
\qbezier[10](10,20)(15,15)(20,10)
\qbezier[10](90,20)(95,15)(100,10)
\qbezier[10](110,20)(115,15)(120,10)
\qbezier[10](130,20)(135,15)(140,10)
\qbezier[20](20,30)(30,20)(40,10)
\qbezier[30](30,40)(45,25)(60,10)
{\thicklines
\put(10,50){\line(1,1){10}}
\put(10,50){\line(-1,1){10}}
\put(10,70){\line(1,-1){10}}
\put(10,70){\line(-1,-1){10}}
\put(10,45){\vector(0,-1){10}}
\put(0,10){\line(1,0){160}}
\put(0,10){\line(1,1){10}}
\put(10,20){\line(1,-1){10}}
\put(20,10){\line(1,1){20}}
\put(40,30){\line(1,-1){20}}
\put(60,10){\line(1,1){10}}
\put(70,20){\line(1,-1){10}}
\put(80,10){\line(1,1){20}}
\put(100,30){\line(1,-1){10}}
\put(110,20){\line(1,1){10}}
\put(120,30){\line(1,-1){10}}
\put(130,20){\line(1,1){10}}
\put(140,30){\line(1,-1){20}}
}
\put(10,20){\line(1,1){30}}
\put(40,50){\line(1,-1){30}}
\qbezier[5](0,4)(0,7)(0,10)
\qbezier[5](10,4)(10,7)(10,10)
\qbezier[5](20,4)(20,7)(20,10)
\qbezier[5](80,4)(80,7)(80,10)
\qbezier[5](160,4)(160,7)(160,10)
\put(-1,-3){$\scriptscriptstyle 0$}
\put(8,-3){$\scriptscriptstyle 1$}
\put(19,-3){$\scriptscriptstyle 2$}
\put(78,-3){$\scriptscriptstyle 8$}
\put(146,-3){$\scriptscriptstyle L=16$}
\end{picture}
\parbox{15cm}{
\caption{(Color online) A desorption event.
The incoming tile at site 1 triggers  an
avalanche of $5$ tiles, which are shaded. All  the shaded tiles are
removed in the desorption event.}
}
\end{center}
\label{fig6}
\end{figure}

Notice that the adsorption and desorption rates were taken equal,
this is the RPM at the Razumov-Stroganov point. The model was studied in
detail in the case of Dyck path configurations ($L$ even)
also when the adsorption and desorption rates are different and the phase diagram
of the model was obtained  (see Refs.\cite{GNPR,ALR},
for a review see \cite{AR2}). The Razumov-Stroganov point is special in two
ways. Firstly, the Hamiltonian which gives the continuous time evolution of
the system can be mapped into an XXZ spin one-half quantum chain which is
integrable \cite{SB}. The finite-size scaling limit of the
 Hamiltonian spectrum is given by $c = 0$
Virasoro characters ($c$ is the central charge of the Virasoro algebra) and
hence the system is conformal invariant. Secondly, the probability
distribution function (PDF) describing the stationary state of the system
for finite systems has remarkable combinatorial properties  \cite{RS1,RS2} which allows one to
obtain exact results for physical observables.

The RPM was also considered in the $L$ odd case  and it was shown that
the defect  (see Figure\,3) makes L\'{e}vy flights and behaves like a
"relativistic" random walker
( dispersion relation $\langle x^2\rangle\sim  t^2$) \cite{AR1}. In the
present paper we are going to present more results which will show some
surprising properties of this model.

The PDF describing the stationary states is expressed
  in terms of Dyck
paths only. The dynamics  seen in the space of defect configurations is
interesting: it describes the pair annihilation of two defects in an
unquenched random media. The average density of defects decreases like $1/t$ \cite{AR1}.

We are going to extend the RPM model acting on ballot paths by making the
first site  active  (site $0$ in Figures\,4 and 5). We will define in this way
the RPM in the presence of a wall (RPMW).

The evolution of the system (Monte Carlo steps) is given by the following
rules. With a probability $P_i = 1/(L+a-1)$ a tile from the gas hits the site
$i,\;\; i = 1,\dots ,L-1$. The changes of the interface produced by the hits are
the same as in the RPM model. With a probability $P_0 = a/(L+a-1)$, a half-tile
hits the site $0$.
The boundary rate is equal to $a$, as opposed to the bulk rates which are equal to 1.
If the slope $s_0= h_1-h_0$ is equal to 1, the half-tile
gets adsorbed (see Figure\,7). If $s_0 = -1$, the half-tile is reflected.
Notice that the adsorption process is local. This is not the case if one
uses the defect configurations picture (see Appendix A).
\begin{figure}[t]
\begin{center}
\begin{picture}(180,110)(0,-10)
{\green{
\qbezier[1](10,66)(10.5,66.5)(11,67)
\qbezier[2](10,64)(11,65)(12,66)
\qbezier[3](10,62)(11.5,63.5)(13,65)
\qbezier[4](10,60)(12,62)(14,64)
\qbezier[5](10,58)(12.5,60.5)(15,63)
\qbezier[6](10,56)(13,59)(16,62)
\qbezier[7](10,54)(13.5,57.5)(17,61)
\qbezier[8](10,52)(14,56)(18,60)
\qbezier[1](10,106)(10.5,106.5)(11,107)
\qbezier[2](10,104)(11,105)(12,106)
\qbezier[3](10,102)(11.5,103.5)(13,105)
\qbezier[4](10,100)(12,102)(14,104)
\qbezier[5](10,98)(12.5,100.5)(15,103)
\qbezier[6](10,96)(13,99)(16,102)
\qbezier[7](10,94)(13.5,97.5)(17,101)
\qbezier[8](10,92)(14,96)(18,100)}}
\qbezier[30](10,30)(25,45)(40,60)
\qbezier[40](10,10)(30,30)(50,50)
\qbezier[30](30,10)(45,25)(60,40)
\qbezier[20](50,10)(60,20)(70,30)
\qbezier[30](70,10)(85,25)(100,40)
\qbezier[20](90,10)(100,20)(110,30)
\qbezier[10](110,10)(115,15)(120,20)
\qbezier[10](130,10)(135,15)(140,20)
\qbezier[20](10,30)(20,20)(30,10)
\qbezier[40](10,50)(30,30)(50,10)
\qbezier[50](20,60)(45,35)(70,10)
\qbezier[20](70,30)(80,20)(90,10)
\qbezier[30](80,40)(95,25)(110,10)
\qbezier[10](120,20)(125,15)(130,10)
{\thicklines
\put(10,90){\line(1,1){10}}
\put(10,110){\line(1,-1){10}}
\put(10,90){\line(0,1){20}}
\put(10,88){\vector(0,-1){10}}
\put(10,70){\line(1,-1){10}}
\put(10,50){\line(0,1){20}}
\put(10,10){\line(1,0){140}}
\put(10,10){\line(0,1){40}}
\put(20,60){\line(1,1){10}}
\put(30,70){\line(1,-1){40}}
\put(70,30){\line(1,1){20}}
\put(90,50){\line(1,-1){30}}
\put(120,20){\line(1,1){10}}
\put(130,30){\line(1,-1){20}}
}
\put(10,50){\line(1,1){10}}
\qbezier[5](10,4)(10,7)(10,10)
\qbezier[5](20,4)(20,7)(20,10)
\qbezier[5](70,4)(70,7)(70,10)
\qbezier[5](120,4)(120,7)(120,10)
\qbezier[6](150,4)(150,7)(150,10)
\put(8,-3){$\scriptscriptstyle 0$}
\put(19,-3){$\scriptscriptstyle 1$}
\put(68,-3){$\scriptscriptstyle 6$}
\put(117,-3){$\scriptscriptstyle 11$}
\put(136,-3){$\scriptscriptstyle L=14$}
\end{picture}
\begin{picture}(180,110)(0,-10)
{\green{
\qbezier[1](10,46)(10.5,46.5)(11,47)
\qbezier[2](10,44)(11,45)(12,46)
\qbezier[3](10,42)(11.5,43.5)(13,45)
\qbezier[4](10,40)(12,42)(14,44)
\qbezier[5](10,38)(12.5,40.5)(15,43)
\qbezier[6](10,36)(13,39)(16,42)
\qbezier[7](10,34)(13.5,37.5)(17,41)
\qbezier[8](10,32)(14,36)(18,40)
\qbezier[8](12,50)(16,54)(20,58)
\qbezier[8](13,49)(17,53)(21,57)
\qbezier[8](14,48)(18,52)(22,56)
\qbezier[8](15,47)(19,51)(23,55)
\qbezier[8](16,46)(20,50)(24,54)
\qbezier[8](17,45)(21,49)(25,53)
\qbezier[8](18,44)(22,48)(26,52)
\qbezier[8](19,43)(23,47)(27,51)
\qbezier[8](20,42)(24,46)(28,50)
\multiput(0,0)(10,-10){3}{
\qbezier[8](22,60)(26,64)(30,68)
\qbezier[8](23,59)(27,63)(31,67)
\qbezier[8](24,58)(28,62)(32,66)
\qbezier[8](25,57)(29,61)(33,65)
\qbezier[8](26,56)(30,60)(34,64)
\qbezier[8](27,55)(31,59)(35,63)
\qbezier[8](28,54)(32,58)(36,62)
\qbezier[8](29,53)(33,57)(37,61)
\qbezier[8](30,52)(34,56)(38,60)}
\qbezier[8](52,75)(56,79)(60,83)
\qbezier[8](53,74)(57,78)(61,82)
\qbezier[8](54,73)(58,77)(62,81)
\qbezier[8](55,72)(59,76)(63,80)
\qbezier[8](56,71)(60,75)(64,79)
\qbezier[8](57,70)(61,74)(65,78)
\qbezier[8](58,69)(62,73)(66,77)
\qbezier[8](59,68)(63,72)(67,76)
\qbezier[8](60,67)(64,71)(68,75)
}}
\qbezier[30](10,30)(25,45)(40,60)
\qbezier[40](10,10)(30,30)(50,50)
\qbezier[30](30,10)(45,25)(60,40)
\qbezier[20](50,10)(60,20)(70,30)
\qbezier[30](70,10)(85,25)(100,40)
\qbezier[20](90,10)(100,20)(110,30)
\qbezier[10](110,10)(115,15)(120,20)
\qbezier[10](130,10)(135,15)(140,20)
\qbezier[20](10,30)(20,20)(30,10)
\qbezier[40](10,50)(30,30)(50,10)
\qbezier[50](20,60)(45,35)(70,10)
\qbezier[20](70,30)(80,20)(90,10)
\qbezier[30](80,40)(95,25)(110,10)
\qbezier[10](120,20)(125,15)(130,10)
{\thicklines
\put(50,75){\line(1,1){10}}
\put(50,75){\line(1,-1){10}}
\put(60,65){\line(1,1){10}}
\put(60,85){\line(1,-1){10}}
\put(60,63){\vector(0,-1){10}}
\put(10,10){\line(1,0){140}}
\put(10,10){\line(0,1){20}}
\put(60,40){\line(1,-1){10}}
\put(70,30){\line(1,1){20}}
\put(90,50){\line(1,-1){30}}
\put(120,20){\line(1,1){10}}
\put(130,30){\line(1,-1){20}}
\put(10,30){\line(1,1){20}}
\put(30,50){\line(1,-1){20}}
\put(50,30){\line(1,1){10}}
}
\put(10,30){\line(0,1){20}}
\put(10,50){\line(1,1){20}}
\put(30,70){\line(1,-1){30}}
\qbezier[5](10,4)(10,7)(10,10)
\qbezier[5](20,4)(20,7)(20,10)
\qbezier[5](70,4)(70,7)(70,10)
\qbezier[5](120,4)(120,7)(120,10)
\qbezier[6](150,4)(150,7)(150,10)
\put(8,-3){$\scriptscriptstyle 0$}
\put(19,-3){$\scriptscriptstyle 1$}
\put(68,-3){$\scriptscriptstyle 6$}
\put(117,-3){$\scriptscriptstyle 11$}
\put(136,-3){$\scriptscriptstyle L=14$}
\end{picture}
\parbox{15cm}{
\caption{\small (Color online) The adsorption of a half-tile at the first site
and desorption of a layer touching the boundary
for the
ballot path shown in Figure\,4.}
}
\end{center}
\label{fig7}
\end{figure}
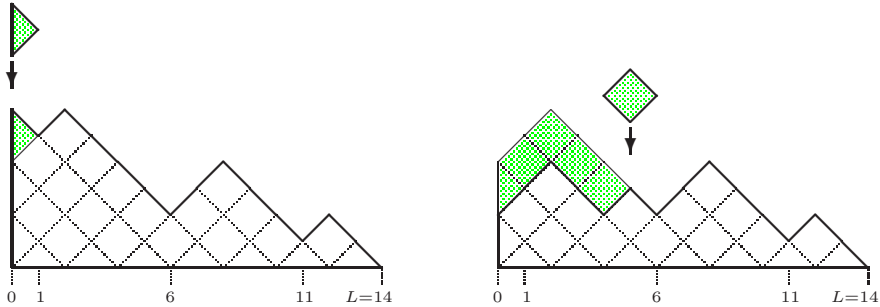

The RPMW in its time continuous version, can be obtained in a simple way
using a Hamiltonian expressed in terms of generators of the one-boundary
Temperley-Lieb algebra. This is explained in Section 5. Similar to the
RPM model, the stationary states of the RPMW have magical combinatorial
properties for finite values of $L$ \cite{MNGB,P,GR}. New expressions and relations are
summarized in Section 3 and derived in Sections 7 and 8. Moreover, the
Hamiltonian can be mapped onto an integrable XXZ spin one-half quantum
chain and the finite-size scaling limit of the Hamiltonian
spectrum is known \cite{GNPyR}. It can be expressed in terms of Virasoro characters
therefore, similar to the RPM, the RPMW is conformal invariant. The
consequences of this observation will be presented in Section 4.

\section{Some exact results and some conjectures}

In this section we present some exact results which will be
discussed in Sections 7 and 8. These results besides being interesting on
their own,  will be used in Section 4  to derive critical exponents and
to check the data obtained in
Monte Carlo simulations. We consider the RPM with and without a wall at
the Razumov-Stroganov point and the boundary rate $a$.

We first introduce two useful functions $S(L,n)$ and $P(L,n,m)$. They are
defined as follows:
\ba
\nonumber
S(L,n)&=&
2^{^{-\lfloor n^2/ 4\rfloor}}
\prod_{p=1}^{n}{1\over (2p-1)!!}\,
\prod_{p=0}^{\lfloor{n-1\over 3}\rfloor}{\bigl(L-\lfloor{n+p\over 2}\rfloor-p-1\bigr)!\over
\bigl(L-2n+3p\bigr)!}\times
\\
\lb{pascal-solved}
&&\hspace{48mm}\prod_{p=0}^{\lfloor{n-2\over 3}\rfloor}{\bigl(2L+2n-6p-3\bigr)!!\over
\bigl(2L-2\lfloor{n+p\over 2}\rfloor+4p+1\bigr)!! }\, ,\hspace{20mm}
\\[3mm]
\nonumber
P(L,n,m)& =&  2^{\lfloor{m-1\over 2}\rfloor}\, S(L,n-1)\,
{(2n-m)!(n-1)!\over m!(n-m)!(2n-1)!}\times
\\
\lb{P-Lnm}
&&\hspace{5mm}
{(L-2n+m-1)!(L-n)!(2L-2n+2m-1)!!\over (L-2n-1)!(L-n+\lfloor{m\over 2}\rfloor)!
(2L-2n+2\lfloor{m+1\over 2}\rfloor-1)!!}\, .
\ea
The expressions (\ref{pascal-solved}) have first appeared in \cite{MNGB,P}.
The expressions  for $$S(L)\,=\, S(L,\lfloor (L-1)/2\rfloor)$$
appear also in the problem of enumeration of vertically
symmetric $A^V_p$ and vertically and horizontally symmetric $A^{VH}_p$  $~p\times p~$
alternating sign matrices (a known topic in combinatorics \cite{B,K,R}):
\ba
\lb{S-2n}
S(2N)
& =& A^V_{2N+1}\, =\,
\prod_{0\leq i\leq N-1} \frac{(3i+2)(6i+3)!(2i+1)!}{(4i+2)!(4i+3)!}\, ,\\
\lb{S-2n-1}
S(2N-1)
&=&{A^{VH}_{4N-1}/ A^V_{2N-1}}\, =\,
{A^{VH}_{4N+1}/ A^V_{2N+1}}\, =\,
\prod_{\rule{0pt}{3mm}0\leq i\leq N-1}
\!\!{(3i+1)(6i)!(2i)!\over (4i)!(4i+1)!}\, .
\ea

\medskip
We next fix the notations to be used below: for the RPMW of a size $L$ and  boundary rate $a$
an observable  $\Phi(\dots )$ is denoted as $\Phi^{(a)}_L(\dots )$. For the RPM
of a size $L$ the same function is denoted as $\Phi_L(\dots )$. Notice the obvious relation
$\Phi^{(0)}_L(\dots)=\Phi_L(\dots)$.
\medskip

We now state the following results for the stationary states of the
RPM and RPMW.
\medskip

In the  RPM, the
probability of having $k$ clusters for a system of size $L$ is:
\be
\lb{P_L(k)}
P_L(k)\, =\, {1\over S(L)}P(L,\lfloor{\textstyle \frac{L-1}{2}}\rfloor,k-1)\, .
\ee
For $L$ even this result coincides with a conjecture made by de Gier \cite{G}.

In the presence of a wall (RPMW), the probabilities to have
$k$ clusters for a system of size $L$ are:
\ba
\lb{P1B-Lodd}
\hspace{-15mm}
\begin{array}{c}
L \mbox{\small ~odd,}\\
a \mbox{\small ~arbitrary:}
\end{array} &&
P_L^{(a)}(k)  \, =\,P_L(k)\, =\,
{\textstyle\frac{1}{S(L)}}\,P(L,{\textstyle \frac{L-1}{2}},k-1),\qquad
\\[1mm]
\hspace{-15mm}
\lb{P1B-Leven}
\begin{array}{c}
L \mbox{\small ~even,}\\
a=1 :
\end{array} &&
P_L^{(1)}(k)  \, =\,
{\textstyle\frac{1}{S(L+1)}} \Bigl\{P(L+1,{\textstyle\frac{L}{2}},k)
\, +\, P(L+1,{\textstyle\frac{L}{2}}-1,k-1)\Bigr\} .
\ea
Note the remarkable fact that for $L$ odd,  the probabilities to have $k$
clusters for the RPM with or without a wall are the same.
This statement is valid for any value of $a$.
Another
connection between  these two systems (with or without wall for $L$
odd) will be presented below. The relation (\ref{P1B-Leven}) for $L$ even is valid only for $a=1$.

Knowing the probabilities to find $k$ clusters, one can compute the average
number of clusters $\langle k\rangle_L$ and $\langle k\rangle^{(a)}_L$,
for the RPM and RPMW, respectively. We give here only the
large $L$ limit behavior of these numbers:
\ba
\lb{k(L)}
\langle k\rangle_L &\stackrel{\raisebox{2pt}{$\scriptscriptstyle L\rightarrow\infty$}}{\longrightarrow}&
\left\{
\begin{array}{ll}
{\Gamma(1/3)\sqrt{3}\over 2\pi}
L^{2/3} \approx 0.738488\, L^{2/3}, &  \mbox{($L$ even)}
\\[3mm]
\alpha\, L^{2/3}, &          \mbox{($L$ odd)}
\end{array}
\right.
\\[3mm]
\lb{k(L)1B}
\langle k\rangle^{(a)}_L&\stackrel{\raisebox{2pt}{$\scriptscriptstyle L\rightarrow\infty$}}{\longrightarrow}&
\quad\,\alpha\, L^{2/3},     \qquad \mbox{($L$ even or odd, $a$ arbitrary)}
\ea
where approximants give for $\alpha$ the value $\alpha\approx 0.5056(3)$.
Observe that in all cases the average size of a cluster increases like
$L^{1/3}$. Notice that for the RPM for $L$ even one has the largest number of
clusters. The explanation is very simple. In the case $L$ odd or in the
presence of the wall, the first cluster near the active boundary has a size of the
order of $L$ (see the comment after eq.(\ref{symm-Pf-Leven})) which leaves less available space for the other clusters,

We now consider the probability densities $F^{(a)}_L(x)$ to have the leftmost  (first)
cluster end at a distance $x$ from the wall for both $L$ even and odd. These
probability densities  are related to the probability density $D_L(x)$ to
have the leftmost cluster end (which coincides with the position of the defect)
at the point $x$ for the RPM for $L$ odd:
\ba
\lb{PfPd-Lodd}
\begin{array}{c}
L \mbox{\small ~odd,}\\
a \mbox{\small ~arbitrary:}
\end{array} &&
F^{(a)}_L(x)\, =\,
D_L(x)    ,\qquad\qquad x = 1,3,...,L,
\\[1mm]
\lb{PfPd-Leven}
\begin{array}{c}
L \mbox{\small ~even,}\\
a=1 :
\end{array} &&
F^{(1)}_L(x)\, =\,D_{L+1}(x+1) ,\quad\; x = 0,2,...,L.
\ea
Moreover, \emph{the conditional probabilities to have the first cluster at distance $x$
and any given configuration to the right of it are the same in the two models
for any value of $a$ if $L$ is odd, and for $a=1$ if $L$ is even.}
In particular, the densities of the contact points (definded on page \pageref{contactpoint}
and illustrated in Figures 1, 2 and 4) at a distance $x$ in the RPMW and in the RPM,
${N}^{(a)}_{L}(x)$ and ${N}_L(x)$, respectively, satisfy the relations
\ba
\lb{n_c-Lodd}
\begin{array}{c}
L \mbox{\small ~odd,}\\
a \mbox{\small ~arbitrary:}
\end{array} &&
N^{(a)}_L(x)\, =\,
N_L(x)    ,\qquad\;\qquad x = 1,3,...,L,
\\[1mm]
\lb{n_c-Leven}
\begin{array}{c}
L \mbox{\small ~even,}\\
a=1 :
\end{array} &&
N^{(1)}_L(x)\, =\,N_{L+1}(x+1) ,\;\;\quad x = 0,2,...,L.
\ea

As discussed in Appendix A, for $L$ odd, in the RPM to a one-step
Dyck path corresponds  the one defect picture in the configuration
space. The probability density to have the  defect at a distance
$x$ is obviously the same as at $(L-x)$, $~x=\frac{1}{2},
\frac{5}{2},\dots ,(L-\frac{1}{2})$. This observation implies the
following symmetry relations: \ba \lb{symm-Pf-Lodd}
\begin{array}{c}
L \mbox{\small ~odd,}\\
a \mbox{\small ~arbitrary:}
\end{array} &&
F^{(a)}_L(x)\, =\,
F^{(a)}_L(L+1-x)   ,\qquad x = 1,3,...,L,
\\[1mm]
\lb{symm-Pf-Leven}
\begin{array}{c}
L \mbox{\small ~even,}\\
a=1 :
\end{array} &&
F^{(1)}_L(x)\, =\,F^{(1)}_L(L-x) ,\qquad\qquad x = 0,2,...,L.
\ea
The symmetry relations (\ref{symm-Pf-Lodd}) and  (\ref{symm-Pf-Leven})
are surprising since one could expect
that the probability density function should be biased towards the wall.
Another important consequence is that the average size of the first
cluster is $L/2$. This is in contrast with the RPM for $L$ even when it is of
order $L^{1/3}$ (see Section 4).

Before closing this section we give a conjecture for the RPMW with the boundary rate
$a=1$.
The fraction of the interface where adsorption can take place
(this is the average number of local minima of the interface divided by $L$)
is
\be
\lb{a(L)}
A^{(1)}_L \, =\,
\left\{
\begin{array}{ll}
{\displaystyle \frac{6L^2+8L-5}{4(2L+1)(2L+3)}},&  \mbox{($L$ even)}
\\[6mm]
{\displaystyle \frac{(6L^2+14L+9)(L-1)}{4L(2L+1)(2L+3)}},& \mbox{($L$ odd)}
\end{array}.
\right.
\ee
These expressions were checked up to $L=9$.  It is amusing to note that
simple guesses are possible for $A^{(1)}_L$. A similar situation occurred also for
the RPM \cite{GNPR}. The expressions (\ref{a(L)}) will be used in
Section 4 to characterize the avalanches occurring in the model.

\section{Density profiles in the raise and peel models with and without a wall in
the stationary states}

As mentioned in the previous sections, in the continuum limit, the RPM and
the RPMW are conformal invariant and are described by a $c = 0$
conformal field theory \cite{SB,GNPyR}. If the central charge of the Virasoro algebra
vanishes, the scaling indices (highest weights of the irreducible
representations of the Virasoro algebra) are:
\be
\Delta_{p,q} = \frac{(3p-2q)^2 - 1}{24},
\lb{4.1}
\ee
where $p$ and $q$ are nonnegative integers. This implies that $\Delta_{p,q}$ is either an
integer or it is equal to $1/3$ plus an integer.

A local operator $\phi$ is characterized by its scaling dimension $X$ and
conformal spin $s$:
\be
\lb{4.2}
X = \Delta_{p,q} + \bar{\Delta}_{p',q'},\qquad s = \Delta_{p,q} - \bar{\Delta}_{p',q'}.
\ee

If $x$ is the distance from the origin (which coincides with $i$ in the
discrete versions described in Section 2, $i = 0$ corresponds to $x = 0$) and
$L$ is the size of the system,
in the scaling limit, one expects \cite{BX} the following density profile in the
case of symmetric boundaries "$r$":
\be
\lb{4.3}
\langle \phi(x/L)\rangle_{r,r}\, =\, \frac{C}{\bigl[L \sin({\textstyle \frac{\pi x}{L}})\bigr]^X},
\ee
where $C$ is a constant.

The profile of a primary operator vanishes  if its conformal
spin is not zero \cite{BX} therefore one should have
\be
\lb{4.4}
X = 2 \Delta_{p,q}.
\ee

The fact that one can make predictions for the functional dependence of
the density profiles in stationary states is a consequence of the
conformal invariance of the Hamiltonian which gives the time evolution of
the stochastic process.

If the two boundaries are different ($r \neq s$), one expects \cite{BX}:
\be
\lb{4.5}
\langle\phi(x/L)\rangle_{r,s}\, =\,
\frac{\Phi_{r,s}(\cos({\textstyle \frac{\pi x}{L}}))}{\bigl[L
\sin({\textstyle \frac{\pi x}{L}})\bigr]^X},
\ee
where the  function $\Phi_{r,s}$ is calculable.

Before discussing density profiles, it is instructive to have under our
eyes typical profiles for finite lattices. For the RPM, $L$ even, typical
profiles are shown in Ref.\cite{ALR}. In the case of the RPMW and finite $L$, the
profiles  depend on the boundary rate $a$. In Figure 8 we show for $a= 1$,
a typical configuration for $L = 128$. One notices that $h_0 = 4$ which
implies that in the corresponding defect   configuration one
has 4 defects (see Appendix A). One has a large leftmost cluster of size 124. In this paper
we will not discuss the profile of the heights (for the RPM ($L$ even) this
was done in full detail in \cite{ALR}) but we will show that the average height at the
origin behaves like $\langle h_0\rangle \sim \ln(L)$.

\begin{figure}[t]
\begin{center}
\begin{picture}(260,100)
\put(0,0){\epsfxsize=260pt\epsfbox{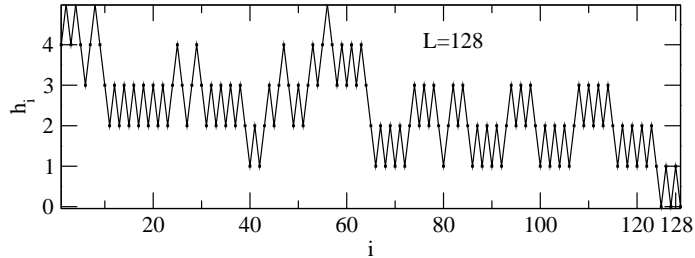}}
\end{picture}
\caption{\small Typical configuration in the stationary state for the RPMW with
a boundary rate $a = 1$. The system has a size $L = 128$. $h_i$ is the height
at site $i$.}
\end{center}
\end{figure}

We discuss several density profiles.
\medskip

\ni
{\bf 1)}~ {\em Density of contact points.}
\medskip

{\bf a)}~ {\em The RPM model ($L$ even). Density of contact points in Dyck path
configurations.}
\medskip

Since the average number of contact points is known (see (\ref{k(L)})), if
(\ref{4.3}) is valid, we have to choose $X = 1/3$.
 The functional dependence
of the density is therefore also fixed as well as the constant $C$ in (\ref{4.3}):
\be
\lb{4.6}
C\, =\, -\,\frac{\sqrt{3}}{6\pi^{5/6}}\,\Gamma(\textstyle{-\frac{1}{6}})\,\simeq\, 0.753149\, .
\ee

In order to check the prediction of conformal invariance, using
Monte Carlo simulations on large lattices, we have measured the density of
contact points $N_L(x)$. In Figure\,9 we show $N_L(x)[L\sin(\pi x/L)]^{1/3}/C$
for several lattice sizes. If the prediction (\ref{4.3}) is correct, one should
obtain the value 1 for this quantity. This is indeed the case. (Data with
poorer statistics but indicating the same result were shown already in \cite{ALR}).

\begin{figure}[t]
\begin{center}
\begin{picture}(260,170)
\put(0,0){\epsfxsize=200pt\epsfbox{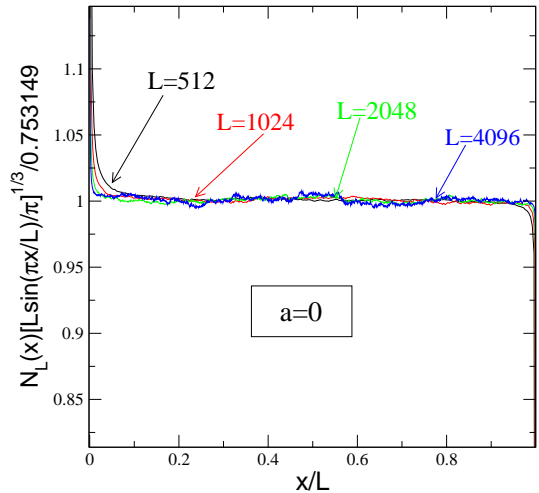}}
\end{picture}
\parbox{15cm}{
\caption{\small (Color online) The RPM for L even. The density of contact points $N_L(x)$
multiplied by $[L\sin(\pi x/L)/\pi]^{1/3}/0.753149$ as a function of $x/L$ for $L = 512$,
1024, 2048 and 4096.}
}
\end{center}
\end{figure}

One does not need to find the density of contact points for the RPM in
the case $L$ odd since it coincides (see Section 3) with the one observed in
the RPMW for a boundary rate $a = 1$ which is going to be discussed next.
\medskip

{\bf b)}~{\em The RPMW model. Density of contact points in ballot path configurations.}
\medskip

In this case the boundary conditions at $x = 0$ and $x = L$ do not coincide
since for $x = 0$ one has the wall and for $x = L$, one has $h_L = 0$, fixed.
For this reason one expects an expression like the one given by (\ref{4.5}) \cite{BX}
with the function  $\Phi_{w,0} (\cos(\pi x/L))$
to be determined (here $w$ corresponds to the wall, $0$ corresponds to $h_L=0$).

\begin{figure}[t]
\begin{center}
\begin{picture}(260,180)
\put(0,0){\epsfxsize=200pt\epsfbox{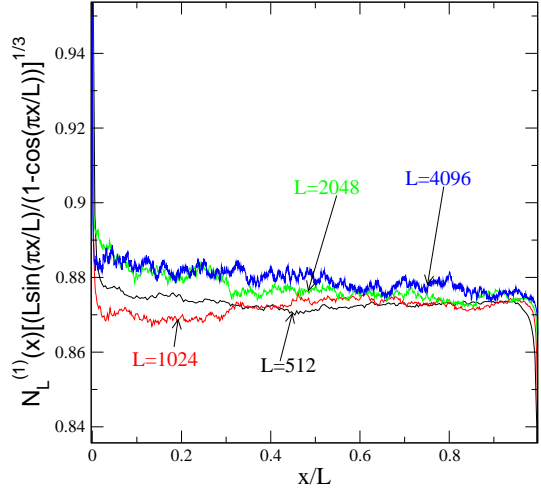}}
\end{picture}
\parbox{15cm}{
\caption{\small (Color online) The density of contact points at $x$ for a system of size $L$,
$N_L^{(1)}(x)$ multiplied by $[L\sin(\pi x/L)/(1-\cos(\pi x/L))]^{1/3}$  for a boundary
rate  $a = 1$ as a function of $x/L$ for different lattice sizes $L$ (even) $= 512$,
1024, 2048 and 4096.}
}
\end{center}
\end{figure}

We start with the case in which the boundary rate is $a = 1$. Since the
average number of clusters is known exactly (see (\ref{k(L)1B})), the
exponent $X$ in (\ref{4.3}) is $X = 1/3$. Monte Carlo simulations on large
lattices suggest the following simple expressions for $\Phi_{w,0} (\cos(\pi x/L))$:
\be
\lb{4.7}
\Phi_{w,0}\, =\, \beta\, \bigl[1 - \cos({\textstyle \frac{\pi x}{L}})\bigr]^{1/3},
\ee
where
\be
\lb{4.8}
\beta = \sqrt{3}\, \alpha \simeq 0.8757(8).
\ee
This number is compatible with the data shown in Figure\,10.

In order to make sure that the finite-size behavior of the density of
contact points is indeed  a consequence of conformal invariance, in
Figure\,11, we give the density of contact points $N_L^{(a)}(x)$ multiplied by
$[L\sin(\pi x/L)/(1-\cos(\pi x/L))]^{1/3}$ for different boundary rates  $a > 1$, and
a fixed, large value of $L$. One notices that within errors the value of $\beta$
in eq.(\ref{4.7}) stays unchanged. We repeated the study for $a < 1$ and for odd
values of $L$. For $L$ odd, as expected, we observe no $a$ dependence of the density profiles.

\begin{figure}[t]
\begin{center}
\begin{picture}(260,180)
\put(0,0){\epsfxsize=200pt\epsfbox{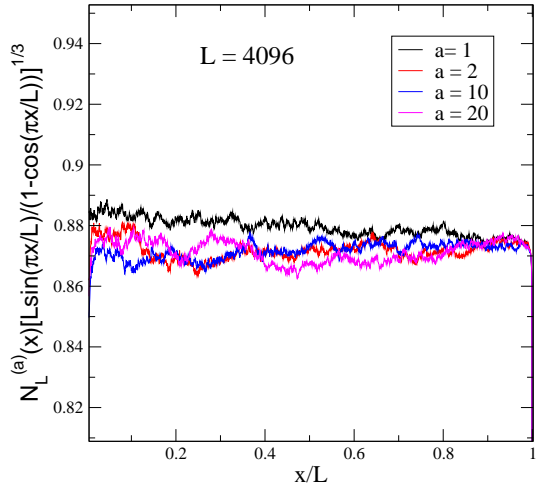}}
\end{picture}
\parbox{15cm}{
\caption{\small (Color online) Density of contact points $N_L^{(a)}(x)$ multiplied by
$[(L\sin(\pi x/L)/(1-\cos(\pi x/L))]^{1/3}$ as a function of $x/L$ for $L$ (even) $= 4096$ and
input rates $a =$ 1, 2, 5, 10 and 20.}
}
\end{center}
\end{figure}

It is known that the finite-size scaling limit of the Hamiltonian spectrum which gives the
time evolution of the system is independent of the boundary rate $a$ \cite{GNPyR}.
Consequently one could expect the space one-point function in the scaling limit
to be also independent of $a$. This is indeed the case.

To sum up, the density of contact points in the presence of a wall, has
for any boundary rate $a$ and for both $L$ even and odd, the expression:
\be
\lb{4.9}
N_L^{(a)} (x)\, \simeq\, 0.8757 \left[
\frac{1-\cos({\textstyle \frac{\pi x}{L}})}{L\sin({\textstyle \frac{\pi x}{L}})}\right]^{1/3}.
\ee

The same expression describes the density of contact points for the RPM ($L$ odd)
defined on one-step Dyck paths.

From (\ref{4.9}) one learns that for small values of $x$ (large $L$) one has very
few clusters. This can be explained by the existence of a large leftmost
cluster (see Figure\,8 and a quantitative argument below). At the other end of
the system $\, x = L - y$, $\, y$ small , the density of clusters is larger than in
the Dyck path configurations (a factor of $0.8757\cdot {2}^{1/3}\simeq 1.103$ compared with
$0.753$).

Up to now we have not  discussed   the exponent $X = 1/3$ and we are going to see
something unexpected. According to (\ref{4.4}) one gets a scaling
index $\Delta_{p,q} = 1/6$. This would imply that one has to chose $p$ and $q$ in
equation (\ref{4.1}) such that
\be
\lb{4.10}
(3p-2q)^2 = 5,
\ee
which is not possible for any rational values of $p$ and $q$. On the other
hand if one would take $\Delta_{p,q} = 1/3$ and $\bar{\Delta}_{p',q'} = 0$ in eq.(\ref{4.2}) (both
values compatible with (\ref{4.1})) one would get $X = 1/3$ and $s = 1/3$, a
non-zero value of the conformal spin and therefore a vanishing density
profile. We have not found  an explanation for this puzzle. We are going to see
that there is also a second one.
\medskip

\ni
{\bf 2)}~ {\em Density of defects.}
\medskip

We consider the RPMW in the defect configurations (see Appendix A and
Figure\,5). We have to stress that in the ballot path configurations, defects
are nonlocal observables the same way as contact points are nonlocal
observables in defect configurations. One should add that if one studies
the XXZ quantum chain \cite{NRG}, which has the same spectrum as the RPMW, both
contact points and defects are nonlocal observables. In the case of the
quantum chain the natural observable could be  the local
magnetization $\langle\sigma^z_i\rangle$ \cite{A}. One should keep in mind that in the spin basis
one looses the probabilistic interpretation of the ground state of the
Hamiltonian.

In the defect configurations, the RPMW Hamiltonian acts in the following
way. The clusters evolve as   in the RPM, adjacent defects ($D$) annihilate,
defects hop over the adjacent clusters pealing them and moving them as a result
of the hopping. What we have described up to now can be seen as a
pair annihilation process
\be
\lb{4.11}
D + D \rightarrow\emptyset
\ee
in an unquenched disordered media. This is
the situation if the boundary rate $a = 0$ and one has no
defects in the stationary state. If the boundary rate is not zero, one
injects defects in the system which compensate for the losses through
annihilation and in the stationary states one finds defects and is interested
in the density of defects profiles.

For convenience, if a defect is at the point $(x + 1/2)$, $x$ an integer, we will
shift it at $x$.

We expect the density of defects $D^{(a)}_L(x)$ to have an expression like (\ref{4.3}).
As opposed to the density of contact points $X$ is unknown, since the total
number of defects, equal to the average height
 at the origin $\langle h_0\rangle$ in the ballot path
configurations, is not known.

We have observed that one has to consider separately the scaling
properties of the density of defects on even and odd sites (denoted by
$D^{(a)\,\rm e}_{L}(x)$ and $D^{(a)\,\rm o}_{L}(x)$ respectively). Monte Carlo
simulations on large lattices have shown that
the scaling dimension of the
local operator corresponding to the defects is:
\be
\lb{4.12}
X = 1,
\ee
and therefore one has:
\be
\lb{4.13}
D_L^{(a)\, \rm e/o}(x) =
\frac{G^{(a)\,\rm e/o}(\cos({\textstyle \frac{\pi x}{L}}))}{L\sin({\textstyle \frac{\pi x}{L}})}.
\ee

\begin{figure}[t]
\begin{center}
\begin{picture}(260,180)
\put(0,0){\epsfxsize=200pt\epsfbox{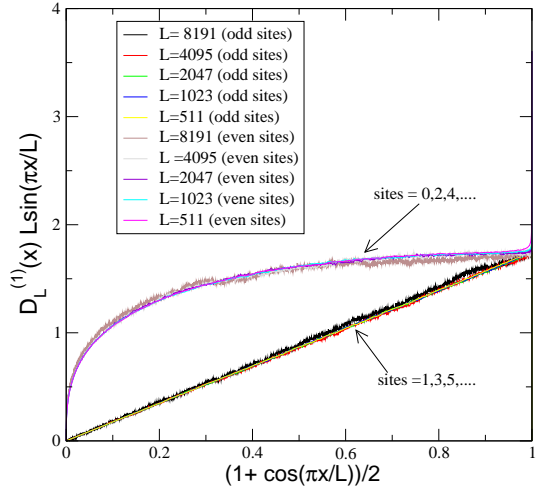}}
\end{picture}
\parbox{15cm}{
\caption{\small (Color online) The density of defects $D^{(1)}_L(x)$ multiplied by $L\sin(\pi x/L)$ as a
function of $(1 + \cos(\pi x/L))/2$ for different lattice sizes. The profiles
for even and odd sites are shown separately.}
}
\end{center}
\end{figure}

In Figure12 one can see the scaling functions for an input rate $a = 1$.
Surprisingly, $G^{(1)\,\rm o}$ has a simple expression but not $G^{(1)\, \rm e}$. One observes that
$G^{(1)\,\rm e}(1) = G^{(1)\,\rm o}(1) \simeq 1.75$. This implies that near the source we have the universal behavior:
\be
\lb{4.14}
D^{(1)}_L(x)\, \simeq\, \frac{1.75}{\pi x}\, \simeq\, 0.577\,\frac{1}{x}
\ee
for both even and odd sites,
and that the total number of defects, equal to $\langle h_{0}\rangle_L^{(1)}$ in the ballot path
picture, increases logarithmically with $L$. A careful counting gives:
\be
\lb{4.15}
\langle h_0\rangle_L^{(1)}\,\simeq\, 0.467\, +\, 0.577 \ln L\, .
\ee

We have checked that the results presented above are independent of the
boundary rate $a$ as expected from conformal invariance. We now try to
understand the exponent $X = 1$ (see (\ref{4.12})). This would imply (see (\ref{4.1}) and
(\ref{4.2})):
\be
\lb{4.16}
\Delta_{p,q}\, =\, 1/2\quad \mbox{and}\quad (3p-2q)^2\, =\, 13,
\ee
which as  in the case of the density of contact points, is not possible.
Similar to the previous case, one could take $\Delta_{p,q} = 1$, $\bar{\Delta}_{p',q'} = 0$, both
values compatible with (\ref{4.1}) and get a nonzero conformal spin.
 Notice that the $1/x$ fall off (\ref{4.14}) of $D^{(1)}_L(x)$ is the mean-field result for
the case in which the hopping is local and symmetric and the input of
particles is also local \cite{CRL}. The same behavior is obtained when
fluctuations are taken into account \cite{HRS}. In our case
neither the source acts locally nor is the symmetric hopping local.
\medskip

\ni
{\bf 3)}~ {\em Probability density for the first cluster.}
\medskip

{\bf a)}~ {\em The RPM ($L$ even).}
\medskip

We consider the probability $F_L(x)$ to have the first cluster end at the
point $x$ ({\it i. e.} the second contact point at $x$ in a
Dyck path). In Figure~13 we show how this function scales:
\be
\lb{4.17}
F_L(x) \simeq f({\textstyle \frac{x}{L}})/x^{5/3}\, .
\ee
This implies that the first cluster increases like $3.3\, L^{1/3}$, similar to
the average cluster size $1/0.738488\, L^{1/3} \simeq 1.354118\,  L^{1/3}$
(see (\ref{k(L)})).
No surprises here.

\begin{figure}[t]
\begin{center}
\begin{picture}(260,180)
\put(0,0){\epsfxsize=200pt\epsfbox{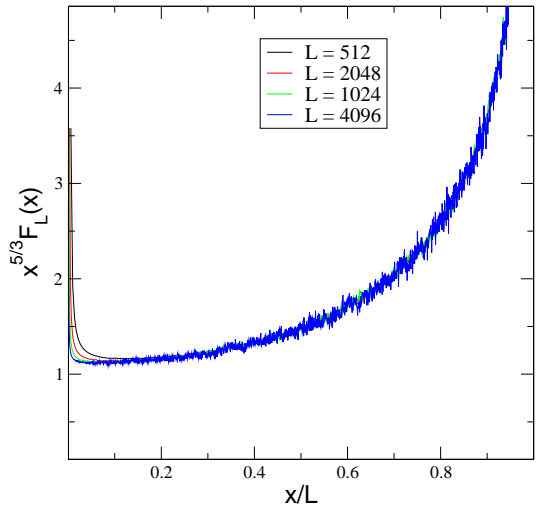}}
\end{picture}
\parbox{15cm}{
\caption{\small (Color online) The probability $F_L(x)$ to have the first cluster ending  at $x$ in the
RPM ($L$ even) multiplied by $x^{5/3}$, as a function of $x/L$.}
}
\end{center}
\end{figure}

\medskip

{\bf b)}~ {\em The RPM and the RPMW for $L$ odd.}
\medskip

This case is much more interesting. Firstly let us recall that
$F^{(a)}_L(x)$ coincides with the probability of having the first cluster end at $x$ in the
one-step Dyck path configurations in the RPM model ($L$ odd) (see (\ref{PfPd-Lodd})
and (\ref{PfPd-Leven})).
The latter coincides with the of having  the defect at $(x + 1/2)$ in
the one-defect configurations. This suggests that $F^{(a)}_L(x,L)$ might have
other properties.

Let us consider the following, normalized to 1, probability density
function:
\be
\lb{4.18}
p(x,L)\, =\, \frac{C}{L\bigl[\sin({\textstyle \frac{\pi x}{L}})\bigr]^{1/3}},
\ee
where
\be
\lb{4.19}
C\,  =\, 2\sqrt{\pi}\ \Gamma({\textstyle \frac{5}{6}})/\Gamma({\textstyle \frac{1}{3}})\, \simeq\, 1.4936\, .
\ee

In Figure\,14 we show that in the scaling limit, $F^{(a)}_L(x)$ stays a probability
density distribution given by (\ref{4.18}) for any boundary rate $a$. This is
unexpected and should be
understood. Note the reappearance of the exponent $1/3$ in (\ref{4.18})

Notice that the first cluster is either small or very big
($p(x,L) = p(L-x,L)$) and that the average size of the first cluster is $L/2$ (this is
a linear dependence as opposed to the $L^{1/3}$ dependence in the RPM with
 $L$ even).

\begin{figure}[t]
\begin{center}
\begin{picture}(260,180)
\put(0,0){\epsfxsize=200pt\epsfbox{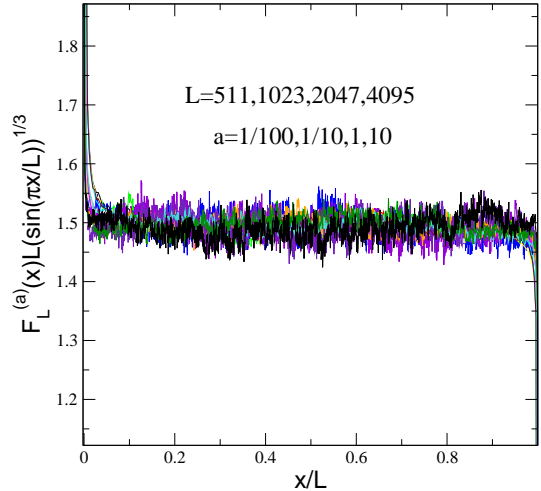}}
\end{picture}
\parbox{15cm}{
\caption{\small (Color online) The probability distribution function to have the first cluster ending
at $x$, for a system of size $L$, $F^{(a)}_L(x)$ multiplied by $L(\sin(\pi x/L)^{1/3}$ as a
function of $x/L$ for different values of $L$ and boundary rates $a$.}
}
\end{center}
\end{figure}

How one can derive a probability density  (\ref{4.18}) from conformal field theory
is not clear to us.

One of the most interesting aspects of our study is the deep connection between the RPM
for $L$ odd (one-step Dyck paths) and the RPMW model in the defect configurations
(corresponding to ballot paths). The rightmost defect behaves like the single defect
which defines the position of the leftmost cluster in the RPM (one-step Dyck paths
picture). Like the single defect, the rightmost defect makes L\'{e}vy flights and behaves like a
"relativistic" random walker \cite{AR1}. The physics to the right of the leftmost cluster is
identical in both cases, therefore the density of contact points in the two models is the same.
The difference between the two models is all in the first cluster.
In the RPMW model one has defects in the first cluster, while in the RPM
there are no defects.
The content of the first cluster in the RPM can be easier understood in the one-defect picture
in which the time-evolution operator acts in the left-right symmetric way.

Before closing this section, let us discuss shortly the implication of
the conjectures (\ref{a(L)}) for $A_L^{(1)}$, the average number  of local minima of the
interface divided by $L$ in the RPMW. In the large $L$ limit one obtains
\be
\lb{4.20}
 A_L^{(1)}\,
 \stackrel{\raisebox{2pt}{$\scriptscriptstyle L\rightarrow\infty$}}{\longrightarrow}\,
 3/8\, .
\ee
This number is identical with the one observed in the RPM ($L$ even) \cite{GNPR}.
This implies (see \cite{GNPR}) that the average number of tiles desorbed in
avalanches is again $1.5$. This coincidence suggests that most of the
desorption processes take place within the clusters.
The different cluster structures in the RPM and the RPMW, do not
affect most of the avalanches. This does not imply that the rare large
avalanches \cite{ALR} are the same  in the two models.

\section{The raise and peel model with and without a wall in continuous time.
Temperley-Lieb algebras}

In this section we  show how the RPM and RPMW  described
in Section 2 were obtained. We also give the necessary background to
derive the exact results given in Section 3 and derived in Sections 7 and 8.

\medskip
The continuous time evolution of a system composed of the states
$\alpha = 1,2,\ldots, n$
with probabilities $P_\alpha(t)$ is given by a master equation that can be interpreted
as an imaginary time Schr\"odinger equation:
\begin{equation}
\label{5.1}
\frac{d}{dt} P_\alpha(t) = -\sum_\beta H_{\alpha,\beta} P_\beta(t),
\end{equation}
where the Hamiltonian $H$ is an $n\times n$ intensity matrix: $H_{\alpha,\beta}$
nonpositive ($\alpha \neq \beta$)
 and $\sum_\alpha H_{\alpha,\beta} = 0$. $-H_{\alpha,\beta}$ is the rate for the transition
$\ket \beta \rightarrow \ket \alpha$. The ground-state eigenvector of the system
$\ket0$, $H \ket0 = 0$, gives
the probabilities in the stationary state:
\begin{equation}
\label{5.2}
\ket0 = \sum_\alpha P_\alpha \ket \alpha,\;\;\;\;\;\; P_\alpha = \lim_{t \to \infty}  P_\alpha(t).
\end{equation}
 The normalization factor of the unnormalized probabilities $P_\alpha$ is
$\langle 0 |\, 0\rangle$ where
\begin{equation}
\label{5.3}
\bra0 = \sum_\alpha \bra \alpha, \;\;\;\;\;  \bra0 H = 0.
\end{equation}
Since $H$ is an intensity matrix, the real parts of the $n$ eigenvalues $E_\alpha$
are nonnegative and, if complex, the eigenvalues come in conjugate pairs.
The eigenvalue zero is not degenerate.

For our present purposes we need the following observation \cite{AR2}: if $g_i$
($i=1,\dots ,p$) are generators of a semigroup algebra, then:
\be
\lb{5.4}
H = \sum_{i=1}^p a_i(1-g_i)
\ee
acting from the left in the vector space defined by the regular
representation
or by a left ideal of the algebra, is an intensity matrix ($a_i$ are
non-negative real numbers).

In the following we are going to use the Temperley-Lieb (TL) and the
one-boundary Temperley-Lieb semigroup algebras.

The Temperley-Lieb algebra ${\mathcal T}_L$ \cite{TL} is
given by a set of $(L-1)$ generators $e_i$, $i=1,\ldots,L-1$, subject to  relations
\be
\lb{TL}
e_i^2 = \tau\, e_i, \qquad e_ie_{i\pm1}e_i = e_i, \qquad
e_ie_j = e_je_i\;\;\; \mbox{if}\;\; |i-j| >1.
\label{TLdef}
\ee
Here $\tau$ is the parameter of the algebra which for our applications we put equal
to one: $\tau =1$. We notice that with this choice
all structure constants of the algebra become units and the TL algebra becomes
a semigroup algebra
(see \cite{AR2}).

The one-boundary extension of the Temperley-Lieb algebra ${\mathcal T}_L^{(1B)}$ (also known under the name of
blob algebra \cite{MS}) is obtained by adding a boundary generator
$e_0$ satisfying relations
\be
\lb{1bTL}
e_0^2 = \omega\, e_0,\qquad e_1e_0e_1 = e_1,
\qquad
e_0e_i = e_ie_0 \quad\forall\, i >1.
\ee
We fix again the parameter $\omega =1$ and get a semigroup
algebra.

For our purposes it is convenient to represent the generators $e_i$ and the
boundary generator $e_0$ as, respectively, tiles and a half-tile:
\be
\lb{pic1}
\raisebox{9pt}{$e_i\, =\;\;$}
\begin{picture}(25,30)(0,7)
{\thicklines
\put(0,20){\line(1,1){10}}
\put(0,20){\line(1,-1){10}}
\put(20,20){\line(-1,1){10}}
\put(20,20){\line(-1,-1){10}}
}
\qbezier[25](10,0)(10,20)(10,40)
\put(12,-2){$\scriptstyle i$}
\end{picture}
\raisebox{9pt}{$,\qquad\qquad e_0\, =\;\;$}
\begin{picture}(25,30)(0,7)
{\thicklines
\put(0,10){\line(0,1){20}}
\put(10,20){\line(-1,1){10}}
\put(10,20){\line(-1,-1){10}}
}
\put(2,-2){$\scriptstyle 0$}
\qbezier[25](0,0)(0,20)(0,40)
\end{picture}
\ee
The tiles and the half-tile are placed on  labeled vertical lines
and they can move freely along the lines unless they meet other
(half-)tiles:
$$
\begin{picture}(100,45)(0,2)
{\thicklines
\put(0,5){\line(0,1){20}}
\put(10,15){\line(-1,1){10}}
\put(10,15){\line(-1,-1){10}}
\put(10,35){\line(1,1){10}}
\put(10,35){\line(1,-1){10}}
\put(30,35){\line(-1,1){10}}
\put(30,35){\line(-1,-1){10}}
\put(80,12){\line(1,1){10}}
\put(80,12){\line(1,-1){10}}
\put(100,12){\line(-1,1){10}}
\put(100,12){\line(-1,-1){10}}
\put(50,25){\line(1,1){10}}
\put(50,25){\line(1,-1){10}}
\put(70,25){\line(-1,1){10}}
\put(70,25){\line(-1,-1){10}}
}
\qbezier[30](0,0)(0,20)(0,50)
\qbezier[30](10,0)(10,25)(10,50)
\qbezier[30](20,0)(20,25)(20,50)
\qbezier[30](30,0)(30,25)(30,50)
\qbezier[30](40,0)(40,25)(40,50)
\qbezier[30](50,0)(50,25)(50,50)
\qbezier[30](60,0)(60,25)(60,50)
\qbezier[30](70,0)(70,25)(70,50)
\qbezier[30](80,0)(80,25)(80,50)
\qbezier[30](90,0)(90,25)(90,50)
\qbezier[30](100,0)(100,25)(100,50)
\put(0,-7){$\scriptscriptstyle 0$}
\put(10,-7){$\scriptscriptstyle 1$}
\put(20,-7){$\scriptscriptstyle 2$}
\put(30,-7){$\scriptscriptstyle 3$}
\put(38,-7){$\scriptscriptstyle \dots$}
\put(47,-7){$\scriptscriptstyle i\!-\!1$}
\put(60,-7){$\scriptscriptstyle i$}
\put(67,-7){$\scriptscriptstyle i\!+\!1$}
\put(87,-7){$\scriptscriptstyle \dots$}
\end{picture}
\quad \raisebox{15pt}{=}\quad
\begin{picture}(100,45)(0,2)
{\thicklines
\put(0,25){\line(0,1){20}}
\put(10,35){\line(-1,1){10}}
\put(10,35){\line(-1,-1){10}}
\put(10,12){\line(1,1){10}}
\put(10,12){\line(1,-1){10}}
\put(30,12){\line(-1,1){10}}
\put(30,12){\line(-1,-1){10}}
\put(80,35){\line(1,1){10}}
\put(80,35){\line(1,-1){10}}
\put(100,35){\line(-1,1){10}}
\put(100,35){\line(-1,-1){10}}
\put(50,22){\line(1,1){10}}
\put(50,22){\line(1,-1){10}}
\put(70,22){\line(-1,1){10}}
\put(70,22){\line(-1,-1){10}}
}
\qbezier[30](0,0)(0,20)(0,50)
\qbezier[30](10,0)(10,25)(10,50)
\qbezier[30](20,0)(20,25)(20,50)
\qbezier[30](30,0)(30,25)(30,50)
\qbezier[30](40,0)(40,25)(40,50)
\qbezier[30](50,0)(50,25)(50,50)
\qbezier[30](60,0)(60,25)(60,50)
\qbezier[30](70,0)(70,25)(70,50)
\qbezier[30](80,0)(80,25)(80,50)
\qbezier[30](90,0)(90,25)(90,50)
\qbezier[30](100,0)(100,25)(100,50)
\put(0,-7){$\scriptscriptstyle 0$}
\put(10,-7){$\scriptscriptstyle 1$}
\put(20,-7){$\scriptscriptstyle 2$}
\put(30,-7){$\scriptscriptstyle 3$}
\put(38,-7){$\scriptscriptstyle \dots$}
\put(47,-7){$\scriptscriptstyle i\!-\!1$}
\put(60,-7){$\scriptscriptstyle i$}
\put(67,-7){$\scriptscriptstyle i\!+\!1$}
\put(87,-7){$\scriptscriptstyle \dots$}
\end{picture}
$$
Multiplication in the algebra corresponds to a simultaneous
placement of several (half-)tiles on the same picture and
an order of the product corresponds to moving the tiles downwards.
Thus, the picture above corresponds to the commutation relations in (\ref{TL})
and (\ref{1bTL}).
The remaining relations in (\ref{TL}) and (\ref{1bTL}) are equivalent to the following
pictures
\be
\lb{rule-1}
\begin{picture}(40,45)(0,2)
{\thicklines
\put(10,35){\line(1,1){10}}
\put(10,35){\line(1,-1){10}}
\put(30,35){\line(-1,1){10}}
\put(30,35){\line(-1,-1){10}}
\put(10,15){\line(1,1){10}}
\put(10,15){\line(1,-1){10}}
\put(30,15){\line(-1,1){10}}
\put(30,15){\line(-1,-1){10}}
}
\qbezier[30](0,0)(0,25)(0,50)
\qbezier[30](10,0)(10,25)(10,50)
\qbezier[30](20,0)(20,25)(20,50)
\qbezier[30](30,0)(30,25)(30,50)
\qbezier[30](40,0)(40,25)(40,50)
\put(-2,-7){$\scriptscriptstyle \dots$}
\put(7,-7){$\scriptscriptstyle i\!-\!1$}
\put(20,-7){$\scriptscriptstyle i$}
\put(27,-7){$\scriptscriptstyle i\!+\!1$}
\put(37,-7){$\scriptscriptstyle \dots$}
\end{picture}
\quad \raisebox{18pt}{=} \quad
\begin{picture}(40,45)(0,2)
{\thicklines
\put(10,35){\line(1,1){10}}
\put(10,35){\line(1,-1){10}}
\put(30,35){\line(-1,1){10}}
\put(30,35){\line(-1,-1){10}}
\put(0,25){\line(1,1){10}}
\put(0,25){\line(1,-1){10}}
\put(20,25){\line(-1,1){10}}
\put(20,25){\line(-1,-1){10}}
\put(10,15){\line(1,1){10}}
\put(10,15){\line(1,-1){10}}
\put(30,15){\line(-1,1){10}}
\put(30,15){\line(-1,-1){10}}
}
\qbezier[30](0,0)(0,25)(0,50)
\qbezier[30](10,0)(10,25)(10,50)
\qbezier[30](20,0)(20,25)(20,50)
\qbezier[30](30,0)(30,25)(30,50)
\qbezier[30](40,0)(40,25)(40,50)
\put(-2,-7){$\scriptscriptstyle \dots$}
\put(7,-7){$\scriptscriptstyle i\!-\!1$}
\put(20,-7){$\scriptscriptstyle i$}
\put(27,-7){$\scriptscriptstyle i\!+\!1$}
\put(37,-7){$\scriptscriptstyle \dots$}
\end{picture}
\quad \raisebox{18pt}{=}\quad
\begin{picture}(40,45)(0,2)
{\thicklines
\put(10,35){\line(1,1){10}}
\put(10,35){\line(1,-1){10}}
\put(30,35){\line(-1,1){10}}
\put(30,35){\line(-1,-1){10}}
\put(20,25){\line(1,1){10}}
\put(20,25){\line(1,-1){10}}
\put(40,25){\line(-1,1){10}}
\put(40,25){\line(-1,-1){10}}
\put(10,15){\line(1,1){10}}
\put(10,15){\line(1,-1){10}}
\put(30,15){\line(-1,1){10}}
\put(30,15){\line(-1,-1){10}}
}
\qbezier[30](0,0)(0,25)(0,50)
\qbezier[30](10,0)(10,25)(10,50)
\qbezier[30](20,0)(20,25)(20,50)
\qbezier[30](30,0)(30,25)(30,50)
\qbezier[30](40,0)(40,25)(40,50)
\put(-2,-7){$\scriptscriptstyle \dots$}
\put(7,-7){$\scriptscriptstyle i\!-\!1$}
\put(20,-7){$\scriptscriptstyle i$}
\put(27,-7){$\scriptscriptstyle i\!+\!1$}
\put(37,-7){$\scriptscriptstyle \dots$}
\end{picture}
\quad \raisebox{18pt}{=}\quad
\begin{picture}(40,45)(0,2)
{\thicklines
\put(10,25){\line(1,1){10}}
\put(10,25){\line(1,-1){10}}
\put(30,25){\line(-1,1){10}}
\put(30,25){\line(-1,-1){10}}
}
\qbezier[30](0,0)(0,25)(0,50)
\qbezier[30](10,0)(10,25)(10,50)
\qbezier[30](20,0)(20,25)(20,50)
\qbezier[30](30,0)(30,25)(30,50)
\qbezier[30](40,0)(40,25)(40,50)
\put(-2,-7){$\scriptscriptstyle \dots$}
\put(7,-7){$\scriptscriptstyle i\!-\!1$}
\put(20,-7){$\scriptscriptstyle i$}
\put(27,-7){$\scriptscriptstyle i\!+\!1$}
\put(37,-7){$\scriptscriptstyle \dots$}
\end{picture}\;\;,\quad
\ee

\be
\lb{rule-2}
\begin{picture}(40,45)(0,2)
{\thicklines
\put(10,4){\line(0,1){20}}
\put(20,14){\line(-1,1){10}}
\put(20,14){\line(-1,-1){10}}
\put(10,26){\line(0,1){20}}
\put(20,36){\line(-1,1){10}}
\put(20,36){\line(-1,-1){10}}
}
\qbezier[30](10,0)(10,25)(10,50)
\qbezier[30](20,0)(20,25)(20,50)
\qbezier[30](30,0)(30,25)(30,50)
\qbezier[30](40,0)(40,25)(40,50)
\put(10,-7){$\scriptscriptstyle 0$}
\put(20,-7){$\scriptscriptstyle 1$}
\put(30,-7){$\scriptscriptstyle 2$}
\put(37,-7){$\scriptscriptstyle \dots$}
\end{picture}
\quad \raisebox{18pt}{=}
\begin{picture}(40,45)(0,2)
{\thicklines
\put(10,15){\line(0,1){20}}
\put(20,25){\line(-1,1){10}}
\put(20,25){\line(-1,-1){10}}
}
\qbezier[30](10,0)(10,25)(10,50)
\qbezier[30](20,0)(20,25)(20,50)
\qbezier[30](30,0)(30,25)(30,50)
\qbezier[30](40,0)(40,25)(40,50)
\put(10,-7){$\scriptscriptstyle 0$}
\put(20,-7){$\scriptscriptstyle 1$}
\put(30,-7){$\scriptscriptstyle 2$}
\put(37,-7){$\scriptscriptstyle \dots$}
\end{picture}
\;\;  ,\;\;\qquad
\begin{picture}(40,45)(0,2)
{\thicklines
\put(10,15){\line(0,1){20}}
\put(10,35){\line(1,1){10}}
\put(10,35){\line(1,-1){10}}
\put(30,35){\line(-1,1){10}}
\put(30,35){\line(-1,-1){10}}
\put(10,15){\line(1,1){10}}
\put(10,15){\line(1,-1){10}}
\put(30,15){\line(-1,1){10}}
\put(30,15){\line(-1,-1){10}}
}
\qbezier[30](10,0)(10,25)(10,50)
\qbezier[30](20,0)(20,25)(20,50)
\qbezier[30](30,0)(30,25)(30,50)
\qbezier[30](40,0)(40,25)(40,50)
\put(10,-7){$\scriptscriptstyle 0$}
\put(20,-7){$\scriptscriptstyle 1$}
\put(30,-7){$\scriptscriptstyle 2$}
\put(37,-7){$\scriptscriptstyle \dots$}
\end{picture}
\quad \raisebox{18pt}{=}
\begin{picture}(40,47)(0,2)
{\thicklines
\put(10,25){\line(1,1){10}}
\put(10,25){\line(1,-1){10}}
\put(30,25){\line(-1,1){10}}
\put(30,25){\line(-1,-1){10}}
}
\qbezier[30](10,0)(10,25)(10,50)
\qbezier[30](20,0)(20,25)(20,50)
\qbezier[30](30,0)(30,25)(30,50)
\qbezier[30](40,0)(40,25)(40,50)
\put(10,-7){$\scriptscriptstyle 0$}
\put(20,-7){$\scriptscriptstyle 1$}
\put(30,-7){$\scriptscriptstyle 2$}
\put(37,-7){$\scriptscriptstyle \dots$}
\end{picture}\;\; .
\quad\;\;
\vspace{5mm}
\ee

We consider now the following projectors in the TL and the one-boundary TL algebras:
\ba
\raisebox{12pt}{$\left. \begin{array}{r}
\mbox{\small Temperley-Lieb algebra and}\\
\mbox{\small one-boundary  TL algebra,}
\end{array}\right\}$
$L$ even:}&&
\begin{picture}(82,35)
{\thicklines
\put(0,15){\line(1,1){10}}
\put(0,15){\line(1,-1){10}}
\put(10,25){\line(1,-1){20}}
\put(10,5){\line(1,1){20}}
\put(30,25){\line(1,-1){10}}
\put(30,5){\line(1,1){10}}
\put(60,15){\line(1,-1){10}}
\put(60,15){\line(1,1){10}}
\put(70,5){\line(1,1){10}}
\put(70,25){\line(1,-1){10}}
\put(47.5,14){$\scriptstyle \dots $}
}
\qbezier[20](0,0)(0,15)(0,30)
\qbezier[20](10,0)(10,15)(10,30)
\qbezier[20](20,0)(20,15)(20,30)
\qbezier[20](30,0)(30,15)(30,30)
\qbezier[20](40,0)(40,15)(40,30)
\qbezier[20](60,0)(60,15)(60,30)
\qbezier[20](70,0)(70,15)(70,30)
\qbezier[20](80,0)(80,15)(80,30)
\put(0,-7){$\scriptscriptstyle 0$}
\put(10,-7){$\scriptscriptstyle 1$}
\put(20,-7){$\scriptscriptstyle 2$}
\put(30,-7){$\scriptscriptstyle 3$}
\put(40,-7){$\scriptscriptstyle 4$}
\put(52,-7){$\scriptscriptstyle \ldots$}
\put(65,-7){$\scriptscriptstyle L\! -\! 1$}
\put(79,-7){$\scriptscriptstyle L$}
\end{picture}
\quad \raisebox{12pt}{$=\;\;\,  e_1 e_3 \dots e_{L-1}$}\qquad
\nonumber
\\[5mm]
\raisebox{12pt}{Temperley-Lieb algebra,~~~ $L$ odd:}&&
\begin{picture}(92,30)
{\thicklines
\put(10,15){\line(1,-1){10}}
\put(10,15){\line(1,1){10}}
\put(20,25){\line(1,-1){20}}
\put(20,5){\line(1,1){20}}
\put(40,5){\line(1,1){10}}
\put(40,25){\line(1,-1){10}}
\put(70,15){\line(1,-1){10}}
\put(70,15){\line(1,1){10}}
\put(80,25){\line(1,-1){10}}
\put(80,5){\line(1,1){10}}
\put(57.5,14){$\scriptstyle \dots $}
}
\qbezier[20](0,0)(0,15)(0,30)
\qbezier[20](10,0)(10,15)(10,30)
\qbezier[20](20,0)(20,15)(20,30)
\qbezier[20](30,0)(30,15)(30,30)
\qbezier[20](40,0)(40,15)(40,30)
\qbezier[20](50,0)(50,15)(50,30)
\qbezier[20](70,0)(70,15)(70,30)
\qbezier[20](80,0)(80,15)(80,30)
\qbezier[20](90,0)(90,15)(90,30)
\put(0,-7){$\scriptscriptstyle 0$}
\put(10,-7){$\scriptscriptstyle 1$}
\put(20,-7){$\scriptscriptstyle 2$}
\put(30,-7){$\scriptscriptstyle 3$}
\put(40,-7){$\scriptscriptstyle 4$}
\put(50,-7){$\scriptscriptstyle 5$}
\put(62,-7){$\scriptscriptstyle \ldots$}
\put(75,-7){$\scriptscriptstyle L\! -\! 1$}
\put(89,-7){$\scriptscriptstyle L$}
\end{picture}
\quad \raisebox{12pt}{$=\;\;\,  e_2 e_4 \ldots e_{L-1}$}\qquad
\nonumber
\\[5mm]
\raisebox{12pt}{one-boundary  TL algebra~~~ $L$ odd:}&&
\begin{picture}(92,30)
{\thicklines
\put(0,5){\line(0,1){20}}
\put(0,25){\line(1,-1){20}}
\put(0,5){\line(1,1){20}}
\put(20,25){\line(1,-1){20}}
\put(20,5){\line(1,1){20}}
\put(40,5){\line(1,1){10}}
\put(40,25){\line(1,-1){10}}
\put(70,15){\line(1,-1){10}}
\put(70,15){\line(1,1){10}}
\put(80,25){\line(1,-1){10}}
\put(80,5){\line(1,1){10}}
}
\qbezier[20](0,0)(0,15)(0,30)
\qbezier[20](10,0)(10,15)(10,30)
\qbezier[20](20,0)(20,15)(20,30)
\qbezier[20](30,0)(30,15)(30,30)
\qbezier[20](40,0)(40,15)(40,30)
\qbezier[20](50,0)(50,15)(50,30)
\qbezier[20](70,0)(70,15)(70,30)
\qbezier[20](80,0)(80,15)(80,30)
\qbezier[20](90,0)(90,15)(90,30)
\put(0,-7){$\scriptscriptstyle 0$}
\put(10,-7){$\scriptscriptstyle 1$}
\put(20,-7){$\scriptscriptstyle 2$}
\put(30,-7){$\scriptscriptstyle 3$}
\put(40,-7){$\scriptscriptstyle 4$}
\put(50,-7){$\scriptscriptstyle 5$}
\put(62,-7){$\scriptscriptstyle \ldots$}
\put(75,-7){$\scriptscriptstyle L\! -\! 1$}
\put(89,-7){$\scriptscriptstyle L$}
\end{picture}
\quad \raisebox{15pt}{$=\quad e_0 e_2 e_4 \ldots e_{L-1}$}\qquad
\nonumber
\ea
We multiply the projectors by all elements of the
respective algebras from the left (that is, dropping the (half-)tiles
downwards) and generate in this way
left ideals (left-invariant subspaces in the algebras).
The projectors are non-invertible
and therefore the ideals are smaller then the algebras themselves.
The dimensions of the resulting ideals are: $C_L$ (see eq.\,(\ref{2.2})) for the TL algebra
for $L$ even, $C_{L+1}$ for the TL algebra for $L$ odd, and
${L\choose \lfloor L/2 \rfloor}\,$ for{\smallskip} the one-boundary TL algebra.
These dimensions coincide with the numbers of the Dyck paths, one-step Dyck paths and
the ballot paths, respectively. One can easily see a correspondence between
the paths and the pictures of words in the ideals.
The correspondence is illustrated  below.
\ba
\raisebox{25pt}{Dyck path:}&&
\begin{picture}(80,55)(0,-10)
\put(0,15){\line(1,0){80}}
{\thicklines
\put(0,15){\line(1,1){10}}
\put(10,25){\line(1,-1){10}}
\put(20,15){\line(1,1){20}}
\put(40,35){\line(1,-1){10}}
\put(50,25){\line(1,1){10}}
\put(60,35){\line(1,-1){20}}
}
\qbezier[10](0,0)(0,7.5)(0,15)
\qbezier[10](10,0)(10,7.5)(10,15)
\qbezier[10](20,0)(20,7.5)(20,15)
\qbezier[10](30,0)(30,7.5)(30,15)
\qbezier[10](40,0)(40,7.5)(40,15)
\qbezier[10](50,0)(50,7.5)(50,15)
\qbezier[10](60,0)(60,7.5)(60,15)
\qbezier[10](70,0)(70,7.5)(70,15)
\qbezier[10](80,0)(80,7.5)(80,15)
\put(0,-7){$\scriptscriptstyle 0$}
\put(10,-7){$\scriptscriptstyle 1$}
\put(20,-7){$\scriptscriptstyle 2$}
\put(30,-7){$\scriptscriptstyle 3$}
\put(50,-7){$\scriptscriptstyle \dots$}
\put(78,-7){$\scriptscriptstyle L=8$}
\end{picture}
\quad \raisebox{25pt}{$\Leftrightarrow$}\quad
\begin{picture}(80,52)(0,-10)
{\thicklines
\put(0,15){\line(1,1){10}}
\put(10,25){\line(1,-1){20}}
\put(0,15){\line(1,-1){10}}
\put(10,5){\line(1,1){30}}
\put(40,35){\line(1,-1){30}}
\put(30,25){\line(1,-1){20}}
\put(30,5){\line(1,1){30}}
\put(50,5){\line(1,1){20}}
\put(70,5){\line(1,1){10}}
\put(60,35){\line(1,-1){20}}
}
\qbezier[25](0,0)(0,20)(0,40)
\qbezier[25](10,0)(10,20)(10,40)
\qbezier[25](20,0)(20,20)(20,40)
\qbezier[25](30,0)(30,20)(30,40)
\qbezier[25](40,0)(40,20)(40,40)
\qbezier[25](50,0)(50,20)(50,40)
\qbezier[25](60,0)(60,20)(60,40)
\qbezier[25](70,0)(70,20)(70,40)
\qbezier[25](80,0)(80,20)(80,40)
\put(0,-7){$\scriptscriptstyle 0$}
\put(10,-7){$\scriptscriptstyle 1$}
\put(20,-7){$\scriptscriptstyle 2$}
\put(30,-7){$\scriptscriptstyle 3$}
\put(40,-7){$\scriptscriptstyle 4$}
\put(50,-7){$\scriptscriptstyle 5$}
\put(60,-7){$\scriptscriptstyle 6$}
\put(70,-7){$\scriptscriptstyle 7$}
\put(77,-7){$\scriptscriptstyle L=8$}
\end{picture}
\quad \raisebox{25pt}{$=\quad e_4 e_6 e_1 e_3 e_5 e_7$}\qquad
\nn
\\[7mm]
\raisebox{25pt}{One-step Dyck path:}&&
\begin{picture}(65,50)(0,-10)
\put(0,15){\line(1,0){70}}
{\thicklines
\put(0,15){\line(0,1){10}}
\put(0,25){\line(1,1){20}}
\put(20,45){\line(1,-1){20}}
\put(40,25){\line(1,1){10}}
\put(50,35){\line(1,-1){20}}
}
\qbezier[10](0,0)(0,7.5)(0,15)
\qbezier[10](10,0)(10,7.5)(10,15)
\qbezier[10](20,0)(20,7.5)(20,15)
\qbezier[10](30,0)(30,7.5)(30,15)
\qbezier[10](40,0)(40,7.5)(40,15)
\qbezier[10](50,0)(50,7.5)(50,15)
\qbezier[10](60,0)(60,7.5)(60,15)
\qbezier[10](70,0)(70,7.5)(70,15)
\put(0,-7){$\scriptscriptstyle 0$}
\put(10,-7){$\scriptscriptstyle 1$}
\put(20,-7){$\scriptscriptstyle 2$}
\put(30,-7){$\scriptscriptstyle 3$}
\put(48,-7){$\scriptscriptstyle \dots$}
\put(68,-7){$\scriptscriptstyle L=7$}
\end{picture}
\qquad \raisebox{25pt}{$\Leftrightarrow$}\quad
\begin{picture}(80,50)(0,-10)
{\thicklines
\put(0,25){\line(1,1){20}}
\put(0,25){\line(1,-1){20}}
\put(10,35){\line(1,-1){30}}
\put(10,15){\line(1,1){20}}
\put(20,45){\line(1,-1){40}}
\put(20,5){\line(1,1){30}}
\put(50,35){\line(1,-1){20}}
\put(40,5){\line(1,1){20}}
\put(60,5){\line(1,1){10}}
}
\qbezier[30](0,0)(0,25)(0,50)
\qbezier[30](10,0)(10,25)(10,50)
\qbezier[30](20,0)(20,25)(20,50)
\qbezier[30](30,0)(30,25)(30,50)
\qbezier[30](40,0)(40,25)(40,50)
\qbezier[30](50,0)(50,25)(50,50)
\qbezier[30](60,0)(60,25)(60,50)
\qbezier[30](70,0)(70,25)(70,50)
\put(0,-7){$\scriptscriptstyle 0$}
\put(10,-7){$\scriptscriptstyle 1$}
\put(20,-7){$\scriptscriptstyle 2$}
\put(30,-7){$\scriptscriptstyle 3$}
\put(40,-7){$\scriptscriptstyle 4$}
\put(50,-7){$\scriptscriptstyle 5$}
\put(60,-7){$\scriptscriptstyle 6$}
\put(67,-7){$\scriptscriptstyle L=7$}
\end{picture}
\quad \raisebox{25pt}{$=\quad e_2 e_1 e_3 e_5 e_2 e_4 e_6$}\qquad
\nn
\ea
\ba
\raisebox{15pt}{Ballot path:}&&
\begin{picture}(70,65)
\put(0,15){\line(1,0){70}}
{\thicklines
\put(0,15){\line(0,1){30}}
\put(0,45){\line(1,1){10}}
\put(10,55){\line(1,-1){30}}
\put(40,25){\line(1,1){10}}
\put(50,35){\line(1,-1){20}}
}
\qbezier[10](0,0)(0,7.5)(0,15)
\qbezier[10](10,0)(10,7.5)(10,15)
\qbezier[10](20,0)(20,7.5)(20,15)
\qbezier[10](30,0)(30,7.5)(30,15)
\qbezier[10](40,0)(40,7.5)(40,15)
\qbezier[10](50,0)(50,7.5)(50,15)
\qbezier[10](60,0)(60,7.5)(60,15)
\qbezier[10](70,0)(70,7.5)(70,15)
\put(0,-7){$\scriptscriptstyle 0$}
\put(10,-7){$\scriptscriptstyle 1$}
\put(20,-7){$\scriptscriptstyle 2$}
\put(30,-7){$\scriptscriptstyle 3$}
\put(48,-7){$\scriptscriptstyle \dots$}
\put(68,-7){$\scriptscriptstyle L=7$}
\end{picture}
\qquad \raisebox{15pt}{$\Leftrightarrow$}\quad
\begin{picture}(80,55)
{\thicklines
\put(0,5){\line(0,1){40}}
\put(0,45){\line(1,1){10}}
\put(0,25){\line(1,1){20}}
\put(0,25){\line(1,-1){20}}
\put(0,45){\line(1,-1){40}}
\put(0,5){\line(1,1){30}}
\put(20,5){\line(1,1){30}}
\put(10,55){\line(1,-1){50}}
\put(40,5){\line(1,1){20}}
\put(50,35){\line(1,-1){20}}
\put(60,5){\line(1,1){10}}
}
\qbezier[36](0,0)(0,25)(0,60)
\qbezier[36](10,0)(10,25)(10,60)
\qbezier[36](20,0)(20,25)(20,60)
\qbezier[36](30,0)(30,25)(30,60)
\qbezier[36](40,0)(40,25)(40,60)
\qbezier[36](50,0)(50,25)(50,60)
\qbezier[36](60,0)(60,25)(60,60)
\qbezier[36](70,0)(70,25)(70,60)
\put(0,-7){$\scriptscriptstyle 0$}
\put(10,-7){$\scriptscriptstyle 1$}
\put(20,-7){$\scriptscriptstyle 2$}
\put(30,-7){$\scriptscriptstyle 3$}
\put(40,-7){$\scriptscriptstyle 4$}
\put(50,-7){$\scriptscriptstyle 5$}
\put(60,-7){$\scriptscriptstyle 6$}
\put(67,-7){$\scriptscriptstyle L=7$}
\end{picture}
\quad \raisebox{15pt}{$=\quad e_1 e_0 e_2 e_1 e_3 e_5 e_0 e_2 e_4 e_6$}\qquad
\nn
\ea

\ni
Obviously, the projectors correspond to the substrates in the Figures
1, 2 and 4 (see Sec.2).

The Hamiltonian governing the time fluctuations of the RPM and the RPMW are given by the following expressions
(see (\ref{5.4}))
\be
\lb{Hs}
H_L\, =\, \sum_{i=1}^{L-1} (1- e_i)\, , \qquad
H_L^{(a)}\, =\, a(1-e_0)\, +\, \sum_{i=1}^{L-1} (1- e_i)\, ,
\ee
where $a$ is the boundary rate.

We now notice that the multiplication of the words from the ideals
by the generators $e_i$ and $e_0$ drawn in pictures according to the rules
(\ref{rule-1}) and  (\ref{rule-2}) coincides with the elementary processes
(adsorption, reflection and desorption) of the RPM and the RPMW models.

The groundstate eigenvectors of the two  intensity matrices
(\ref{Hs}) have remarkable combinatorial properties \cite{RS1,RS2,MNGB,P,GR,BGN,PRGN}.
New combinatorial aspects of these groundstate eigenvectors are going
to be shown in Sections  7 and 8
and will be used to derive the results enumerated in Section 3.

One can use the representation of the generators $e_0$, $e_i$
in terms of the Pauli matrices to derive
exact results.  In this representation   $H_L$ and $H_L^{(a)}$
are given by  XXZ spin one-half quantum chain Hamiltonians with non-Hermitean
boundary terms \cite{GNPyR}. From  the finite-size scaling limit of
the Hamiltonian eigenspectra
we can derive the conformal properties of the systems \cite{SB,NRG}.

\section{Hexagon recurrence relations}

In this section we want to show that the expressions (\ref{pascal-solved}) and
(\ref{P-Lnm}) for $S(L,n)$ and $P(L,n,m)$ are solutions of two recurrence relations. They can be
 seen as generalizations of the recurrence relations obtained
from the Pascal triangle. This section is pure mathematics. The connection with
 the RPM and RPMW is going to be made in Sections 7 and 8.
\medskip

\subsection{Pascal's hexagon}
Let us place numbers  on  the sites of a trigonal lattice in such a way
that any six numbers, occupying the vertices of an elementary hexagon as shown on the figure below,
\ba
\lb{hexagon}
\begin{array}{c}
\begin{picture}(20,80)(0,-40)
\put(30,-3.5){$a$}
\put(-38,-3.5){ {$a'$}}
\put(11,27){ {$b'$}}
\put(-23,27){ $c$}
\put(-23,-34){ $b$}
\put(11,-34){ {$c'$}}
\put(-3,-11){\circle*{2}}
\put(-71,-11){\circle*{2}}
\put(65,-11){\circle*{2}}
\put(-37,-11){\circle*{5}}
\put(31,-11){\circle*{5}}
\put(-54,20){\circle*{2}}
\put(48,20){\circle*{2}}
\put(-20,20){\circle*{5}}
\put(14,20){\circle*{5}}
\put(-54,-41){\circle*{2}}
\put(48,-41){\circle*{2}}
\put(-20,-41){\circle*{5}}
\put(14,-41){\circle*{5}}
\put(-88,20){\circle*{2}}
\put(-88,20){\vector(1,0){25}}
\put(-88.0,20){\vector(-1,-2){10}}
\put(-83,23){$\overrightarrow{x}$}
\put(-106,10){$\overrightarrow{y}$}
\end{picture}
\end{array}
\ea
satisfy the relation
\be
\lb{pascal}
c'\, c\, =\, a'\, a\, +\,  b'\, b\, .
\ee
Introducing a coordinate frame $\{\overrightarrow{x},\overrightarrow{y}\}$ on the lattice
(see figure above) and denoting $S(L,n)$ the number associated to a vertex
$(L\overrightarrow{x}+n\overrightarrow{y})$ we can
write the relation (\ref{pascal}) in the form
\be
\lb{hexagonS}
S(L+1,n+1)\, S(L-1,n-1) \, =\, S(L-1,n)\, S(L+1,n) + S(L,n-1)\, S(L,n+1)  \, .
\ee
This equation was obtained in \cite{DJM} as a discretized version of the Boussinesq equation.
It is a particular two-dimensional reduction of the three-dimensional discrete Hirota equation
(see \cite{H,Z} and
references therein) which is also known as the octahedron recurrence in combinatorics (see \cite{S} and
references therein).

We treat (\ref{hexagonS}) as a recurrence relation which gives the number $S(L+1,n+1)$
(corresponding to $c'$ in the figure)
in terms of the numbers placed in the two upper rows
(respectively, $a$, $a'$, $c$ and $b'$) and
at the left neighbour site from the same row (resp., $b$).
We solve the recurrence relation by  moving rightwards in a row and downwards row by row.
Taking the following set of initial data
\be
\lb{initial0}
S(L,-1)\, =\, S(L,0)\, =\, 1\, , \qquad S(2n-1,n)\, =\, 0\quad\forall\, n>0.
\ee
we can calculate numbers $S(L,n)$ in a sector $n\geq -1$, $L\geq 2n-1$.
As far as we know these boundary conditions were not considered before.

The values of $S(L,n)$ for small $L$ and $n$ are given in  Figure 15.
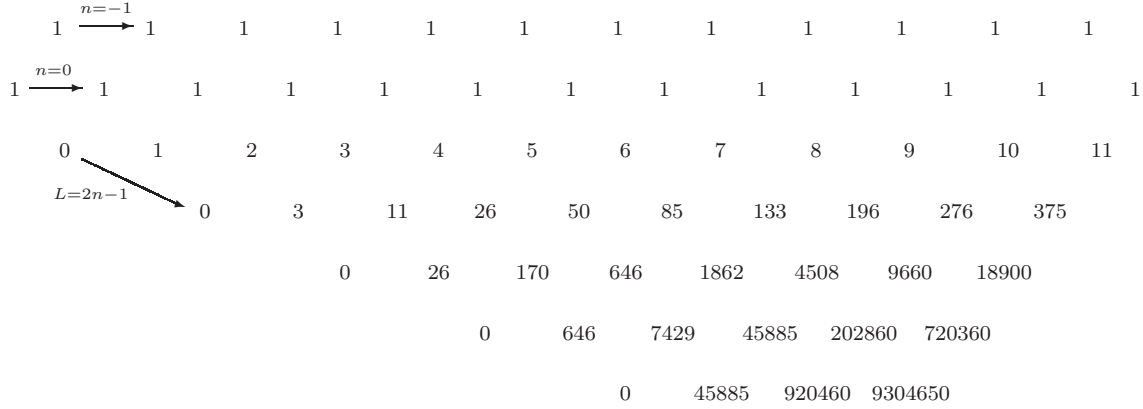
\begin{figure}
\ba
\nonumber
\hspace{-15mm}
\begin{tabular}{ccccccccccccccccccccccccc}
\begin{picture}(25,25)
%
\qbezier[200](50,-47)(68,-55.5)(86,-64)
\put(86,-64){\vector(3,-1){3}}
\put(39,-62){$\scriptscriptstyle L=2n-1$}
\put(30,-20){\vector(1,0){20}}
\put(49,3.5){\vector(1,0){20}}
\put(49,8.5){$\scriptscriptstyle n=-1$}
\put(32,-15){$\scriptscriptstyle n=0$}
\end{picture}
&
\scriptsize 1\,&\,&\scriptsize 1\,&\,
&\scriptsize 1\,&\,&\scriptsize 1\,&\,&
\scriptsize 1\,&\,&\scriptsize 1\,&\,
&\scriptsize 1\,&\,&\scriptsize 1\,\,&\,&
\scriptsize 1\,&\,&\scriptsize 1\,&\,
&\scriptsize 1\,&\, &\scriptsize 1\,
\\[3mm]
\quad\quad\scriptsize 1&\,&
\scriptsize 1\,&\,&\scriptsize {1}\,&\,&\scriptsize {1}\,&\,
&\scriptsize {1}\,&\,&
\scriptsize {1}\,&\,&\scriptsize {1}\,&\,
&\scriptsize {1}\,&\,&\scriptsize {1}\,&\,&
\scriptsize {1}\,&\,&\scriptsize {1}\,&\,
&\scriptsize {1}\,&\, &\scriptsize {1}\,
\\[3mm]
&
$\lefteqn{\scriptstyle 0}$&&
$\lefteqn{\scriptstyle 1}$&&
$\lefteqn{\scriptstyle {2}}$&&
$\lefteqn{\scriptstyle {3}}$&&
$\lefteqn{\scriptstyle {4}}$&&
$\lefteqn{\scriptstyle {5}}$&&
$\lefteqn{\scriptstyle {6}}$&&
$\lefteqn{\scriptstyle {7}}$&&
$\lefteqn{\scriptstyle {8}}$&&
$\lefteqn{\scriptstyle {9}}$&&
$\lefteqn{\scriptstyle {10}}$&&
$\lefteqn{\scriptstyle {11}}$&
\\[3mm]
&&&&
$\lefteqn{\scriptstyle 0}$&&
$\lefteqn{\scriptstyle 3}$&&
$\lefteqn{\scriptstyle {11}}$&&
$\lefteqn{\!\scriptstyle {26}}$&&
$\lefteqn{\!\scriptstyle {50}}$&&
$\lefteqn{\!\scriptstyle {85}}$&&
$\lefteqn{\!\!\scriptstyle {133}}$&&
$\lefteqn{\!\!\scriptstyle {196}}$&&
$\lefteqn{\!\!\scriptstyle {276}}$&&
$\lefteqn{\!\!\scriptstyle {375}}$
\\[3mm]
&&&&&&&
$\lefteqn{\scriptstyle 0}$&&
$\lefteqn{\!\scriptstyle 26}$&&
$\lefteqn{\!\!\scriptstyle {170}}$&&
$\lefteqn{\!\!\scriptstyle {646}}$&&
$\lefteqn{\!\!\!\scriptstyle {1862}}$&&
$\lefteqn{\!\!\!\scriptstyle {4508}}$&&
$\lefteqn{\!\!\!\scriptstyle {9660}}$&
&$\lefteqn{\!\!\!\!\scriptstyle {18900}}$&
\\[3mm]
&&&&&&&&&&
$\lefteqn{\scriptstyle 0}$&&
$\lefteqn{\!\!\scriptstyle 646}$&&
$\lefteqn{\!\!\!\scriptstyle {7429}}$&&
$\lefteqn{\!\!\!\!\scriptstyle {45885}}$&&
$\lefteqn{\!\!\!\!\!\scriptstyle {202860}}$&&
$\lefteqn{\!\!\!\!\!\scriptstyle {720360}}$&&
\\[3mm]
&&&&&&&&&&&&&
$\lefteqn{\scriptstyle 0}$&&
$\lefteqn{\!\!\!\!\scriptstyle 45885}$&&
$\lefteqn{\!\!\!\!\!\scriptstyle {920460}}$&&
$\lefteqn{\!\!\!\!\!\!\scriptstyle {9304650}}$&&
\\[1mm]
\end{tabular}
\ea
\caption{\small Pascal's hexagon rule.}
\label{fig8}
\end{figure}
This figure and the recursive procedure described above reminds
of the famous Pascal's triangle rule --- an arrangement of the
binomial coefficients in a triangle.
A closer inspection of the solution adds more arguments in favour of such
an analogy.
First, we notice that all numbers on Figure 15 are integers, which is not
trivial since solving the recurrence relation (\ref{hexagonS}) one obtains ratios.\footnote{
For a general discussion of such phenomena see
Ref.\cite{FZ}.}
Moreover, all the numbers in Figure\,15 can be factorized into relatively small
primes (compared to the numbers themselves).
For example $S(9,4)$, $S(10,5)$ and $S(12,5)$ are given  respectively by:
$$
7429= 17\cdot 19\cdot 23,\quad
45885=3\cdot 5\cdot 7\cdot 19\cdot 23, \quad
9304650 = 2\cdot 3^2\cdot 5^2\cdot 23\cdot 29\cdot 31 .
$$
This  factorization property suggests an idea to look for an expression of $S(L,n)$ in terms of
factorials, just as it is the case for the binomial coefficients.
The solution of the recurrence relation (6.3), with initial
condition (6.4),  is given by  (\ref{pascal-solved}) \cite{P}.
\medskip

One can look for a more general
solution of the Pascal's hexagon recurrence (\ref{hexagonS}) using
the initial conditions depending now on a parameter $a$:
\be
\lb{initial-a}
S^{(a)}(L,-1) =1\, ,\quad S^{(a)}(L,0)\, =\, a^{\delta_{L,0}} , \quad S^{(a)}(2n-1,n)\, =\, 0\quad\forall\, n>0.
\ee
Remarkably again, all the quantities $S^{(a)}(L,n)$ in the sector
$n\geq -1$, $L\geq 2n-1$, turn out to be
polynomials in variable $a$ with
integer coefficients. These polynomials are  particular examples of
the polynomial solutions considered in the Appendix of
Ref.~\cite{P} ($S^{(a)}(L,n)=F_{L,n}(x=1,y=a)$ in the notation
of \cite{P}).
The polynomials $S^{(a)}(L,n)$ can be also observed in the
stationary states of the RPMW with an arbitrary boundary rate $a$ (see
Sec.\,8).
Obviously, one has $S^{(1)}(L,n) = S(L,n)$.

\subsection{Split-hexagon}
We now introduce a vertical direction $\overrightarrow{z}$
in the picture (\ref{hexagon})
and put numbers $P(L,n,m)$
into correspondence with the vertices
$(L\overrightarrow{x}+n\overrightarrow{y}+m\overrightarrow{z})$
of the $3d$ lattice (see Figure 16).
\begin{figure}[h]
\begin{center}
\begin{picture}(220,125)(-110,-60)
\qbezier[45](35,-60)(35,-5)(35,30)
\qbezier[30](70,13)(70,43)(70,73)
\qbezier[9](70,-14)(70,-7)(70,3)
\qbezier[45](-30,-30)(-30,15)(-30,60)
{\red{ \put(35,-56){\circle*{2}} \put(35,-41){\circle*{2}}
\put(35,-26){\circle*{2}} \put(35,-11){\circle*{2}}
\put(35,4){\circle*{4}} \put(35,19){\circle*{2}}
\put(-30,-26){\circle*{2}}
\put(-30,-11){\circle*{2}}
\put(-30,4){\circle*{2}}
\put(-30,19){\circle*{2}}
\put(-30,34){\circle*{4}}
\put(-30,49){\circle*{2}}
\put(70,-11){\circle*{2}}
\put(70,2){\circle*{2}}
\put(70,19){\circle*{2}}
\put(70,34){\circle*{4}}
\put(70,49){\circle*{2}}
\put(70,64){\circle*{2}}
}}
\put(-24,24){ {$\scriptstyle S(L-1,n-1)$}}
\put(-23,-36){ $\scriptstyle S(L,n+1)$}
\put(77,-7){ $\scriptstyle S(L+1,n)$}
\put(33,6.5){ {$\scriptscriptstyle P(L+1,n+1,m)$}}
\put(-76,37){ {$\scriptscriptstyle P(L-1,n,m)$}}
\put(68,37){ {$\scriptscriptstyle P(L,n-1,m-2)$}}
\put(-20,-41){\circle*{5}}
\put(25,-11){\circle*{2}}
\put(70,19){\circle{5}}
\put(35,-41){\circle{5}}
\put(80,-11){\circle*{5}}
\put(125,19){\circle*{2}}
\put(90,-41){\circle*{2}}
\put(135,-11){\circle*{2}}
\put(-75,-41){\circle*{2}}
\put(-30,-11){\circle{5}}
\put(15,19){\circle*{5}}
\put(-130,-41){\circle*{2}}
\put(-85,-11){\circle*{2}}
\put(-40,19){\circle*{2}}
\put(-140,-11){\circle*{2}}
\put(-95,19){\circle*{2}}
\qbezier[25](-20,-41)(7.5,-41)(35,-41)
\qbezier[25](35,-41)(57.5,-26)(80,-11)
\qbezier[12](80,-11)(77.8,-4)(75.5,3)
\qbezier[5](72,13)(71,16)(70,19)
\qbezier[25](-20,-41)(-25,-26)(-30,-11)
\qbezier[25](-30,-11)(-7.5,4)(15,19)
\qbezier[7](15,19)(23.5,19)(30,19)
\qbezier[14](40,19)(55,19)(70,19)
\put(-140,-11){\vector(1,0){25}}
\put(-140,-11){\vector(0,1){25}}
\put(-140.0,-11){\vector(-3,-2){20}}
\put(-129,-8){$\footnotesize \overrightarrow{x}$}
\put(-159,-15){$\footnotesize \overrightarrow{y}$}
\put(-153,2){$\footnotesize \overrightarrow{z}$}
\end{picture}
\parbox{15cm}{
\caption{
\small Split hexagon: the numbers $S(L+1,n+1)$, $S(L-1,n)$ and $S(L,n-1)$
split, respectively, into the parts $P(L+1,n+1,m)$, $P(L-1,n,m)$ and $P(L,n-1,m)$,
$~m=0,1,\dots .$ }
}
\end{center}
\label{fig9}
\end{figure}
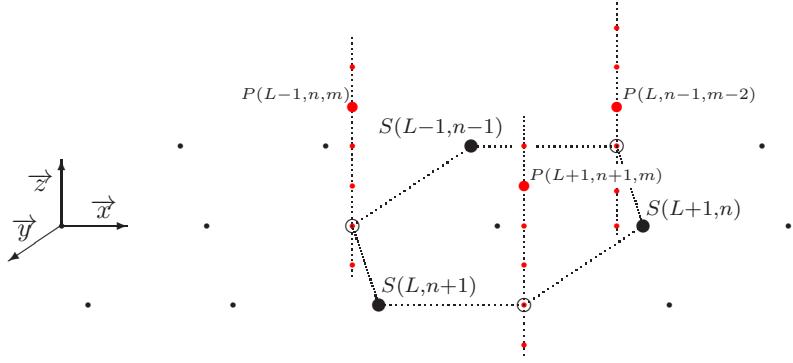

\ni
We impose the following relations on $P(L,n,m)$'s:
\ba
\lb{partition}
 \sum_{m=0}^n P(L,n,m)& =& S(L,n)\, ,
\\[1mm]
\lb{initial1} \qquad P(L,n,-k)&=&P(L,n,n+k)=0 \;\;\;\forall k>0\, ,\; n\geq 0\, ,
\ea
and the equations
\ba
\nn
\lefteqn{\hspace{-15mm}P(L+1,n+1,m)\, S(L-1,n-1)}&&
\\[2mm]
\lb{split-hexagon}
&&= P(L-1,n,m)\, S(L+1,n) + P(L,n-1,m-2)\, S(L,n+1)\, .
\ea
Summing over $m$ the relations (\ref{split-hexagon}) one obtains
the hexagon relations (\ref{hexagonS}), therefore we call (\ref{split-hexagon}) the
{\em split-hexagon} relations.

Taking into account (\ref{partition}), the Eq.~(\ref{split-hexagon})
defines recurrence relations for the $P(L,n,m)$'s.
Indeed, using (\ref{initial1}) and the initial conditions (c.f. with the initial data (\ref{initial0}))
\be
\lb{initial2}
P(L,0,m)\, =\, \delta_{m,0}\, , \quad
P(L,-1,m)\, =\, \delta_{m,-1}\, , \quad\mbox{\rm and}\quad
P(2n,n,m)\, =\, 0\;\;\;\forall\, m>1,
\ee
one can calculate the numbers $P(L,n,m)$ starting from the row $n=1$ downwards and
moving rightwards from the point $L=2n$ in a row. For each given pair $L$ and $n$ one calculates
$P(L,n,m)$ for all $m=0,1,\dots n$.

We notice that the split-hexagon recursion is a kind of "bootstrap" procedure as after
each step (for example, the calculation of $P(L,n+1,m)$ for all $m=0,1,\dots ,n+1$) one has
to find a coefficient for the equations which are solved at the next step of the recursion
(that is, the coefficient $S(L,n+1)=\sum_{m=0}^{n+1} P(L,n+1,m)$ in the equations for $P(L+1,n+1,m)$).
We also remark that there is no need to set the boundary values
$$
P(2n,n,0) = S(2n,n-1) \quad\mbox{\rm and}\quad P(2n,n,1)=0
$$
as they are completely defined by the data from the previous $(n-1)$-th row  only
(the last term in the split-hexagon relation (\ref{split-hexagon})
vanishes for $m=0$ or $1$).

The relations (\ref{split-hexagon}) are new and can
be generalized changing the condition (\ref{partition}). The continuous version
of (\ref{split-hexagon}) is not yet known.


The solution of the split-hexagon recursion is given by integer numbers with simple
factorization properties, e.g.,
$$
P(12,4,2)=2041020=2^2\cdot 3^2\cdot 5\cdot 17\cdot 23\cdot 29\, , \qquad
P(12,4,3)=2281140=2^2\cdot 3^2\cdot 5\cdot 19\cdot 23\cdot 29\, .
$$
This suggests the $P(L,n,m)$'s can be written as a product of factorials.
One can show that the expression given by (\ref{P-Lnm})
for $P(L,n,m)$ is the solution of the split-hexagon recurrence relations
(\ref{partition}) and  (\ref{split-hexagon})
with the boundary conditions  (\ref{initial1}) and (\ref{initial2}).

\section{Combinatorics in the stationary states of the RPM}

We have shown in the last section that the quantities $P(L,n,m)$ are solutions of the
split-hexagon recurrence relations (\ref{split-hexagon}) and we gave their
expression (\ref{P-Lnm}). We now
make the connection with the stationary states of the RPM.\footnote{
Actually it was by observing the combinatorial
properties of the RPM's stationary states
that we were led to the hexagon and
the split-hexagon relations.
}
\medskip

The stationary state of the RPM for a system of size $L$
is described by the eigenvector ${\ket 0}_L$ of the
operator $H_L$ (\ref{Hs}) corresponding to eigenvalue zero.
It is a linear combination of the words in the Dyck path representation (see Sec.\,5)
defined up to a common constant factor. A suitable normalization for ${\ket 0}_L$
is to set the coefficient of the maximal Dyck path (see figures below) to be equal to 1.
\ba
\nn
\hspace{-15mm}
\raisebox{10pt}{Maximal  Dyck paths:}&&\hspace{5mm}
\begin{picture}(280,45)(0,10)
{\green{
\qbezier[1](180,28)(180.5,28.5)(181,29)
\qbezier[2](180,26)(181,27)(182,28)
\qbezier[3](180,24)(181.5,25.5)(183,27)
\qbezier[4](180,22)(182,24)(184,26)
\qbezier[4](181,21)(183,23)(185,25)
\qbezier[3](183,21)(184.5,22.5)(186,24)
\qbezier[2](185,21)(186,22)(187,23)
\qbezier[1](187,21)(187.5,21.5)(188,22)
\multiput(0,0)(20,0){2}{
\qbezier[8](193,21)(197,25)(201,29)
\qbezier[7](195,21)(198.5,24.5)(202,28)
\qbezier[6](197,21)(200,24)(203,27)
\qbezier[5](199,21)(201.5,23.5)(204,26)
\qbezier[4](201,21)(203,23)(205,25)
\qbezier[3](203,21)(204.5,22.5)(206,24)
\qbezier[2](205,21)(206,22)(207,23)
\qbezier[1](207,21)(207.5,21.5)(208,22)}
\multiput(0,0)(20,0){3}{
\qbezier[8](3,21)(7,25)(11,29)
\qbezier[7](5,21)(8.5,24.5)(12,28)
\qbezier[6](7,21)(10,24)(13,27)
\qbezier[5](9,21)(11.5,23.5)(14,26)
\qbezier[4](11,21)(13,23)(15,25)
\qbezier[3](13,21)(14.5,22.5)(16,24)
\qbezier[2](15,21)(16,22)(17,23)
\qbezier[1](17,21)(17.5,21.5)(18,22)}
}}
\put(0,20){\line(1,0){60}}
\put(180,20){\line(1,0){50}}
{\thicklines
\put(180,30){\line(0,-1){10}}
\put(0,20){\line(1,1){30}}
\put(30,50){\line(1,-1){30}}
\put(180,30){\line(1,1){20}}
\put(200,50){\line(1,-1){30}}}
\put(80,20){for $L$ even,}
\put(250,20){for $L$ odd.}
\qbezier[10](10,30)(15,25)(20,20)
\qbezier[10](20,20)(25,25)(30,30)
\qbezier[10](30,30)(35,25)(40,20)
\qbezier[10](40,20)(45,25)(50,30)
\qbezier[10](180,30)(185,25)(190,20)
\qbezier[10](190,20)(195,25)(200,30)
\qbezier[10](200,30)(205,25)(210,20)
\qbezier[10](210,20)(215,25)(220,30)
\qbezier[5](0,14)(0,17)(0,20)
\qbezier[5](10,14)(10,17)(10,20)
\qbezier[5](60,14)(60,17)(60,20)
\put(-1.8,7){$\scriptscriptstyle 0$}
\put(8.5,7){$\scriptscriptstyle 1$}
\put(48,7){$\scriptscriptstyle L=6$}
\qbezier[5](180,14)(180,17)(180,20)
\qbezier[5](190,14)(190,17)(190,20)
\qbezier[5](230,14)(230,17)(230,20)
\put(178.2,7){$\scriptscriptstyle 0$}
\put(188.5,7){$\scriptscriptstyle 1$}
\put(219,7){$\scriptscriptstyle L=5$}
\end{picture}
\ea
This is the smallest coefficient in the vector ${\ket 0}_L$ and the
corresponding profile is the least probable in the stationary state.
With this choice, all the weights of the different Dyck paths in the ground-state
eigenvectors are integer numbers which can be related to an enumeration of loops in
a fully packed loops  ice-model \cite{BGN}.
The most probable stationary configuration is the substrate shown as a dashed region on
the figures above.\footnote{Intuitively, the more reflection points (local maxima)
one has in a profile, the larger is its contribution to the stationary state.}

\begin{figure}[t]
\begin{center}
\begin{picture}(200,40)
{\green{
\multiput(0,0)(20,0){8}{
\qbezier[8](2,11)(6,15)(10,19)
\qbezier[7](4,11)(7.5,14.5)(11,18)
\qbezier[6](6,11)(9,14)(12,17)
\qbezier[5](8,11)(10.5,13.5)(13,16)
\qbezier[4](10,11)(12,13)(14,15)
\qbezier[3](12,11)(13.5,12.5)(15,14)
\qbezier[2](14,11)(15,12)(16,13)
\qbezier[1](16,11)(16.5,11.5)(17,12)}
\multiput(0,0)(20,0){6}{\multiput(0,0)(1,-1){9}{\multiput(0,20)(10,10){2}{\qbezier[8](12,0)(16,4)(20,8)}}}
\multiput(120,0)(1,-1){9}{\multiput(0,20)(10,10){1}{\qbezier[8](12,0)(16,4)(20,8)}}
}}
{\linethickness{0.2pt}
\put(-10,30){\line(1,0){150}}
\put(20,40){\line(1,0){110}}
\put(20,90){\line(1,0){60}}
\put(-10,10){\line(1,0){10}}
}
\put(-7,20){\vector(0,1){10}}
\put(-7,20){\vector(0,-1){10}}
\put(23,65){\vector(0,1){25}}
\put(23,65){\vector(0,-1){25}}
\put(8,62){$\scriptscriptstyle n=5$}
\put(-22,17){$\scriptscriptstyle h=2$}
\qbezier[50](30,40)(55,66)(80,90)
\qbezier[50](80,90)(105,65)(130,40)
%
{\red{\multiput(0,0)(20,0){5}{\put(40,30){\circle*{5}}}
}}
{\thicklines
\put(0,10){\line(1,0){160}}
\put(0,10){\line(1,1){30}}
\put(30,40){\line(1,-1){10}}
\put(40,30){\line(1,1){10}}
\put(50,40){\line(1,-1){10}}
\put(60,30){\line(1,1){10}}
\put(70,40){\line(1,-1){10}}
\put(80,30){\line(1,1){10}}
\put(90,40){\line(1,-1){10}}
\put(100,30){\line(1,1){10}}
\put(110,40){\line(1,-1){10}}
\put(100,30){\line(1,1){10}}
\put(110,40){\line(1,-1){10}}
\put(120,30){\line(1,1){10}}
\put(130,40){\line(1,-1){30}}
}
\qbezier[5](0,4)(0,7)(0,10)
\qbezier[5](10,4)(10,7)(10,10)
\qbezier[5](160,4)(160,7)(160,10)
\put(-1,-3){$\scriptscriptstyle 0$}
\put(8,-3){$\scriptscriptstyle 1$}
\put(146,-3){$\scriptscriptstyle L=16$}
\end{picture}
\begin{picture}(200,40)
{\green{
\multiput(0,0)(20,0){8}{
\qbezier[8](2,11)(6,15)(10,19)
\qbezier[7](4,11)(7.5,14.5)(11,18)
\qbezier[6](6,11)(9,14)(12,17)
\qbezier[5](8,11)(10.5,13.5)(13,16)
\qbezier[4](10,11)(12,13)(14,15)
\qbezier[3](12,11)(13.5,12.5)(15,14)
\qbezier[2](14,11)(15,12)(16,13)
\qbezier[1](16,11)(16.5,11.5)(17,12)}
\multiput(0,0)(20,0){6}{\multiput(0,0)(1,-1){9}{\multiput(0,20)(10,10){2}{\qbezier[8](12,0)(16,4)(20,8)}}}
\multiput(120,0)(1,-1){9}{\multiput(0,20)(10,10){1}{\qbezier[8](12,0)(16,4)(20,8)}}
}}
{\linethickness{0.2pt}
\put(-10,30){\line(1,0){150}}
\put(20,40){\line(1,0){110}}
\put(20,90){\line(1,0){60}}
\put(-10,10){\line(1,0){10}}
}
\put(-7,20){\vector(0,1){10}}
\put(-7,20){\vector(0,-1){10}}
\put(23,65){\vector(0,1){25}}
\put(23,65){\vector(0,-1){25}}
\put(8,62){$\scriptscriptstyle n=5$}
\put(-22,17){$\scriptscriptstyle h=2$}
{\red{\put(80,30){\circle*{5}}
\put(120,30){\circle*{5}}
}}
\qbezier[50](30,40)(55,66)(80,90)
\qbezier[50](80,90)(105,65)(130,40)
{\thicklines
\put(0,10){\line(1,0){160}}
\put(0,10){\line(1,1){50}}
\put(50,60){\line(1,-1){30}}
\put(80,30){\line(1,1){20}}
\put(100,50){\line(1,-1){20}}
\put(120,30){\line(1,1){10}}
\put(130,40){\line(1,-1){30}}
}
\qbezier[5](0,4)(0,7)(0,10)
\qbezier[5](10,4)(10,7)(10,10)
\qbezier[5](160,4)(160,7)(160,10)
\put(-1,-3){$\scriptscriptstyle 0$}
\put(8,-3){$\scriptscriptstyle 1$}
\put(146,-3){$\scriptscriptstyle L=16$}
\end{picture}
\begin{picture}(200,120)(0,-10)
{\green{
\multiput(0,0)(20,0){8}{
\qbezier[8](2,11)(6,15)(10,19)
\qbezier[7](4,11)(7.5,14.5)(11,18)
\qbezier[6](6,11)(9,14)(12,17)
\qbezier[5](8,11)(10.5,13.5)(13,16)
\qbezier[4](10,11)(12,13)(14,15)
\qbezier[3](12,11)(13.5,12.5)(15,14)
\qbezier[2](14,11)(15,12)(16,13)
\qbezier[1](16,11)(16.5,11.5)(17,12)}
\multiput(0,0)(20,0){6}{\multiput(0,0)(1,-1){9}{\multiput(0,20)(10,10){2}{\qbezier[8](12,0)(16,4)(20,8)}}}
\multiput(120,0)(1,-1){9}{\multiput(0,20)(10,10){1}{\qbezier[8](12,0)(16,4)(20,8)}}
}}
{\linethickness{0.2pt}
\put(-10,30){\line(1,0){150}}
\put(20,40){\line(1,0){110}}
\put(20,90){\line(1,0){60}}
\put(-10,10){\line(1,0){10}}
}
\put(-7,20){\vector(0,1){10}}
\put(-7,20){\vector(0,-1){10}}
\put(23,65){\vector(0,1){25}}
\put(23,65){\vector(0,-1){25}}
\put(8,62){$\scriptscriptstyle n=5$}
\put(-22,17){$\scriptscriptstyle h=2$}
\qbezier[50](30,40)(55,66)(80,90)
\qbezier[50](80,90)(105,65)(130,40)
{\thicklines
\put(0,10){\line(1,0){160}}
\put(0,10){\line(1,1){40}}
\put(40,50){\line(1,-1){10}}
\put(50,40){\line(1,1){20}}
\put(70,60){\line(1,-1){20}}
\put(90,40){\line(1,1){10}}
\put(100,50){\line(1,-1){10}}
\put(110,40){\line(1,1){10}}
\put(120,50){\line(1,-1){40}}
}
\qbezier[5](0,4)(0,7)(0,10)
\qbezier[5](10,4)(10,7)(10,10)
\qbezier[5](160,4)(160,7)(160,10)
\put(-1,-3){$\scriptscriptstyle 0$}
\put(8,-3){$\scriptscriptstyle 1$}
\put(146,-3){$\scriptscriptstyle L=16$}
\end{picture}
\parbox{15cm}{\small
\caption{(Color online) Examples of Dyck paths of the size $L=16$ satisfying
the selection criteria for $h=2$ ($\Leftrightarrow\; n=5$).
The number $m$ of return points at the height $h=2$ (red points on the pictures)
is equal, respectively, to $5$, $2$, and $0$.
}}
\end{center}
\vspace{-5mm}
\label{fig10}
\end{figure}
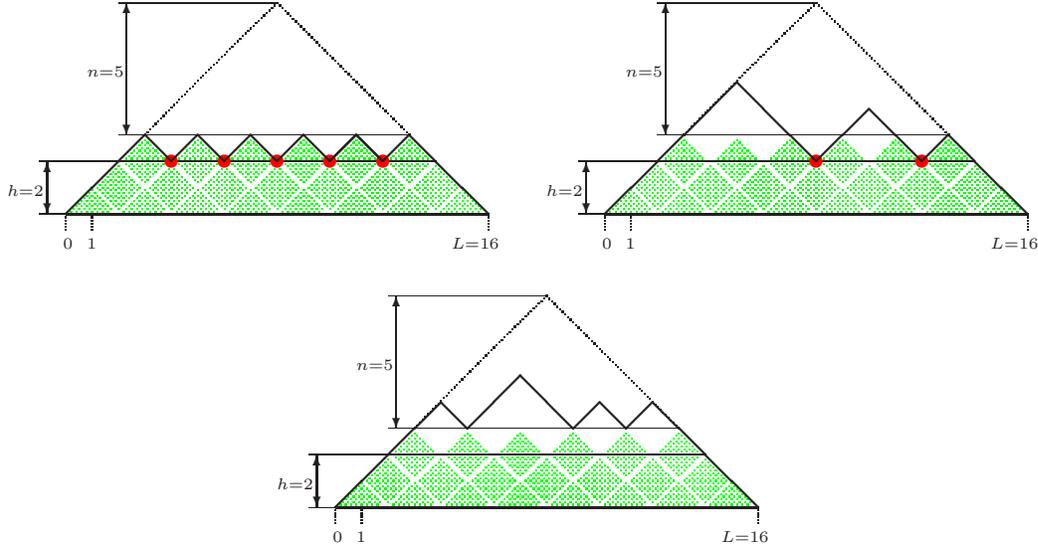

Consider a Dyck path with $L$ sites. Let $h$ be a
number from the list
$$0,\,1,\,\dots ,\, \lfloor \frac{L-1}{2}\rfloor\, ,$$
such that
the Dyck path has no local  minima  (return points) below the height $h$. (See Figure\,17).
Let us denote by $m$ ($m\geq 0$) the number of returns at the height $h$ for a given path.
We consider all the configurations which have $m$ return points at the height $h$
and the remaining ones  higher.
\vspace{1mm}

\ni
{\bf Conjecture.~}
\lb{conj}
{\em
The sum of  all the coefficients in
${\ket {\hspace{.5pt}0}}_L$
corresponding to these configurations is given by $P(L,n,m)$. Here
\be
\lb{7.1}
n = \lfloor \frac{L-1}{2}\rfloor - h \, , \quad   n=0,1,\dots ,\lfloor \frac{L-1}{2}\rfloor\, ,
\quad\mbox{and}\quad  m=0,1,\dots ,n\, .
\ee
$n$ counts the height from the top of the triangle instead of its basis.
}
\vspace{1mm}

This conjecture was checked for lattices of size $L\leq 13$.
There are more consistency checks for this conjecture:
\begin{itemize}
\item
One clearly has $P(L,n+1,0) = \sum_{m=0}^{n}P(L,n,m)=S(L,n)$ (see (\ref{partition})),
which is the sum of  all the coefficients in ${\ket 0}_L$  corresponding to the configurations
with local minima at values higher or equal to $h=\left(\lfloor (L-1)/2\rfloor -n\right)$ and is given by (\ref{pascal-solved}).
This expression was conjectured earlier and checked independently on small lattices
in Refs.~\cite{MNGB,P}.
\item
More importantly, it was
proven analytically in Refs.~\cite{DiF,PZJ} that
$S(L)$ (see (\ref{S-2n}) and (\ref{S-2n-1})) coincides with the normalization factor
(sum of all the weights) of ${\ket 0}_L$:
\be
\lb{normalization-RPM}
\raisebox{-1.8pt}{$_L$}\!{\langle 0 |\, 0\rangle}_{\! L}\, =\,
S(L)\, =\, S(L,\lfloor(L-1)/2\rfloor)\, .\hspace{20mm}
\ee

\end{itemize}

The expression for
the probability of having $k$ clusters for a system of size $L$
given in Sec.~3 (see (\ref{P_L(k)})) is an obvious corollary of the conjecture.

\section{Clusters in the stationary states of the RPMW and  RPM}

In this section we prove
the expressions
(\ref{P1B-Lodd}) and  (\ref{P1B-Leven}) for the probabilities to have $k$ clusters for a system of size $L$
as well the relations
(\ref{PfPd-Lodd})--(\ref{n_c-Leven})
for the probability densities
observed in the RPMW.
To this end we make use
of the algebraic background (the TL algebras described in Section 6), and of the
conjecture presented in Section 7.

\medskip

The stationary state of the RPMW for a system of size $L$
is described by the ground state  eigenvector  of the
operator $H^{(a)}_L$ (\ref{Hs}) which
is a linear combination of the words in the ballot
path representation (see Sec.~5) with coefficients (weights)
depending on the boundary rate $a$.
We  use a notation ${\ket 0}_L^{(a)}$ for this
eigenvector
and, if necessary, we explicitly show the value of $a$ in the notation, say
${\ket 0}_{L}^{(1)}$ for $a=1$.
A suitable normalization for ${\ket 0}_L^{(a)}$
is to set the coefficient of the maximal ballot path (see figure below) to be equal to
$a^{\lfloor L/2\rfloor}$.
In case $a=1$ it is the smallest coefficient in ${\ket 0}^{(1)}_{L}$.
\ba
\lb{1b-1}
\mbox{\raisebox{28pt}{Maximal ballot path:}}&&
\begin{picture}(200,55)(0,-10)
{\green{
\qbezier[1](0,8)(0.5,8.5)(1,9)
\qbezier[2](0,6)(1,7)(2,8)
\qbezier[3](0,4)(1.5,5.5)(3,7)
\qbezier[4](0,2)(2,4)(4,6)
\qbezier[4](1,1)(3,3)(5,5)
\qbezier[3](3,1)(4.5,2.5)(6,4)
\qbezier[2](5,1)(6,2)(7,3)
\qbezier[1](7,1)(7.5,1.5)(8,2)
\multiput(-180,-20)(20,0){2}{
\qbezier[8](193,21)(197,25)(201,29)
\qbezier[7](195,21)(198.5,24.5)(202,28)
\qbezier[6](197,21)(200,24)(203,27)
\qbezier[5](199,21)(201.5,23.5)(204,26)
\qbezier[4](201,21)(203,23)(205,25)
\qbezier[3](203,21)(204.5,22.5)(206,24)
\qbezier[2](205,21)(206,22)(207,23)
\qbezier[1](207,21)(207.5,21.5)(208,22)}
}}
\qbezier[10](0,10)(5,5)(10,0)
\qbezier[10](10,0)(15,5)(20,10)
\qbezier[10](20,10)(25,5)(30,0)
\qbezier[10](30,0)(35,5)(40,10)
\put(0,0){\line(1,0){50}}
{\thicklines
\put(0,0){\line(0,1){50}}
\put(0,50){\line(1,-1){50}}
}
\qbezier[5](0,-6)(0,-3)(0,0)
\qbezier[5](10,-6)(10,-3)(10,0)
\qbezier[5](50,-6)(50,-3)(50,0)
\put(-1.8,-13){$\scriptscriptstyle 0$}
\put(8.5,-13){$\scriptscriptstyle 1$}
\put(39,-13){$\scriptscriptstyle L=5$}
\end{picture}
\ea
With this choice, all the weights of the different ballot paths in the ground-state
eigenvectors become polynomials in $a$ with integer coefficients
which can be related to a "weighted" enumeration of loops in
a fully packed loops  ice-model  \cite{GR}.
As it was proven in \cite{PZJ}, in the case $a=1$,
the normalization factor (sum of all weights)
of ${\ket 0}_{L}^{(1)}$
is equal to the number of vertically and horizontally
symmetric alternating sign matrices of a size $(2L+3)$
(see (\ref{S-2n}) and  (\ref{S-2n-1}))
\be
\lb{8.1}
\raisebox{-1pt}{${}_{\;\; L}^{(1)}$}\!{\langle 0 |\, 0\rangle}_{\! L}^{(1)}\, =\,
A^{VH}_{2L+3}\, =\, S(L)\, S(L+1)\, .
\ee
For an arbitrary boundary rate $a$, the normalization factor reads (see \cite{PZJ}, Eq.~(4.8))
\be
\lb{8.1a}
\raisebox{-1pt}{${}_{\;\; L}^{(a)}$}\!{\langle 0 |\, 0\rangle}_{\! L}^{(a)}\, =\,
 S(L)\, S^{(a)}(L+1)\, ,
\ee
Here $S^{(a)}(L)= S^{(a)}(L,\lfloor(L-1)/2\rfloor)$,
and $S^{(a)}(L,n)$ is a polynomial solution of the Pascal's hexagon recurrence
(\ref{hexagonS}) with
the initial conditions (\ref{initial-a}).
\medskip

Now we are going to use homomorphisms from the one-boundary TL algebra  ${\mathcal T}_L^{(1B)}$
to the TL algebras ${\mathcal T}_L$ and ${\mathcal T}_{L+1}$
to establish relations between the
ground state vectors of the RPM models with and without a wall. One has to consider the cases
$L$ even and odd separately.

\subsection{Case of $L$ odd.}

We consider the homomorphism between the one-boundary TL algebra
and the TL algebra (see  (\ref{TL}) and (\ref{1bTL})) of the same size $L$, defined by
\be
\lb{8.2}
{\mathcal T}^{(1B)}_L\, \rightarrow\, {\mathcal T}_L :\qquad
e_0\mapsto 1, \quad e_i\mapsto e_i\;\; \forall\, i=1,\dots , L-1.
\ee
This homomorphism  maps  the intensity matrix of the RPMW ($a$ is arbitrary)
onto the  intensity matrix of the RPM
for the same number of sites $L$ (see (\ref{Hs}))
\be
\lb{8.3}
H_L^{(a)}\, \mapsto\, H_L\, ,
\ee
and so should map their unique groundstate eigenvectors:
\be
\lb{8.4}
{\ket 0}_{L}^{(a)}\, \mapsto\, S^{(a)}(L+1)\, {\ket 0}_{L}\, ,
\ee
where we have computed the scaling constant $S^{(a)}(L+1)$ by comparing the expressions
for normalization factors (\ref{normalization-RPM}) and (\ref{8.1a}).

\begin{figure}
\begin{picture}(320,50)
{\red{
\put(120,10){\circle*{5}}
\put(260,10){\circle*{5}}
\put(360,10){\circle*{5}}
\put(380,10){\circle*{5}}
}}
\multiput(0,0)(120,0){4}{
\put(0,0){\line(0,1){10}}
\put(0,0){\line(1,0){110}}
\qbezier[10](0,10)(5,5)(10,0)
\qbezier[10](10,0)(15,5)(20,10)
\qbezier[10](20,10)(25,5)(30,0)
\qbezier[10](30,0)(35,5)(40,10)
\qbezier[10](60,10)(65,5)(70,0)
\qbezier[10](70,0)(75,5)(80,10)
\put(0,10){\line(1,1){10}}
\put(10,20){\line(1,-1){10}}
\put(20,10){\line(1,1){10}}
{\thicklines
\put(30,20){\line(1,-1){20}}
\put(50,0){\line(1,1){20}}
\put(70,20){\line(1,-1){20}}
\put(90,0){\line(1,1){10}}
\put(100,10){\line(1,-1){10}}
}}
{\thicklines
\put(0,10){\line(1,1){10}}
\put(10,20){\line(1,-1){10}}
\put(20,10){\line(1,1){10}}
\put(130,20){\line(1,-1){10}}
\put(140,10){\line(1,1){10}}
}
\qbezier[10](240,30)(245,25)(250,20)
\qbezier[10](250,20)(255,25)(260,30)
\qbezier[10](360,30)(365,25)(370,20)
\qbezier[10](360,30)(365,35)(370,40)
\qbezier[10](370,20)(375,25)(380,30)
{\blue{
\put(121,16.5){$\ast$}
\put(241,16.5){$\ast$}
\put(361,16.5){$\ast$}
\put(361,36.5){$\ast$}
\put(257.5,16.5){$\dag$}
\put(247.5,26.5){$\dag$}
\put(377.5,16.5){$\dag$}
\put(367.5,26.5){$\dag$}
}}
{\thicklines
\put(120,30){\line(1,-1){10}}
\put(240,30){\line(1,1){10}}
\put(250,40){\line(1,-1){20}}
\put(360,50){\line(1,-1){30}}
}
\put(120,10){\line(0,1){20}}
\put(240,10){\line(0,1){20}}
\put(360,10){\line(0,1){40}}
\qbezier[5](0,-6)(0,-3)(0,0)
\qbezier[5](10,-6)(10,-3)(10,0)
\qbezier[5](20,-6)(20,-3)(20,0)
\qbezier[5](110,-6)(110,-3)(110,0)
\put(-1,-13){$\scriptscriptstyle 0$}
\put(9,-13){$\scriptscriptstyle 1$}
\put(19,-13){$\scriptscriptstyle 2$}
\put(96,-13){$\scriptscriptstyle L=11$}
\end{picture}
\\[1mm]
$\underbrace{\hspace{165mm}}$
\\[3mm]
\\
\begin{picture}(200,40)(-185,-10)
{\linethickness{0.2pt}
\put(-25,10){\line(1,0){65}}
}
\put(-24,12){$\scriptstyle h=1$}
{\red{
\put(0,10){\circle*{5}}
\put(20,10){\circle*{5}}
}}
\put(0,0){\line(0,1){10}}
\put(0,0){\line(1,0){110}}
\qbezier[10](0,10)(5,5)(10,0)
\qbezier[10](10,0)(15,5)(20,10)
\qbezier[10](20,10)(25,5)(30,0)
\qbezier[10](30,0)(35,5)(40,10)
\qbezier[10](60,10)(65,5)(70,0)
\qbezier[10](70,0)(75,5)(80,10)
{\thicklines
\put(0,10){\line(1,1){10}}
\put(10,20){\line(1,-1){10}}
\put(20,10){\line(1,1){10}}
\put(30,20){\line(1,-1){20}}
\put(50,0){\line(1,1){20}}
\put(70,20){\line(1,-1){20}}
\put(90,0){\line(1,1){10}}
\put(100,10){\line(1,-1){10}}
}
\qbezier[5](0,-6)(0,-3)(0,0)
\qbezier[5](10,-6)(10,-3)(10,0)
\qbezier[5](20,-6)(20,-3)(20,0)
\qbezier[5](110,-6)(110,-3)(110,0)
\put(-1,-13){$\scriptscriptstyle 0$}
\put(9,-13){$\scriptscriptstyle 1$}
\put(19,-13){$\scriptscriptstyle 2$}
\put(96,-13){$\scriptscriptstyle L=11$}
\end{picture}
\begin{center}
\parbox{15cm}{\small
\caption{Four different ballot paths (shown on top)
map to the same one-step Dyck path (shown below) under the homomorphism (\ref{8.2}).
Marked by asterisk "$\ast$" are the half-tiles which one erases during the mapping.
Marked by dagger "$\dag$" are the tiles  which one then removes using the rule (\ref{rule-1}).
There are $r=2$ reflection points at height $h=1$ in the first cluster
of the one-step Dyck path (see marked points).
Therefore, one finds $2^r=4$ different ballot paths  in the preimage.
The starting points for the layers of tiles are marked on the pictures of
ballot paths.
}}
\end{center}
\lb{map-Lodd}
\end{figure}
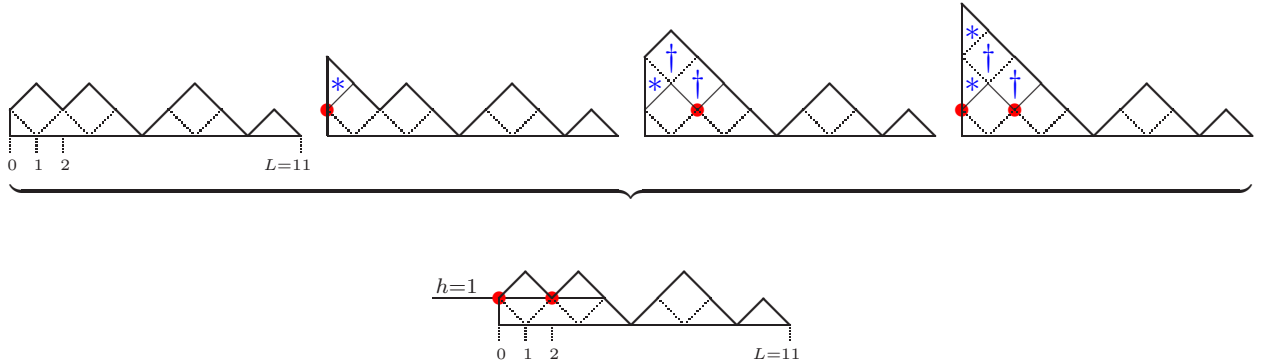

Let us consider in more details an action of the map (\ref{8.4}) on a ballot path.
Graphically one can realize this mapping using the  following procedure (see Figure~18): one first
erases all the half-tiles from the picture of the ballot path and then, one reduces the
resulting expression using the rules (\ref{rule-1})
(actually one needs only the last equality from (\ref{rule-1})).
There can be several ballot paths mapping onto the same one-step Dyck path under this
procedure.
One can reconstruct the corresponding ballot paths by dressing the
one-step Dyck path with the layers of tiles starting at any of its return point (local minima) at height $h=1$ in the first cluster and ending by a
half-tile at the wall (see Figure\,18).
  Assuming there are $r$ such return points in the one-step Dyck path,
one can construct $2^r$ different "dressed" ballot paths.

We observe that the profiles of all the ballot paths mapping under (\ref{8.2})
onto the same Dyck path coincide except for their leftmost clusters.
They are identical to the profile of the corresponding Dyck path, again, with an exception of its
leftmost cluster.
We now notice that,
according to (\ref{8.4}), in the RPMW ($L$ odd) the stationary weights of
the ballot paths mapping to the same Dyck path sum up (modulo the constant factor)
$S^{(a)}(L+1)$ to the stationary weight of the corresponding Dyck path in the RPM (same size $L$).
These two observations justify the following statement.
\vspace{2mm}

\ni
{\bf Proposition 1.\,}{\em
Consider the RPM and the RPMW for the same number of sites $L$ ($L$ odd) and any boundary rate $a$.
For both models
the stationary probabilities to have the first cluster at distance $x$
and any given configuration to the right of it are the same.
}
\vspace{2mm}

In particular, for $L$ odd,
the relations (\ref{PfPd-Lodd}) and  (\ref{n_c-Lodd}) are valid and
the conjecture  announced  on page \pageref{conj} implies the equality
(\ref{P1B-Lodd}).

\subsection{Case of $L$ even.}

In this case we restrict our consideration to the case $a=1$ only.
We consider another homomorphism mapping the one-boundary TL algebra
of size $L$
onto the TL algebra of size $(L+1)$:
\be
\lb{8.5}
{\mathcal T}^{(1B)}_L\, \rightarrow\, {\mathcal T}_{L+1} :\qquad
e_0\mapsto e_1, \quad e_i\mapsto e_{i+1}\;\; \forall\, i=1,\dots , L-1.
\ee
This homomorphism  maps  the intensity matrix of the RPMW, $a=1$,
to  the intensity matrix of the RPM
\be
\lb{8.8}
H_L^{(1)}\, \mapsto\, H_{L+1}\, ,
\ee
and so should map their unique ground-state eigenvectors:
\be
\lb{8.9}
{\ket 0}_{L}^{(1)}\, \mapsto\, S(L)\, {\ket 0}_{L+1}\, ,
\ee
where we have computed the scaling constant $S(L)$ by comparing the expressions
for their normalization factors (\ref{normalization-RPM}) and (\ref{8.1}).

With respect to the action of homomorphism (\ref{8.5}) the set of even size $L$
ballot paths separates naturally into two parts.

\smallskip\ni
{\bf ~a)\,} The ballot paths which have zero height at the origin, $h_0=0$,
map onto the one-step Dyck paths with  $h_1=0$ (see Figure\,19).
The mapping (\ref{8.5}) on these subsets of paths is a one to one correspondence.

\begin{figure}[t]
\begin{center}
\begin{picture}(120,40)(0,-20)
\put(10,0){\line(1,0){100}}
\put(-13,-10){$\scriptstyle h_0=0$}
\put(10,0){\circle*{3}}
\qbezier[10](20,10)(25,5)(30,0)
\qbezier[10](30,0)(35,5)(40,10)
\qbezier[10](40,10)(45,5)(50,0)
\qbezier[10](50,0)(55,5)(60,10)
{\thicklines
\put(10,0){\line(1,1){20}}
\put(30,20){\line(1,-1){10}}
\put(40,10){\line(1,1){10}}
\put(50,20){\line(1,-1){20}}
\put(70,0){\line(1,1){10}}
\put(80,10){\line(1,-1){10}}
\put(90,0){\line(1,1){10}}
\put(100,10){\line(1,-1){10}}
}
\qbezier[5](10,-6)(10,-3)(10,0)
\qbezier[5](20,-6)(20,-3)(20,0)
\qbezier[5](30,-6)(30,-3)(30,0)
\qbezier[5](110,-6)(110,-3)(110,0)
\put(9,-13){$\scriptscriptstyle 0$}
\put(19,-13){$\scriptscriptstyle 1$}
\put(29,-13){$\scriptscriptstyle 2$}
\put(96,-13){$\scriptscriptstyle L=10$}
\end{picture}
\raisebox{25pt}{$\quad\Rightarrow\quad\qquad$}
\begin{picture}(120,40)(0,-20)
\put(0,0){\line(0,1){10}}
\put(0,0){\line(1,0){10}}
\put(-13,-10){$\scriptstyle h_1=0$}
\put(10,0){\circle*{3}}
\put(10,0){\line(1,0){100}}
\qbezier[10](20,10)(25,5)(30,0)
\qbezier[10](30,0)(35,5)(40,10)
\qbezier[10](40,10)(45,5)(50,0)
\qbezier[10](50,0)(55,5)(60,10)
{\thicklines
\put(0,10){\line(1,-1){10}}
\put(10,0){\line(1,1){20}}
\put(30,20){\line(1,-1){10}}
\put(40,10){\line(1,1){10}}
\put(50,20){\line(1,-1){20}}
\put(70,0){\line(1,1){10}}
\put(80,10){\line(1,-1){10}}
\put(90,0){\line(1,1){10}}
\put(100,10){\line(1,-1){10}}
}
\qbezier[5](10,-6)(10,-3)(10,0)
\qbezier[5](20,-6)(20,-3)(20,0)
\qbezier[5](30,-6)(30,-3)(30,0)
\qbezier[5](110,-6)(110,-3)(110,0)
\put(9,-13){$\scriptscriptstyle 1$}
\put(19,-13){$\scriptscriptstyle 2$}
\put(29,-13){$\scriptscriptstyle 3$}
\put(96,-13){$\scriptscriptstyle L=11$}
\end{picture}
\parbox{15cm}{\small
\caption{A ballot path which has height $0$ at the wall
maps under the homomorphism (\ref{8.5}) to the one-step Dyck path  with
$h_1=0$. There are no other ballot paths mapping to this Dyck path.
}}
\end{center}
\lb{map-Leven1}
\end{figure}
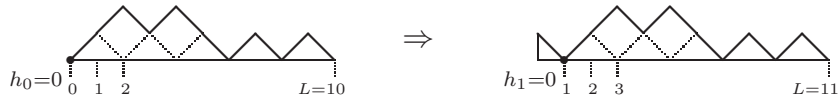

\smallskip\ni
{\bf ~b)\,} The ballot paths which have heights $h_0=2,4,\dots ,\frac{L}{2}\,$ at the origin
map onto the one-step Dyck paths with  $h_1=2$ (see Figure\,20).
Graphically one can realize the map (\ref{8.5}) for such paths in the following
way:  firstly, one completes each half-tile on the picture of the ballot path
to a full tile and then, one reduces the
resulting expression using the rules (\ref{rule-1})
(one needs only the last equality from (\ref{rule-1})).
This map is not a one to one correspondence.
The number of ballot paths  in the preimage
of a given restricted, $h_1=2$, one-step Dyck path is
equal to $2^r$, where $r$ is the number of return points
at height $h=1$ in the first cluster of the one-step Dyck path, the point $h_0=1$
is excluded (see Figure\,20).
The reconstruction of the preimage in this case is obtained again, by dressing the one-step
Dyck path with layers touching the wall (see caption of Figure~20).

\begin{figure}[t]
\begin{picture}(320,60)(0,5)
{\red{\put(144,10){\circle*{5}}
\put(284,10){\circle*{5}}
\put(384,10){\circle*{5}}
\put(404,10){\circle*{5}}
}}
{\blue{
\put(127.5,26.5){$\dag$}
\put(137.5,16.5){$\dag$}
\put(247.5,26.5){$\dag$}
\put(257.5,16.5){$\dag$}
\put(367.5,26.5){$\dag$}
\put(377.5,16.5){$\dag$}
\put(367.5,46.5){$\dag$}
\put(377.5,36.5){$\dag$}
\put(277.5,16.5){$\dag$}
\put(267.5,26.5){$\dag$}
\put(397.5,16.5){$\dag$}
\put(387.5,26.5){$\dag$}
\qbezier[6](0,10)(5,15)(10,20)
\qbezier[6](0,10)(5,5)(10,0)
\qbezier[6](120,10)(125,15)(130,20)
\qbezier[6](120,10)(125,5)(130,0)
\qbezier[6](120,30)(125,35)(130,40)
\qbezier[6](120,30)(125,25)(130,20)
\qbezier[6](240,10)(245,15)(250,20)
\qbezier[6](240,10)(245,5)(250,0)
\qbezier[6](240,30)(245,35)(250,40)
\qbezier[6](240,30)(245,25)(250,20)
\qbezier[6](360,10)(365,15)(370,20)
\qbezier[6](360,10)(365,5)(370,0)
\qbezier[6](360,30)(365,35)(370,40)
\qbezier[6](360,30)(365,25)(370,20)
\qbezier[6](360,50)(365,55)(370,60)
\qbezier[6](360,50)(365,45)(370,40)
}}
\put(10,0){\line(0,1){20}}
\put(10,0){\line(1,0){100}}
\qbezier[10](10,0)(15,5)(20,10)
\qbezier[10](20,10)(25,5)(30,0)
\qbezier[10](30,0)(35,5)(40,10)
\qbezier[10](40,10)(45,5)(50,0)
\qbezier[10](50,0)(55,5)(60,10)
{\thicklines
\put(10,20){\line(1,-1){10}}
\put(20,10){\line(1,1){10}}
\put(30,20){\line(1,-1){10}}
\put(40,10){\line(1,1){10}}
\put(50,20){\line(1,-1){20}}
\put(70,0){\line(1,1){10}}
\put(80,10){\line(1,-1){10}}
\put(90,0){\line(1,1){10}}
\put(100,10){\line(1,-1){10}}
}
\put(130,0){\line(0,1){40}}
\put(130,0){\line(1,0){100}}
\qbezier[10](130,20)(135,25)(140,30)
\qbezier[10](130,20)(135,15)(140,10)
\qbezier[10](140,10)(145,15)(150,20)
\qbezier[10](130,0)(135,5)(140,10)
\qbezier[10](140,10)(145,5)(150,0)
\qbezier[10](150,0)(155,5)(160,10)
\qbezier[10](160,10)(165,5)(170,0)
\qbezier[10](170,0)(175,5)(180,10)
{\thicklines
\put(130,40){\line(1,-1){30}}
\put(160,10){\line(1,1){10}}
\put(170,20){\line(1,-1){20}}
\put(190,0){\line(1,1){10}}
\put(200,10){\line(1,-1){10}}
\put(210,0){\line(1,1){10}}
\put(220,10){\line(1,-1){10}}
}
\put(250,0){\line(0,1){40}}
\put(250,0){\line(1,0){100}}
\qbezier[10](250,20)(255,25)(260,30)
\qbezier[10](250,20)(255,15)(260,10)
\qbezier[20](260,10)(270,20)(280,30)
\qbezier[20](260,30)(270,20)(280,10)
\qbezier[10](280,10)(285,15)(290,20)
\qbezier[10](250,0)(255,5)(260,10)
\qbezier[10](260,10)(265,5)(270,0)
\qbezier[10](270,0)(275,5)(280,10)
\qbezier[10](280,10)(285,5)(290,0)
\qbezier[10](290,0)(295,5)(300,10)
{\thicklines
\put(250,40){\line(1,-1){10}}
\put(260,30){\line(1,1){10}}
\put(270,40){\line(1,-1){40}}
\put(310,0){\line(1,1){10}}
\put(320,10){\line(1,-1){10}}
\put(330,0){\line(1,1){10}}
\put(340,10){\line(1,-1){10}}
}
\put(130,20){\line(1,-1){10}}
\put(140,10){\line(1,1){10}}
\put(250,20){\line(1,-1){10}}
\put(260,10){\line(1,1){10}}
\put(270,20){\line(1,-1){10}}
\put(280,10){\line(1,1){10}}
\put(370,20){\line(1,-1){10}}
\put(380,10){\line(1,1){10}}
\put(390,20){\line(1,-1){10}}
\put(400,10){\line(1,1){10}}
\put(370,0){\line(0,1){60}}
\put(370,0){\line(1,0){100}}
\qbezier[20](370,20)(380,30)(390,40)
\qbezier[10](370,40)(375,45)(380,50)
\qbezier[10](370,20)(375,15)(380,10)
\qbezier[20](380,10)(390,20)(400,30)
\qbezier[30](370,40)(385,25)(400,10)
\qbezier[10](400,10)(405,15)(410,20)
\qbezier[10](370,0)(375,5)(380,10)
\qbezier[10](380,10)(385,5)(390,0)
\qbezier[10](390,0)(395,5)(400,10)
\qbezier[10](400,10)(405,5)(410,0)
\qbezier[10](410,0)(415,5)(420,10)
{\thicklines
\put(370,60){\line(1,-1){60}}
\put(430,0){\line(1,1){10}}
\put(440,10){\line(1,-1){10}}
\put(450,0){\line(1,1){10}}
\put(460,10){\line(1,-1){10}}
}
\qbezier[5](10,-6)(10,-3)(10,0)
\qbezier[5](20,-6)(20,-3)(20,0)
\qbezier[5](30,-6)(30,-3)(30,0)
\qbezier[5](110,-6)(110,-3)(110,0)
\put(9,-13){$\scriptscriptstyle 0$}
\put(19,-13){$\scriptscriptstyle 1$}
\put(29,-13){$\scriptscriptstyle 2$}
\put(96,-13){$\scriptscriptstyle L=10$}
\put(-12,22){$\scriptstyle h_0=2$}
\put(10,20){\circle*{3}}
\put(108,42){$\scriptstyle h_0=4$}
\put(130,40){\circle*{3}}
\put(228,42){$\scriptstyle h_0=4$}
\put(250,40){\circle*{3}}
\put(348,62){$\scriptstyle h_0=6$}
\put(370,60){\circle*{3}}
\end{picture}
\\[3mm]
\phantom{x}
\hspace{-2mm}
$\underbrace{\hspace{165mm}}$
\\[3mm]
\\
\begin{picture}(200,45)(-190,-10)
{\red{\put(20,10){\circle*{5}}
\put(40,10){\circle*{5}}
}}
\put(0,0){\line(0,1){10}}
\put(0,0){\line(1,0){110}}
%
\qbezier[10](0,10)(5,5)(10,0)
\qbezier[10](10,0)(15,5)(20,10)
\qbezier[10](20,10)(25,5)(30,0)
\qbezier[10](30,0)(35,5)(40,10)
\qbezier[10](40,10)(45,5)(50,0)
\qbezier[10](50,0)(55,5)(60,10)
{\thicklines
\put(0,10){\line(1,1){10}}
\put(10,20){\line(1,-1){10}}
\put(20,10){\line(1,1){10}}
\put(30,20){\line(1,-1){10}}
\put(40,10){\line(1,1){10}}
\put(50,20){\line(1,-1){20}}
\put(70,0){\line(1,1){10}}
\put(80,10){\line(1,-1){10}}
\put(90,0){\line(1,1){10}}
\put(100,10){\line(1,-1){10}}
}
\put(-36,13){$\scriptstyle h=1$}
\put(-12,22){$\scriptstyle h_1=2$}
\put(10,20){\circle*{3}}
{\linethickness{0.3pt}
\put(-38,10){\line(1,0){98}}}
\qbezier[5](10,-6)(10,-3)(10,0)
\qbezier[5](20,-6)(20,-3)(20,0)
\qbezier[5](30,-6)(30,-3)(30,0)
\qbezier[5](110,-6)(110,-3)(110,0)
\put(9,-13){$\scriptscriptstyle 1$}
\put(19,-13){$\scriptscriptstyle 2$}
\put(29,-13){$\scriptscriptstyle 3$}
\put(96,-13){$\scriptscriptstyle L=11$}
\end{picture}
\begin{center}
\parbox{15cm}{\small
\caption{Four different ballot paths (shown on top)
map to the same one-step Dyck path (shown below) under the homomorphism (\ref{8.5}).
The half-tiles on the pictures of ballot paths are completed to full tiles.
Marked by dagger "$\dag$" are the tiles  which one removes during the mapping
using the rule (\ref{rule-1}).
There are $r=2$ reflection points at height $h=1$ ($h_0$ excluded) in the first cluster
of the one-step Dyck path (see marked points).
Therefore, one finds $2^r=4$ different ballot paths  in the preimage.
The starting points for the layers of tiles are marked on the pictures of ballot paths.
}}
\end{center}
\lb{map-Leven2}
\end{figure}
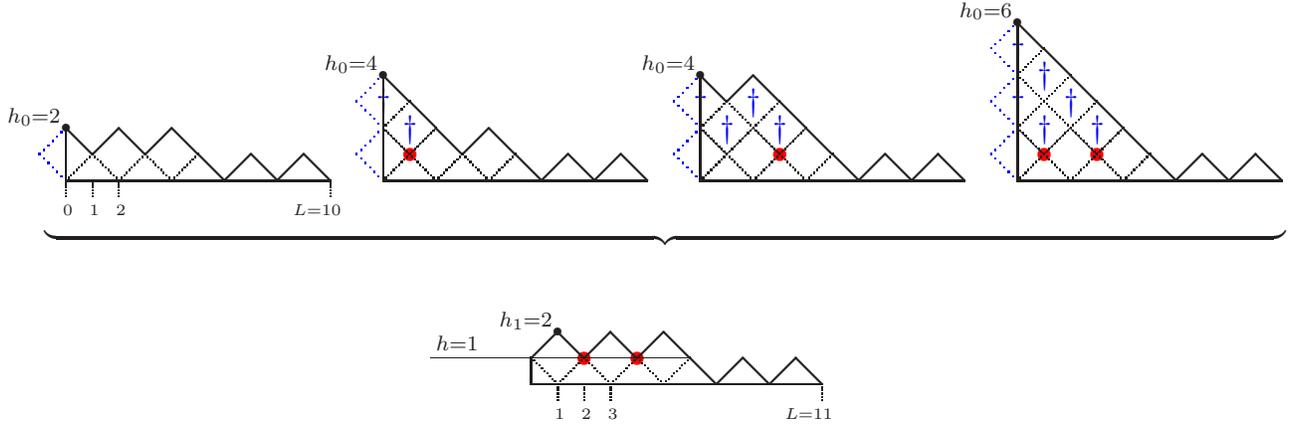

We observe  that in both cases a) and b),
the profiles of the ballot paths, except their leftmost clusters,
are identical to the profiles of the corresponding one-step
Dyck paths. We also notice that the leftmost clusters
become one step larger under the map (\ref{8.5}).
Therefore, taking into account the relation (\ref{8.9})
we conclude:
\vspace{2mm}

\ni
{\bf Proposition 2.\,}{\em
For the  RPMW of even size $L$ and the boundary rate $a=1$,
the stationary probability of having the first cluster at distance $x$
and any given configuration to the right of it is equal to the stationary probability
for the RPM of odd size $(L+1)$,
to have the first cluster at distance $(x+1)$ and the same configuration to the right
of it.
}
\vspace{2mm}

In particular,
the relations (\ref{PfPd-Leven}) and  (\ref{n_c-Leven}) are satisfied.

\medskip
Finally, we  demonstrate that the proposition 2, together with the
conjecture  presented  in page \pageref{conj}
implies the equality (\ref{P1B-Leven}).
Taking into account that the ballot paths with $k$ clusters project under
(\ref{8.5}) to the one-step Dyck paths  with $(k+1)$ clusters in case a) (see Figure\,19)
and with $k$ clusters in case b) (see Figure\,20) one can write
\ba
\nn
\hspace{-10mm}
\mbox{$L$ even:}\qquad
P^{(1)}_L(k) &=&
P_{L+1}(k, h_1=2)\, +\,  P_{L+1}(k+1, h_1=0)
\\[1mm]
\lb{8.10}
&=&
P_{L+1}(k)\, -\, P_{L+1}(k,  h_1=0)\, +\,  P_{L+1}(k+1, h_1=0)\, ,
\ea
where $P_{L+1}(k,{h_1=0(2)})$ are stationary probabilities
to find $k$ clusters and the height \mbox{$h_1=0(2)$} in the first site
for the RPM with the odd number of sites $(L+1)$:
$$
P_{L+1}(k,h_1=0)+P_{L+1}(k,h_1=2)=P_{L+1}(k)\, .
$$
The first term in the sum (\ref{8.10}) is a stationary contribution of the profiles
with $(k-1)$ return points at height $h=0$ ($\Leftrightarrow\; n=L/2$, see Figure\,17)
which is given by the conjecture:
\be
\lb{8.11}
P_{L+1}(k)\, =\, {\textstyle \frac{1}{S(L+1)}}\, P(L+1,{\textstyle \frac{L}{2}},k-1)\, .
\ee
To calculate the last two terms in the sum (\ref{8.10}) we take into account the left-right
symmetry of the one-defect profiles in the "defects" picture of the one-step
Dyck paths (see Appendix A). As it is seen on the following picture
\begin{figure}[h]
\begin{center}
\begin{picture}(300,90)(0,5)
\put(-80,70){\small One-step }
\put(-80,58){\small Dyck paths:}
\put(-80,12){\small  Single defect}
\put(-80,0){\small configurations:}
\put(134,24){\small  left-right }
\put(134,12){\small   symmetry}
{\thicklines
\put(20,0){\line(1,0){80}}
\put(20,0){\line(1,1){20}}
\put(40,20){\line(1,-1){20}}
\put(60,0){\line(1,1){10}}
\put(70,10){\line(1,-1){10}}
\put(80,0){\line(1,1){10}}
\put(90,10){\line(1,-1){10}}
\put(210,0){\line(1,0){80}}
\put(210,0){\line(1,1){10}}
\put(220,10){\line(1,-1){10}}
\put(230,0){\line(1,1){10}}
\put(240,10){\line(1,-1){10}}
\put(250,0){\line(1,1){20}}
\put(270,20){\line(1,-1){20}}
\put(10,60){\line(1,0){90}}
\put(10,60){\line(0,1){10}}
\put(10,70){\line(1,-1){10}}
\put(20,60){\line(1,1){20}}
\put(40,80){\line(1,-1){20}}
\put(60,60){\line(1,1){10}}
\put(70,70){\line(1,-1){10}}
\put(80,60){\line(1,1){10}}
\put(90,70){\line(1,-1){10}}
\put(210,60){\line(1,0){90}}
\put(210,60){\line(0,1){10}}
\put(210,70){\line(1,1){10}}
\put(220,80){\line(1,-1){10}}
\put(230,70){\line(1,1){10}}
\put(240,80){\line(1,-1){10}}
\put(250,70){\line(1,1){20}}
\put(270,90){\line(1,-1){30}}
}
\put(185,70){\line(1,0){105}}
\put(185,72){$\scriptstyle h=1$}
\put(50,35){$\Updownarrow$}
\put(250,35){$\Updownarrow$}
\put(150,0){$\Leftrightarrow$}

\put(230,70){\circle*{4}}
\put(250,70){\circle*{4}}
\put(20,60){\circle*{4}}
\put(60,60){\circle*{4}}
\put(80,60){\circle*{4}}
{\red{
\put(15,0){\circle*{5}}
\put(295,0){\circle*{5}}
}}
\end{picture}
\end{center}
\end{figure}

\ni
in the RPM of the odd size $(L+1)$ the stationary weights of
profiles with $(k-1)$ contact points ({\it i.e.} with $k$ clusters),
one of them is at site 1 ($h_1=0$),
are equal to the stationary weights of  profiles with no
contact points, but with  $(k-2)$ return points at height
$h=1$ ($\Leftrightarrow\; n=\frac{L}{2}-1$, see Figure\,17).
Therefore
\be
\lb{8.12}
P_{L+1}(k, h_1=0)\, =\, {\textstyle \frac{1}{S(L+1)}}\,
P(L+1,{\textstyle \frac{L}{2}}-1,k-2)\, .
\ee
From (\ref{8.10})--(\ref{8.12}) we obtain for $L$ even
$$
P^{(1)}_L(k)\, =\, {\textstyle \frac{1}{S(L+1)}}\, \Bigl\{
P(L+1,{\textstyle \frac{L}{2}},k-1)-P(L+1,{\textstyle \frac{L}{2}}-1,k-2)+
P(L+1,{\textstyle \frac{L}{2}}-1,k-1)
\Bigr\}\, ,
$$
which is equivalent to (\ref{P1B-Leven}) provided one takes into account the identity
$$
P(L+1,{\textstyle \frac{L}{2}},m)-P(L+1,{\textstyle \frac{L}{2}}-1,m-1)=
P(L+1,{\textstyle \frac{L}{2}},m+1)\quad
\mbox{for $L$ even and}\;\forall\,m=0,\dots   ,{\textstyle \frac{L}{2}}\, .
$$

\section{Conclusions}

As far as we know the RPM with and without a wall is the only
example of a one-dimensional fluctuating interface which is conformal
invariant (the central charge of the Virasoro algebra $c = 0$ since in a
stochastic process the ground-state energy is zero for any system size). The
time dependence of average quantities are under control since the
finite-size scaling Hamiltonian spectrum is known (see Refs.~\cite{SB,GNPyR}). For
space dependent phenomena, the situation is more complex since one
doesn't know which critical exponents appear. What we do know is
that, in the finite-size scaling limit, the
functional dependence of density profiles in the stationary
state are fixed once the scaling dimensions of the local operators and the
boundary conditions are specified (see (\ref{4.3}) and (\ref{4.5})). Our aim
was to identify local operators by checking if the expressions (\ref{4.3})
and (\ref{4.5}) are confirmed by the data.

First we have considered the contact-point density profiles in
the RPM and in the RPMW and found the density profiles in the
finite-size scaling limit using Monte Carlo simulations. They have the
expected behavior (see (\ref{4.3}), (\ref{4.6}) and (\ref{4.10})) for an operator with
scaling dimension $X = 1/3$.
This value is puzzling since as explained in Section 4, it is incompatible
with the Kac table for an operator of conformal spin zero
(see (\ref{4.10})).
The material presented in Sections 6, 7 and 8 in which one computes the
probabilities to have $k$ clusters ($k+1$ contact points) in the system
gives a rigorous proof that the exponent is indeed $X = 1/3$.
We would like to qualify what we mean by rigorous. The derivation is based
on finding solutions of the split-hexagon recurrence relations (\ref{split-hexagon}) for
the quantities $P(L,m,n)$. The identification of these quantities with
certain Dyck path configurations as explained in Sec.\,7 remains a
conjecture.

We have also considered the defect density profiles in the RPMW model and
found the expected finite-size scaling function with an exponent $X = 1$
which is also puzzling since it is also incompatible with the Kac table
for a spinless operator (see (\ref{4.16})). We have no rigorous proof that the
exponent is 1 but the physics which follows (see (\ref{4.14})) is very
plausible.

One easy way out of these puzzles is to bring the argument that in a
$c = 0$ theory, everything is possible. We are tempted by another scenario: in
both cases one has the conformal spin $s = X$. In this case both values
$1/3$
and $1$ are given by the Kac table but we run into a different problem: an
operator with conformal spin has a vanishing density. Why this does not
happen in our case, is another puzzle. Possibly there is a simple explanation, but it escaped us.

The model has many interesting properties as discussed in detail in
Section 4. One of them looks to us really unexpected: the connection
between the clusters in the RPM and the parameter dependent (the boundary
rate $a$) RPMW for an odd number of sites. Except for the first cluster,
the two models are identical. The proof of this statement is in  Section 8.
It is perhaps no accident that the function which gives the probability of
finding
 the first cluster at a distance $x$ from the wall has special
properties (see (\ref{4.18})).
The implications of these observations on time dependent average
quantities will be discussed  elsewhere.

We would like to comment about the content of the mathematical part of
this paper. Its relevance goes beyond finding a critical exponent and
mapping of two models. In Section 6 we give solutions for the so
called Pascal's hexagon bilinear recurrence relations corresponding to
certain boundary conditions. We also introduce new bilinear split-hexagon
recurrence relations and solve them for certain boundary conditions. Their
connection to integrable nonlinear differential equations is still to be
found.

\newpage

\section*{Acknowledgements}

We thank Jan de Gier, Vyacheslav Priezzhev, Philippe Ruelle and Alexander Zamolodchikov
for valuable discussions and comments.

The work of P.P. and V.R. was supported by the DFG-RFBR grant
(436~RUS~113/909/0-1(R) and 07-02-91561-a)
and by the grant of the Heisenberg-Landau program.
The work of P.P. was partially supported by the RFBR grant No.~05-01-01086-a.
The work of F.C.A. was partially supported by FAPESP and CNPq (Brazilian
Agencies).

\appendix
\section{Appendix.\\[1mm] The RPM as a pair annihilation process with a source}

In Sec.~2 we have shown that the one-step configurations can be mapped
onto configurations with one defect and clusters (see Figures~2 and 3) and
that the ballot path configurations can be mapped onto defect
configurations. In a defect configuration, clusters start and end on sites
and defects occupy links. We give some simple examples of the mapping.

In Figure~21 we show the three ballot paths for $L = 3$  and in Figure~22 we
show the corresponding three defect configurations. The mapping is shown
in Figures~23 and 24 for $L = 4$.
\begin{figure}[ht!] \label{figa1}
\begin{center}
\begin{picture}(260,150)
\put(0,0){\epsfxsize=150pt\epsfbox{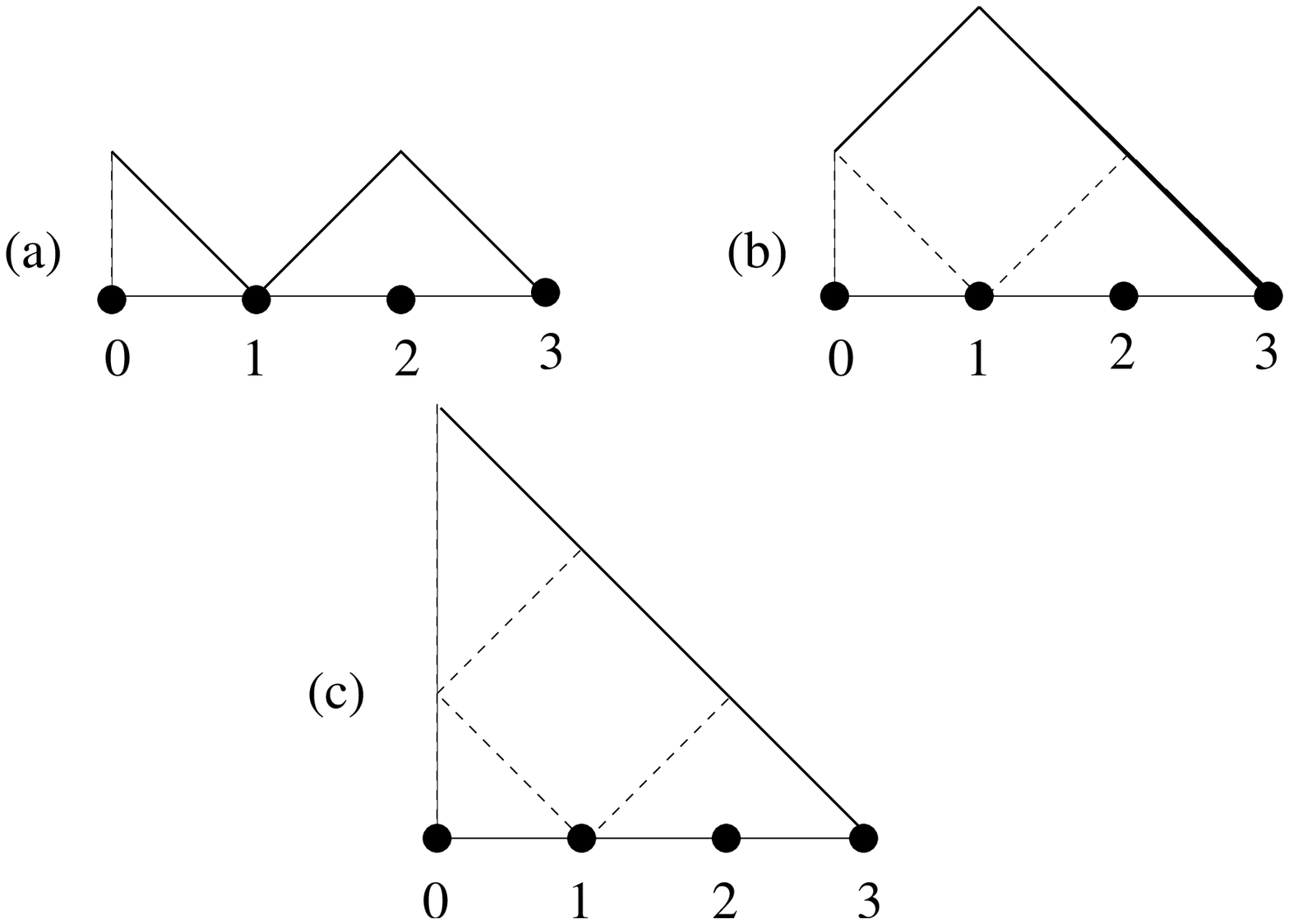}}
\end{picture}
\parbox{15cm}{
\caption{\small The three ballot path
configurations for $L=3$ (four sites).  In the stationary state of the RPMW all configurations are present while
in the
case of the RPM only configurations (a) and (b) (one-step Dyck paths) occur.
The configurations should be compared with these of Figure~22}
}
\end{center}
\end{figure}
\begin{figure}[ht!] \label{figa2}
\begin{center}
\begin{picture}(220,100)
\put(0,0){\epsfxsize=200pt\epsfbox{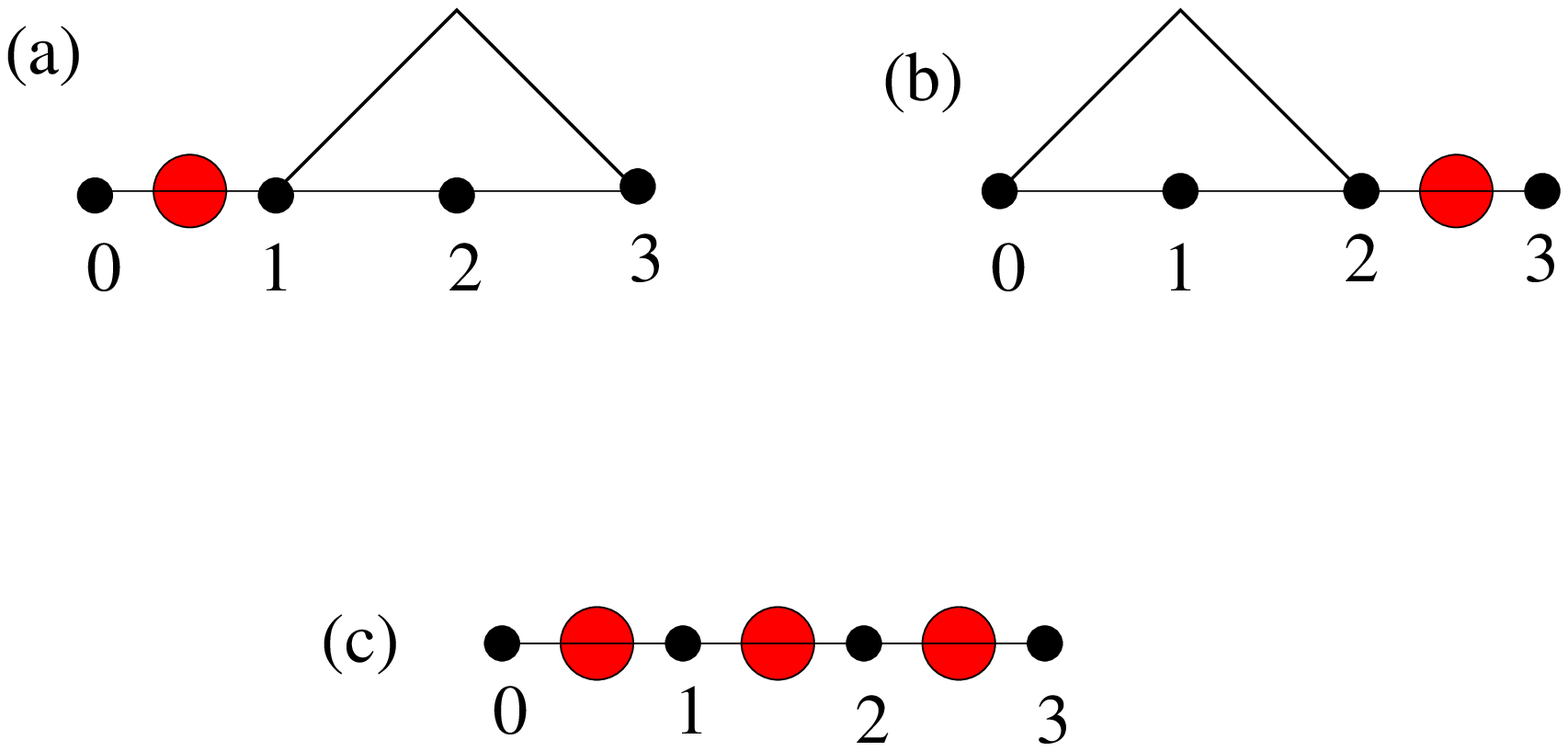}}
\end{picture}
\parbox{15cm}{
\caption{\small (Color online)
The three configurations for $L=3$ (four sites) formed by defects (red circles)
and RSOS clusters. The corresponding ballot path configurations in the RPMW are given
in Figure~21.}
}
\end{center}
\end{figure}
\begin{figure}[ht!] \label{figa3}
\begin{center}
\begin{picture}(260,180)
\put(0,0){\epsfxsize=200pt\epsfbox{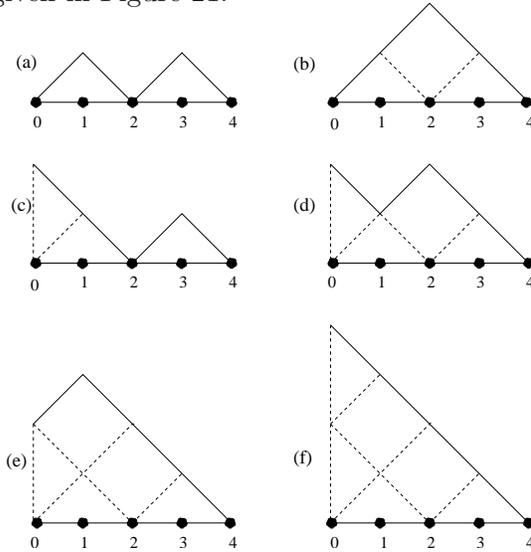}}
\end{picture}
\parbox{15cm}{
\caption{\small The six ballot path configurations for $L=4$ (five sites).
 In the stationary state of the RPMW all configurations are present while
in the
case of the RPM only configurations (a) and (b)  occur.
The configurations should be compared with these of Figure~24}
}
\end{center}
\end{figure}
\begin{figure}[ht!] \label{figa4}
\begin{center}
\begin{picture}(220,130)
\put(0,0){\epsfxsize=200pt\epsfbox{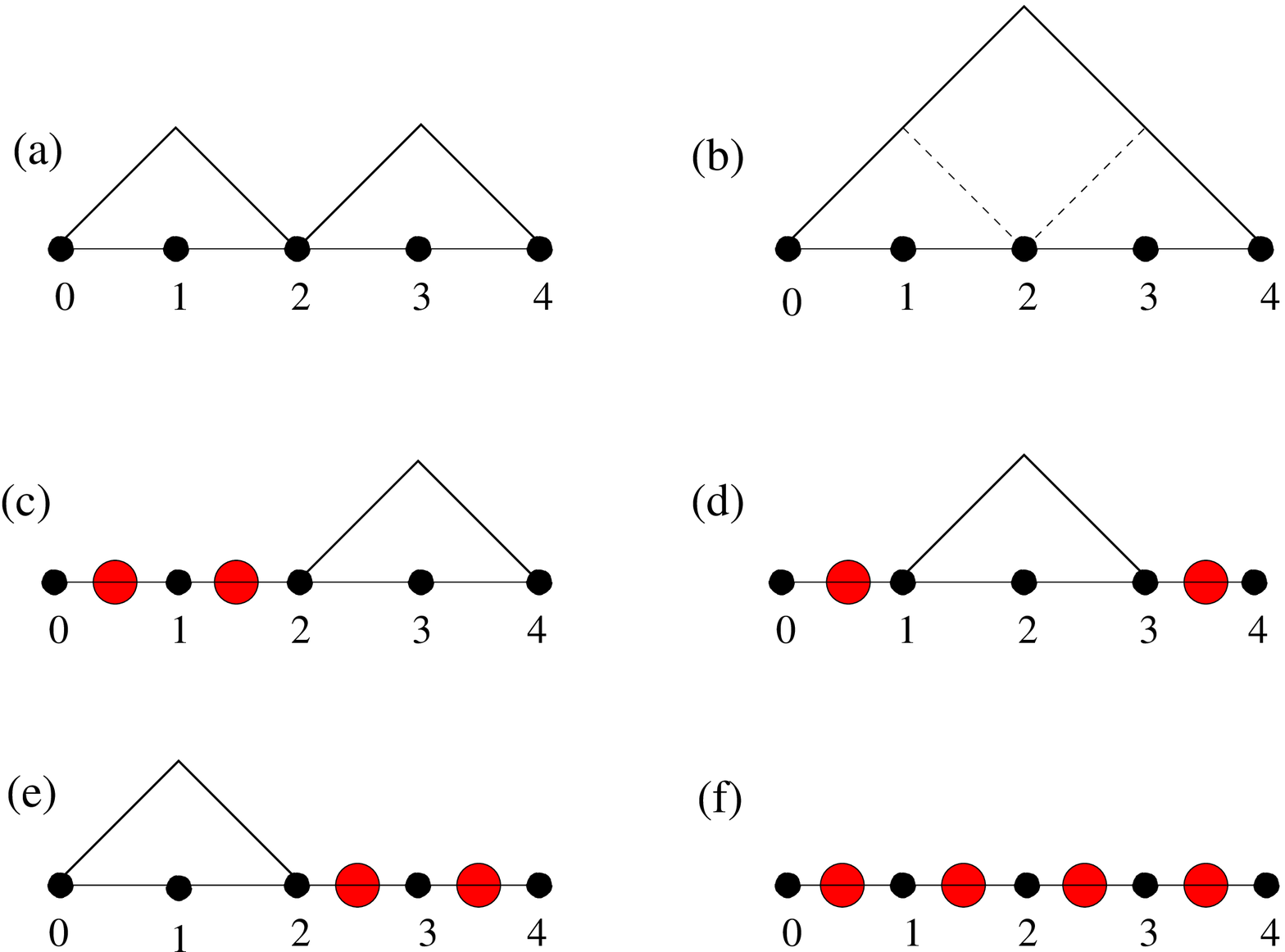}}
\end{picture}
\parbox{15cm}{
\caption{\small (Color online)
The six configurations for $L=4$ (five sites) formed by defects (red circles)
and RSOS clusters. The corresponding ballot path configurations in the RPMW are given
in Figure~23.}
}
\end{center}
\end{figure}
For a  defect configuration  with $L$ even (odd), one has an even
(odd) number of defects.

The dynamics of the RPMW, given in Sec.~2 describing the evolution of the
interface defined by ballot paths, can be translated in the evolution of
the system defined by defect configurations. The evolution of the clusters
is  the same as in the RPM but the defects (impurities) hop over the
adjacent clusters and in the hopping process they peel the clusters. When
two adjacent defects annihilate they are replaced by  a small cluster
"building" the substrate. The source acts nonlocally: it adds a defect on
the first site and another one at the end of the first cluster. We show
now in detail the evolution process.

We consider a rarefied gas of tiles falling on a defect configuration with
$L + 1$ sites. With a probability $p_b = 1/(L-1 +a)$ a tile hits the bulk of
the configuration (the sites $i = 1,\ldots,L-1$).
With a probability $p_s = a/(L -1 +a)$
a tile hits the boundary site $i = 0$. $a$ is called the boundary rate.

The evolution rules depend on where the tile hits the configuration. In
Figure~25 we show the different cases. In the cases {\it a, b} and {\it c}
 the tile
hits a cluster and the rules are the same as in the RPM given in Sec.~2.
We give the rules for the cases  {\it d} and   {\it e} (see Figure~25)

\begin{figure}[ht!]\label{figa5}
\begin{center}
\begin{picture}(260,150)
\put(-40,0){\epsfxsize=350pt\epsfbox{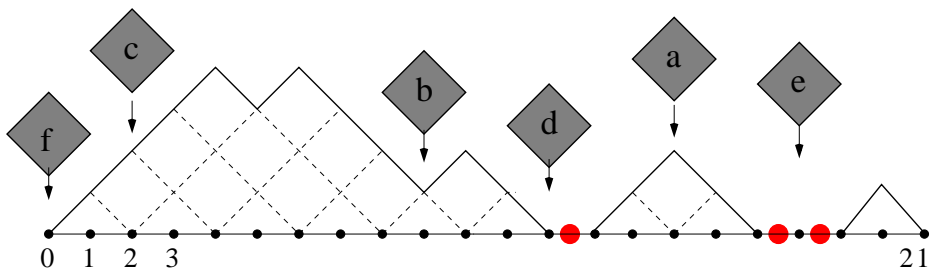}}
\end{picture}
\parbox{15cm}{
\caption{\small (Color online)
One of the
  defect configurations for $L = 21$ (22 sites). There are 3
defects
(red circles on the links) and 3 clusters. Also shown are six tiles (tilted
squares) {\it a - f}
belonging to the gas. When a tile hits the surface the effect is
different in the six cases. }
}
\end{center}
\end{figure}

\begin{itemize}

\item[({\it d})]
The tile hits the site $i$ which is at the right end of a cluster
$h_j>h_{i-c} = h_i = 0$ ($j = i - c + 1,\ldots, i - 1$).
A defect is present on the
link $(i, i + 1)$. The defect
hops on the first left link of the cluster $(i - c, i - c + 1)$ . In order
to create a free link on which the defect hops, part of the outer layer is
desorbed: if $h_{j+1} - h_j$ is negative ($i-c+1 \leq j \leq i-2$) the
 tile,
with the center at coordinates ($j,h_j-1$), evaporates  (see Figure~26a),
afterwards, the remaining cluster moves to the right by one lattice spacing
($h_j\rightarrow h_{j+1}$), as in Figure~26b.
 An alternative way to visualize  the process is shown in
Figure 26b. First a whole layer of the cluster evaporates $h_j \rightarrow h_j- 1$
($j = i- c + 1,\ldots, i - 1$).
 The defect hops from
the link $(i, i + 1)$ to the link $(i - c, i - c + 1)$ and a half-tile cluster
(part of the substrate) is created: $h_{i-1} = 0, h_i = 1, h_{i+1} = 0$.
\end{itemize}
\begin{figure}[ht!] \label{figa6}
\begin{center}
\begin{picture}(260,200)
\put(0,0){\epsfxsize=280pt\epsfbox{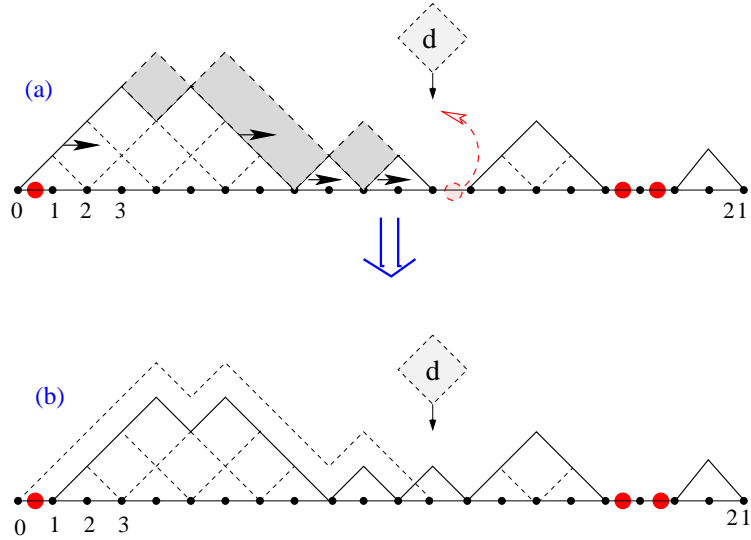}}
\end{picture}
\parbox{15cm}{
\caption{\small (Color online)
The new profile after the tile {\it d} in
Figure~25 has hit the surface at the right end of a cluster.
The defect hops to the left end of the cluster, first peeling a layer of 5 tiles (a), and next, the peeled cluster
 is translated by a lattice spacing unit to the right (b).
}
}
\end{center}
\end{figure}

\begin{figure}[ht!]\label{figa7}
\begin{center}
\begin{picture}(260,80)
\put(0,0){\epsfxsize=280pt\epsfbox{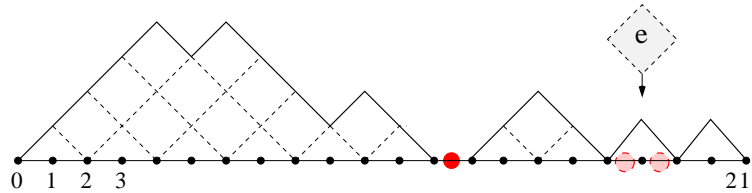}}
\end{picture}
\parbox{15cm}{
\caption{\small (Color online)
The new profile after the tile {\it e} has hit the surface between
two defects. The defects have disappeared and in their place one gets
a new small cluster.}
}
\end{center}
\end{figure}
\begin{itemize}

\item[({\it e})]
The tile hits the site $i$ between two adjacent defects placed on the
links $(i - 1, i)$ and $(i, i + 1)$.
The two
defects annihilate and in their place appears a small cluster ($h_{i-1} = h_{i+1} = 0,
h_i = 1$) (see Figure~27).
\end{itemize}

Up to now we have described the action of the RPM Hamiltonian $H_L$ (see
 (5.10)) on the defect configurations. This model was used in [AR1] to
study the pair annihilation  process of defects in an unquenched disordered
system. We now consider the effect of the boundary operator $e_0$ in the
Hamiltonian $H_L^{(a)}$ (see  (5.10)). This corresponds to the effect of a
tile hitting the boundary site $i = 0$ (case {\it{f}} in Figure 25).
\begin{itemize}

\item[{({\it f})}]
If the tile hits the site $i = 0$ and the link $(0,1)$ is occupied by a
defect, the tile is reflected. If there is no defect on the first
link (this is the case {\it f} in Figure~25), the site $i = 0$ is the left-end of a
cluster $h_j > h_0 = h_c =0$ ($j = 1,\ldots,c-1$) two defects are added to the
configuration (one defect on the link $(0,1)$ another one on the link
$(c-1,c)$) after peeling and moving the first cluster (see Figure~28a).
 If $h_{j+1}-h_j$ is positive ($0 \leq j\leq c-3$) the outer
tile, with center at coordinates ($j+1,h_{j+1}-1$), evaporates, and the
 remaining cluster is
moved to the left by one unit. An alternative way to visualize the effect of the
tile which acts as a source of defects, is to first desorb a layer from
the first cluster  $h_j \rightarrow  h_j -1$ ($j = 1,\ldots, c - 1$).
Next the defects are introduced on the two links which became free as the
result of the desorption process (see Figure~28b).
\end{itemize}

\begin{figure}[ht!]\label{figa8}
\begin{center}
\begin{picture}(260,200)
\put(0,0){\epsfxsize=280pt\epsfbox{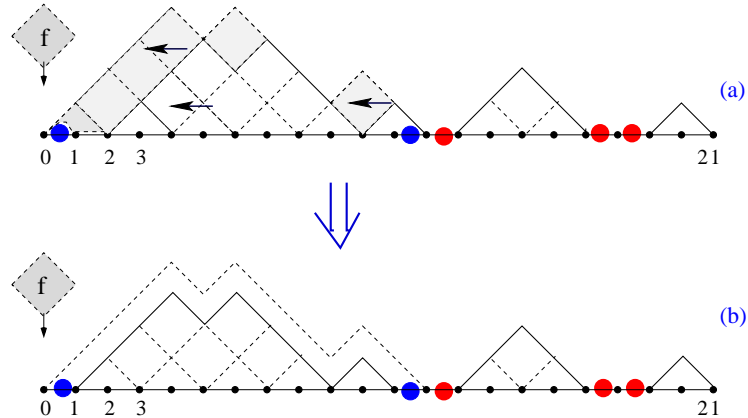}}
\end{picture}
\parbox{15cm}{
\caption{\small (Color online)
The new profile obtained after a tile {\it f} in Figure~25 has hit the
site $i=0$ (the left side of a cluster defined by two
contact points $h_0=h_c=0$, $ c=12$ in Figure~25). One first peels a layer of
the cluster (5 tiles in the figure) (a), moves the remaining cluster by one unit
lattice spacing to the left and then adds two defects (blue circles) (b).
One on the link $(0,1)$  and one on the link $(c-1,c)$.}
}
\end{center}
\end{figure}



\begin{thebibliography}{999}

\bibitem{GNPR} J.~de Gier, B.~Nienhuis, P.A.~Pearce and V.~Rittenberg,
{\em `The raise and peel model of a fluctuating interface'},
J.~Stat.~Phys. {\bf 114} (2004) 1-35
\href{http://arxiv.org/pdf/cond-mat/0301430}{\texttt{arXiv:cond-mat/0301430}}.

\bibitem{ALR}  F.C.~Alcaraz, E.~Levine and  V.~Rittenberg,
{\em `Conformal invariance and its breaking in a stochastic model of a fluctuating interface'}.
J.~Stat.~Mech. (2006) P08003
\href{http://arxiv.org/pdf/cond-mat/0604223}{\texttt{arXiv:cond-mat/0604223}}.

\bibitem{AR2}  F.C. Alcaraz,  and  V.~Rittenberg,
{\em `Different facets of the raise and peel model'}.
J.~Stat.~Mech.  (2007) P07009
\href{http://arxiv.org/pdf/cond-mat/0703725}{\texttt{arXiv:cond-mat/0703725}}.

\bibitem{G} J.~de Gier,
{\em `Loops, matchings and alternating-sign matrices'},
Discr.~Math. {\bf 298} (2005) 365-388
\href{http://arxiv.org/pdf/math.CO/0211285}{\texttt{arXiv:math.CO/0211285}}.

\bibitem{MS}  P.P.~Martin and H.~Saleur, Lett.~Math.~Phys. {\bf 30} (1994)  189-206,
{\em `The blob algebra and the periodic Temperley-Lieb algebra'},
\href{http://arxiv.org/pdf/hep-th/9302094}{\texttt{arXiv:hep-th/9302094}}.

\bibitem{NRG} A.~Nichols, V.~Rittenberg, and J.~de Gier,
{\em `One-boundary Temperley-Lieb algebras in the XXZ and loop models'},
J.~Stat.~Mech.  (2005) P05003
\href{http://arxiv.org/pdf/cond-mat/0411512}{\texttt{arXiv:cond-mat/0411512}}.

\bibitem{Sh} K.~Shelton,
{\em `The Singled Out Game'},
Mathematics Magazine,
{\bf 78} (2005) 15.

\bibitem{AR1} F.C.Alcaraz and V.Rittenberg,
{\em `The pair annihilation reaction D + D $\rightarrow$ 0 in disordered media and conformal invariance'},
Phys.~Rev.~E {\bf 75} (2007) 051110
\href{http://arxiv.org/pdf/cond-mat/0612272}{\texttt{arXiv:cond-mat/0612272}}.

\bibitem{TL}  H.N.V.~Temperley  and  E.H.~Lieb,
{\em `Relations between percolation and colouring problems and other graph theoretical
problems associated with regular planar lattices: some exact results for the percolation problem'},
Proc.\,Roy.\,Soc.\,London\,Ser.\,A {\bf 322} (1971) 251-280.

\bibitem{SB} H.~Saleur and M.~Bauer,
{\em `On some relations between local height probabilities and conformal invariance'},
Nucl.~Phys. {\bf B320} (1989) 591-624.

\bibitem{RS1} A.V.~Razumov  and  Yu.G.~Stroganov,
{\em `Spin chains and combinatorics'},
J.\,Phys.\,A:\,Math.\,Gen. {\bf 34} (2001) 3185-3190
\href{http://arxiv.org/pdf/math.CO/0012141}{\texttt{arXiv:math.CO/0012141}}.

\bibitem{RS2} A.V.~Razumov  and  Yu.G.~Stroganov,
{\em `Combinatorial nature of ground state vector of O(1) loop model},
Theor.\,Math.\,Phys. {\bf 138} (2004) 333-337
\href{http://arxiv.org/pdf/math.CO/0104216}{\texttt{arXiv:math.CO/0104216}}.

\bibitem{MNGB}S.~Mitra, B.~Nienhuis, J.~de Gier and M.T.~Batchelor,
{\em `Exact expressions for correlations in the ground state of the dense O(1) loop model'}.
J.~Stat.~Mech. (2004) P09010
\href{http://arxiv.org/pdf/cond-mat/0401245}{\texttt{arXiv:cond-mat/0401245}}.

\bibitem{P} P.~Pyatov,
{\em `Raise and Peel Models of fluctuating interfaces and combinatorics of Pascal's hexagon'}.
J.~Stat.~Mech. (2004) P09003
\href{http://arxiv.org/pdf/math-ph/0406025}{\texttt{arXiv:math-ph/0406025}}.

\bibitem{GR} J.~de Gier, V.~Rittenberg,
{\em `Refined Razumov-Stroganov conjectures for open boundaries'},
J.~Stat.~Mech. (2004) P09009
\href{http://arxiv.org/pdf/math-ph/0408042}{\texttt{arXiv:math-ph/0408042}}.

\bibitem{GNPyR} J.~de Gier, A.~Nichols, P.~Pyatov and V.~Rittenberg,
{\em `Magic in the spectra of the XXZ quantum chain with boundaries at $\Delta=0$ and $\Delta=-1/2$'},
Nucl.~Phys. {\bf B729} (2005) 387-418
\href{http://arxiv.org/pdf/hep-th/0505062}{\texttt{arXiv:hep-th/0505062}}.

\bibitem{B}  D.M.~Bressoud, {\em `Proofs and Confirmations. The Story
of the Alternating Sign Matrix Conjecture'}, 1999
Cambridge University Press, Cambridge

\bibitem{K}  G.~Kuperberg,{\em `Symmetry classes of alternating-sign-matrices under one roof},
Ann.~of~Math. {\bf 156},~no.\,3 (2002) 835-866
\href{http://arxiv.org/pdf/math.CO/0008184}{\texttt{arXiv:math.CO/0008184}}.

\bibitem{R} D.P.~Robbins,
{\em `Symmetry classes of alternating sign matrices}, 2000
\href{http://arxiv.org/pdf/math.CO/0008045}{\texttt{arXiv:math.CO/0008045}}.

\bibitem{BX} T.W.~Burkhardt and T.~Xue,
{\em `Density profiles in confined critical systems and conformal invariance'},
Phys.~Rev.~Lett. {\bf 66} (1991) 895-898;
Nucl.~Phys. {\bf B354} (1991) 653-665.

\bibitem{A} I.~Affleck,
{\em `Edge magnetic field in the xxz spin-$\frac{1}{2}$ chain'},
J.~Phys.~A: Math.~Gen. {\bf 31} (1998) 2761-2766.

\bibitem{CRL} Z.~Cheng, S.~Redner and F.~Leyvraz,
{\em `Coagulation with a steady point monomer source'},
Phys.~Rev.~Lett. {\bf 62} (1989) 2321-2324.

\bibitem{HRS} H.~Hinrichsen, V.~Rittenberg and H.~Simon,
{\em `Universality properties of the stationary states in the one-dimensional coagulation-diffusion
 model with external particle input'},
J.~Stat.~Phys. {\bf 86} (1997) 1203
\href{http://arxiv.org/pdf/cond-mat/9606088}{\texttt{arXiv:cond-mat/9606088}}.

\bibitem{BGN} M.T.~Batchelor, J.~de Gier, B.~Nienhuis,
{\em `The quantum symmetric XXZ chain at $\Delta=-1/2$, alternating sign matrices and plane partitions'},
J.~Phys.~A:~Math.~Gen. {\bf 34} (2001) L265-L270
\href{http://arxiv.org/pdf/cond-mat/0101385}{\texttt{arXiv:cond-mat/0101385}}.

\bibitem{PRGN} P.A.~Pearce, V.~Rittenberg, J.~de Gier, B.~Nienhuis,
{\em `Temperley-Lieb Stochastic Processes'},
J.~Phys.~A:~Math.~Gen. {\bf 35} (2002) L661-L668
\href{http://arxiv.org/pdf/math-ph/0209017}{\texttt{arXiv:math-ph/0209017}}.

\bibitem{DJM} E.~Date, M.~Jimbo, and T.~Miwa,
{\em `Method for generating discrete soliton equations III},
Journ.~Phys.~Soc.~Japan {\bf 52}, No.2 (1983) 388-393.

\bibitem{H} R.~Hirota,
{\em `Discrete analogue of a generalized Toda equation'},
Journ.~Phys.~Soc.~Japan {\bf 50} (1981) 3785-3791.

\bibitem{Z} A.~Zabrodin,
{\em `A survey of Hirota's difference equations'},
Theor.\,Mat.\,Fiz.  {\bf 113} (1997) 1347-1392
\href{http://arxiv.org/pdf/solv-int/9704001}{\texttt{arXiv:solv-int/9704001}}.

\bibitem{S} D.E.~Speyer,
{\em `Perfect Matchings and the Octahedron Recurrence'},
Journal~of~Algebraic~Combinatorics,
{\bf 25}, no.3 (2007) 309-348
\href{http://arxiv.org/pdf/math.CO/0402452}{\texttt{arXiv:math.CO/0402452}}.

\bibitem{FZ} S.~Fomin  and A.~Zelevinsky,
{\em `The Laurent phenomenon'},
Advances~in~Applied~Mathematics {\bf 28}, no.2 (2002) 119-144
\href{http://arxiv.org/pdf/math.CO/0104241}{\texttt{arXiv:math.CO/0104241}}.


\bibitem{DiF} P.~Di~Francesco,
{\em `Inhomogeneous loop models with open boundaries'},
J.~Phys.~A:~Math.~Gen. {\bf 38} (2005) 6091-6120
\href{http://arxiv.org/pdf/math-ph/0504032}{\texttt{arXiv:math-ph/0504032}}.

\bibitem{PZJ} P.~Zinn-Justin,
{\em `Loop model with mixed boundary conditions, qKZ equation and alternating sign matrices'},
J.~Stat.~Mech. (2007) P01007
\href{http://arxiv.org/pdf/math-ph/0610067}{\texttt{arXiv:math-ph/0610067}}.










\end{thebibliography}
\end{document}